\documentclass[journal]{IEEEtran}
\usepackage{amsmath,amsfonts,amssymb}
\usepackage{adjustbox}
\usepackage[table,xcdraw]{xcolor}
\usepackage{amsthm,mdframed,calc}
\usepackage{mathrsfs}
\usepackage{cite}
\usepackage{textcomp,stfloats,url}
\usepackage{verbatim}
\usepackage{graphicx}
\usepackage{pifont}
\graphicspath{ {./Images/} }
\usepackage{tablefootnote}
\usepackage[usestackEOL]{stackengine}
\usepackage[boxed,linesnumbered,ruled]{algorithm2e}
\usepackage[skip=5pt,font={small},singlelinecheck=off]{caption}
\usepackage{subfig,float}
\usepackage{caption}
\usepackage{cases}

\usepackage{multirow}
\usepackage[draft]{hyperref}
\usepackage[normalem]{ulem}
\usepackage{empheq}
 \usepackage{tikz,lipsum,lmodern}
\usepackage{diagbox}
\usepackage[most]{tcolorbox}
\SetKwInput{KwInput}{Input}               % Set the Input
\SetKwInput{KwOutput}{Output} 
\SetKwComment{Comment}{/* }{ */} 
\SetAlgoCaptionSeparator{.}          % set the Output
\ifCLASSINFOpdf

\else

\fi

\newtheorem{theorem}{Theorem}%[section]
\newtheorem{Proposition}{Proposition}

\newcounter{prob}
\newcommand{\problem}{%
  \refstepcounter{prob}%
  \noindent\textbf{P\theprob.}~
}

\newcommand{\probref}[1]{\textbf{P}\ref{#1}}

\begin{document}
\title{Cooperative Mitigation against Learning-Based Reactive Jammers: Analysis and SDR Validation}

\author{Soumita Hazra and J. Harshan
\thanks{The authors are with the Department of Electrical Engineering, Indian Institute of Technology Delhi, India. Email: Soumita.Hazra@ee.iitd.ac.in, jharshan@ee.iitd.ac.in.}}



\maketitle
%\vspace{-2cm}
\begin{abstract}
Motivated by recent developments in full-duplex radios, cognitive radios, and data-driven signal-processing, we propose a novel class of reactive jamming adversaries wherein the adversary transmits jamming energy on the victim's frequency band while simultaneously monitoring various energy statistics in the network to detect the presence of potential countermeasures, thereby trapping the victim.
These adversaries employ \emph{generalized energy detectors} comprising \emph{statistical detectors}, based on instantaneous and distributional energy metrics, and \emph{data-driven detectors}  employing machine-learning classifiers to learn patterns in the observed energy sequences.
Against such a strong adversary,  we propose a family of cooperative mitigation strategies wherein the victim takes assistance from a helper node, with the strategies tailored to operate under a wide range of latency requirements on victim’s messages and practical radio hardware constraints at helper node.
To provide theoretical guarantees on their efficacy, interesting \textcolor{black}{optimization} problems are formulated on the choice of their underlying parameters, followed by extensive mathematical analyses on their error performance and covertness.
To assess their practical feasibility, we implement the before-deployment and after-deployment setups on a software-defined-radio-based hardware testbed, and to evaluate their detectability on real energy observations, we collect the corresponding datasets to train and test the data-driven machine-learning classifiers employed by adversary.
Both analytical and hardware evaluations show that the proposed strategies cannot be detected with a high-probability under the generalized energy detectors used by adversary.

\end{abstract}


\section{Introduction}
\label{sec:intro}

\IEEEPARstart{J}{amming} \cite{Jamming2022} is a popular form of Denial of Service (DoS) \cite{DOS2} attack on wireless communication wherein an adversary transmits jamming energy over the victim's frequency band such that the signal-to-interference-noise ratio (SINR) at the intended destination is insufficient for reliable communication. \textcolor{black}{Although frequency hopping based anti-jamming solutions have been well studied to mitigate such attacks, they are spectrally inefficient, making them inapplicable to bandwidth-hungry next-generation networks. To circumvent these problems, Reconfigurable Intelligent Surfaces \cite{RIS_Anti_jam4} have recently enabled spectrally-efficient anti-jamming techniques, where reflective surfaces control signal direction and strength to maintain the required SINR under jamming. These solutions are effective against  jammers that passively inject jamming energy aligned with the victim's beam without monitoring the victim's response.}

With the advent of full-duplex radios (FDRs) \cite{SIC2} and cognitive radios \cite{CR2015}, a new class of jamming adversaries called reactive adversaries has gained attention. These adversaries transmit jamming energy over the victim's frequency band while simultaneously monitoring various energy statistics in the network. They monitor these statistics under the assumption that, as a countermeasure to evade the adversary’s attack, the victim may switch to another frequency to communicate with the destination, thereby altering the observed energy statistics on both the jammed band and the new band. As a result, monitoring multiple network frequencies to detect countermeasures on a new band helps the adversary to continue to disrupt the victim-destination link.
For instance, the reactive adversary in \cite{V2,lowlatency} uses an \textit{average energy detector} to detect any drop in the average energy level on the victim’s frequency band, as such a drop indicates that the victim has vacated its band. To generalize this, \cite{ISIT,TVT1} consider a reactive adversary with multi-band sensing capability to detect possible countermeasures on a new band of communication. To achieve this, they use statistical detectors that monitor the Kullback-Leibler Divergence (KLD) between the statistical distribution of received symbols before and after the attack, and the instantaneous energies of the received symbols on all the network frequencies. From the victim’s perspective, designing countermeasures that remain effective while avoiding detection by a reactive adversary is challenging. In this direction, several cooperative countermeasures have been proposed in \cite{V2,lowlatency,ISIT,TVT1}.

Although the statistical detectors employed by the reactive adversary in \cite{V2,lowlatency,ISIT,TVT1} have low-complexity rule-based 
decision mechanisms, they rely on fixed, pre-specified metrics, which could be a limitation in terms of detection capability. In contrast,
it is well-known that Machine-Learning (ML) based data-driven detectors learn all the discriminative features in the data, including features that statistical detectors cannot capture. Despite these developments in data-driven approaches, existing reactive-adversary models rely on statistical detectors and do not account for data-driven detection capabilities, leaving a fundamental gap in how reactive \textcolor{black}{behavior} is modeled. 

\subsection{Motivation and Problem Statement}

Motivated by this gap in existing reactive-adversary models, we consider reactive adversaries that employ generalized energy detectors comprising both statistical and data-driven detectors. Since such adversaries have not been modeled previously, neither there exists countermeasures that are designed to jointly address statistical and data-driven detectors, nor the countermeasures that are designed against statistical detectors \cite{ISIT,TVT1} have been evaluated under data-driven detection models. This reveals a clear gap in countermeasure design and their validation against reactive adversaries with data-driven detection capabilities. In addition to not being validated against adversaries with data-driven detectors, existing cooperative countermeasures, where the victim takes assistance from a helper node in the network, exhibit the following limitations, as also listed in Table \ref{novelty}: 
1) they overlook latency constraints on the victim's messages, which are important for mitigating jamming attacks on mission-critical networks, 
2) they assume ideal hardware at the helper node, which may not be practical,
3) they fail to support high-rate requirements at the helper node, and 
4) they lack hardware validation on software defined radio (SDR) based testbeds, limiting insight into their practical deployability. Thus, towards closing these research gaps, we pose the following research question:
\emph{How can we design cooperative countermeasures against a reactive adversary that is equipped with generalized energy detectors consisting of both statistical detectors and data-driven detectors, such that the resulting countermeasures
1) support variable latency constraints on the victim's messages,
2) use practical hardware at the helper node,
3) support high-rate requirements at the helper node, and
4) are validated using an SDR-based hardware testbed to provide insight into their practical deployability?}

\begin{tiny}
    \begin{table*}[t!]
\centering
\caption{Comparison of existing reactive-adversary models and their corresponding cooperative countermeasures with the generalized energy detectors based reactive-adversary model and the cooperative countermeasure proposed in this work.}
\label{novelty}
\begin{tabular}{|c|cccc|ccc|}
\hline
\multirow{3}{*}{\textbf{References}} & \multicolumn{4}{c|}{\textbf{Generalized Energy Detectors}} & \multicolumn{3}{c|}{\textbf{Countermeasure Design Features}} \\ \cline{2-8} 
 & \multicolumn{3}{c|}{\textbf{Statistical Detectors}}  & \multirow{2}{*}{\textbf{\begin{tabular}[c]{@{}c@{}}Data-Driven\\ Detectors\end{tabular}}} & \multicolumn{1}{c|}{\multirow{2}{*}{\textbf{Low-Latency}}} &  \multicolumn{1}{c|}{\multirow{2}{*}{\textbf{\begin{tabular}[c]{@{}c@{}}Practical\\ FDR\end{tabular}}}} & \multirow{2}{*}{\textbf{\begin{tabular}[c]{@{}c@{}}Hardware\\ Validation\end{tabular}}} \\ \cline{2-4}
 & \multicolumn{1}{c|}{\textbf{\begin{tabular}[c]{@{}c@{}}Average\\ Energy\end{tabular}}} & \multicolumn{1}{c|}{\textbf{KLD-Estimator}} & \multicolumn{1}{c|}{\textbf{\begin{tabular}[c]{@{}c@{}}Instantaneous\\ Energy\end{tabular}}} &  & \multicolumn{1}{c|}{} & \multicolumn{1}{c|}{} &  \\ \hline
\textbf{\cite{V2,V3}} & \multicolumn{1}{c|}{\ding{51}} & \multicolumn{1}{c|}{\ding{55}} & \multicolumn{1}{c|}{\ding{55}} & \ding{55}  & \multicolumn{1}{c|}{\ding{55}} & \multicolumn{1}{c|}{\ding{55}} & \ding{55} \\ \hline
\textbf{\cite{lowlatency}} & \multicolumn{1}{c|}{\ding{51}} & \multicolumn{1}{c|}{\ding{55}} & \multicolumn{1}{c|}{\ding{55}} &  \ding{55}& \multicolumn{1}{c|}{\ding{51}} & \multicolumn{1}{c|}{\ding{51}} & \ding{55} \\ \hline
\textbf{\cite{ISIT,TVT1}} & \multicolumn{1}{c|}{\ding{51}} & \multicolumn{1}{c|}{\ding{51}} & \multicolumn{1}{c|}{\ding{51}} & \ding{55} & \multicolumn{1}{c|}{\ding{55}} & \multicolumn{1}{c|}{\ding{55}} & \ding{55} \\ \hline
\rowcolor[HTML]{34FF34}
\textbf{Proposed work} & \multicolumn{1}{c|}{\ding{51}} & \multicolumn{1}{c|}{\ding{51}} & \multicolumn{1}{c|}{\ding{51}} &\ding{51}  & \multicolumn{1}{c|}{\ding{51}} & \multicolumn{1}{c|}{\ding{51}} & \ding{51} \\ \hline
\end{tabular}
\end{table*}
\end{tiny}

\begin{table*}[t]\caption{\textcolor{black}{In this novelty table, Baseline 1, Baseline 2 and Baseline 3 are used to denote the work done in \cite{V2}, \cite{lowlatency} and \cite{TVT1}, respectively. Also, $M$ denotes the modulation order of the helper node and $q$ represents the number of channel uses.}}\label{Novelty_table}
{%
\centering
\begin{scriptsize}
 \begin{adjustbox}{max width=\textwidth}
\begin{tabular}{|c|c|c|c|c|}
\hline
\textcolor{black}{\textbf{Parameters}}     & \textcolor{black}{\textbf{Baseline 1}}       & \textcolor{black}{\textbf{Baseline 2}} & \textcolor{black}{\textbf{Baseline 3}} & \textcolor{black}{\textbf{Our contribution}} \\ \hline
   \textcolor{black}{Joint spectral-efficiency (bits/sec/Hz)}  & \textcolor{black}{$0.5 (1+\log_2M)$}   &  \textcolor{black}{$0.5 (1+\log_2M)$}   & \textcolor{black}{$0.25 (1+\log_2M)$}  &  \textcolor{black}{$0.5 (0.5+\log_2M)$}     \\ \hline           
 \textcolor{black}{FDRs}  & \textcolor{black}{Impractical} &  \textcolor{black}{Practical  }& 
 \multicolumn{1}{c|}{\textcolor{black}{Impractical}} & 
 \begin{tabular}[c]{@{}l@{}}\textcolor{black}{Considers practical FDRs}\\ \textcolor{black}{(Delay-Tolerant Rate-Three-Fourth Scheme)}
 \end{tabular}   \\ \hline           
\textcolor{black}{ Strict latency constraints on victim's messages}  & \textcolor{black}{Applicable} & \textcolor{black}{Inapplicable}  & \multicolumn{1}{c|}{\textcolor{black}{Inapplicable}} & \begin{tabular}[c]{@{}l@{}}\textcolor{black}{Applicable for strict latency constraints on victim's messages}\\ \textcolor{black}{(Low-Latency Constrained Rate-Three-Fourth Scheme)} \end{tabular}   \\ \hline
      \textcolor{black}{Secret-key overhead} & \textcolor{black}{{0}}  & \textcolor{black}{{$2$  bits/block}}&\multicolumn{1}{c|}{\textcolor{black}{$1+ 2\log_2 M$  bits/block}} & \textcolor{black}{$1+ \log_2 M$  bits/block}        \\ \hline           
      \textcolor{black}{Modulation at helper} &  \multicolumn{3}{c|}{\textcolor{black}{Only $M$-PSK}}   & \textcolor{black}{Arbitrary}      \\ \hline                 \textcolor{black}{Position of helper} &  \multicolumn{3}{c|}{\textcolor{black}{Specific}}   & \textcolor{black}{Arbitrary}      \\ \hline     
       \textcolor{black}{Reliability} &\multirow{3}{*}{\begin{tabular}[c]{@{}l@{}}\textcolor{black}{Inapplicable due} \\ \textcolor{black}{to impractical FDR} \end{tabular}}  & \textcolor{black}{Inferior} &\multicolumn{1}{c|}{\textcolor{black}{Superior (Victim-friendly)}} & \textcolor{black}{Superior (Helper-friendly)}        \\ \cline{1-1} \cline{3-5} 
        \textcolor{black}{Covertness}  &    & \textcolor{black}{Superior} &\multicolumn{1}{c|}{\textcolor{black}{Inferior}} & \textcolor{black}{Inferior}       \\ \cline{1-1} \cline{3-5} 
          \textcolor{black}{Joint reliability and covertness at operating point}  &    & \textcolor{black}{Inferior} &\multicolumn{1}{c|}{\textcolor{black}{Inapplicable}} & \textcolor{black}{Superior }     \\ \hline  
         \textcolor{black}{Spectral-efficiency of helper node} & \multicolumn{2}{c|}{\textcolor{black}{$\log_2M$}}     & \multicolumn{1}{c|}{\textcolor{black}{$0.5\log_2M$}} & \textcolor{black}{$\log_2M$}       \\   \hline   
\textcolor{black}{Decoding complexity (Optimal)}  & \textcolor{black}{$\mathcal{O} (2M)$}   & \textcolor{black}{$\mathcal{O} (4M)$} &\multicolumn{1}{c|}{\textcolor{black}{$\mathcal{O} (2M)$}} & \textcolor{black}{DTRTF-   $\mathcal{O} (2M^2)$,  LLCRTF-   $\mathcal{O} (2M)$}    \\ \hline
\textcolor{black}{Decoding complexity  (Sub-Optimal)}& \textcolor{black}{$\mathcal{O} (2M)$}   & --- &\multicolumn{1}{c|}{\textcolor{black}{$\mathcal{O} (2M)$} }& \textcolor{black}{DTRTF-   $\mathcal{O} (2M)$, LLCRTF-   Not applicable}        \\ \hline
  \begin{tabular}[c]{@{}l@{}}
  \textcolor{black}{Space complexity for decoder}\end{tabular}& \textcolor{black}{$0$} & \textcolor{black}{$0$} &\multicolumn{1}{c|}{\textcolor{black}{$1$}} & \textcolor{black}{DTRTF-   $q$,  LLCRTF-   $0$}    \\ \hline
         
\end{tabular}%
  \end{adjustbox}
\end{scriptsize}

}
\end{table*}

\color{black}

\color{black}

\subsection{Contributions}
\textcolor{black}{
We consider a reactive adversary equipped with FDR, who transmits jamming energy over victim's frequency band, while monitoring network frequencies to detect countermeasures. 
The reactive adversary uses generalized energy detectors comprising statistical detectors based on instantaneous energy metrics and a KLD-estimator, and data-driven detectors that use ML-classifiers to learn discriminative features in the observed energy sequences.
Against such an adversary, we develop two classes of cooperative mitigation strategies, wherein the victim tunes into helper's frequency band, and the two nodes adopt a power-controlled uplink multiple access structure.
Given that statistical detectors have low-complexity rule-based decision mechanisms, the design principles of our proposed strategies are guided by the decision \textcolor{black}{behavior} of statistical detectors,  however, unlike existing strategies that are not validated against data-driven detectors, we validate our proposed strategies against such detectors.
We note that these strategies are designed to ensure that all the research questions posed above are answered 
(see Section \ref{SM}).
}

Our specific contributions are listed below. 
 
1) For the regime \textcolor{black}{when the latency constraints on the victim's messages are relaxed compared to the \emph{reaction-time} of the helper node, we propose an FDR based strategy referred to as the Delay-Tolerant Rate-Three-Fourth (DTRTF) scheme}, wherein the victim and the \textcolor{black}{helper node} cooperatively pour their energies on each other's frequency band to minimally impact the network's  energy statistics (see Section \ref{DTMS}). Subsequently, we formulate an \textcolor{black}{optimization} problem on the energy parameters of  DTRTF scheme to achieve both reliable and covert communication. While the \textcolor{black}{covertness} analysis for DTRTF scheme against  statistical detectors follows from the contributions in \cite{ISIT,TVT1}, the reliability analysis differs due to the multiple-access frame structure. Therefore, we present a comprehensive error analysis of the DTRTF scheme (see Section \ref{DecodingatBob}), \textcolor{black}{assuming the helper uses Phase Shift Keying (PSK) to communicate with the destination.}{\footnote{\textcolor{black}{For an adversary using an instantaneous-energy detector, PSK represents the worst-case covertness scenario due to its constant-symbol energy. In the later section, we also discuss the performance of the proposed strategies for quadrature amplitude modulation.} }

2) To cater to the case \textcolor{black}{when the victim has rigid latency constraints compared to the \emph{reaction-time} of the helper node, we propose the Low-Latency Constrained Rate-Three-Fourth (LLCRTF) scheme which does not require an FDR at the helper node}. Similar to the DTRTF scheme, we formulate \textcolor{black}{optimization} problems to ensure reliable and covert communication with the destination. While the overall framework of LLCRTF resembles DTRTF, its error analysis is inherently different, thus we present the corresponding error analysis for decoding the helper’s and victim’s symbols (see Section \ref{LLCA}).

3) \textcolor{black}{ Using the reliability and the \textcolor{black}{covertness} analyses of the DTRTF and the LLCRTF schemes against statistical detectors,} we present low-complexity algorithms to solve the \textcolor{black}{optimization} problems on their energy parameters. Subsequently, using these solutions, we show that our schemes can facilitate reliable communication for the victim while maintaining the required energy statistics (see Section \ref{CD}). \textcolor{black}{We further use simulations to compare the individual error performance of the helper and the victim under the proposed schemes with that of existing cooperative countermeasures designed for adversaries equipped with statistical detectors.}  Our results indicate that, while such countermeasures are more victim-friendly, the proposed schemes are more helper-friendly in terms of error performance (see Section \ref{SR}). \textcolor{black}{Lastly, we present a generalized formulation of the proposed strategies by considering QAM at the helper node, arbitrary helper placement with respect to the victim and destination, and \textcolor{black}{weighted \textcolor{black}{optimization} settings to study the corresponding reliability and covertness performance under different weighting factors (see Section \ref{Robustness}).} }

\textcolor{black}{4) To demonstrate the practical feasibility of the proposed countermeasures, we implement both the before-countermeasure and after-countermeasure setups on an SDR-based hardware testbed. 
Using this setup, we collect real energy sequences for both configurations and  use them to train the data-driven detectors at adversary implemented using supervised and unsupervised ML-classifiers. Finally, using the trained classifiers, we evaluate the data-driven detection capability at adversary and show that the proposed countermeasures cannot be detected with a high-probability under ML-driven detection mechanisms, \textcolor{black}{ both when countermeasure samples are unavailable and available for training.}
This hardware evaluation complements the statistical-detector analysis by demonstrating covertness under data-driven detectors trained on real energy sequences (see Section \ref{Hardwar_and_ML}).}

\textcolor{black}{ Overall, the novelty of this work is summarized in Table \ref{novelty}. 
Specifically, Table \ref{Novelty_table} presents a quantitative comparison of our contributions with prior works \cite{V2, lowlatency, ISIT, TVT1}.
These tables show that our adversarial model and cooperative strategies encompass a broader feature set.}  Our contributions stem from developments in  FDRs \cite{SIC2}, cognitive radios with FDRs \cite{FDCR2} and jamming attacks on FDRs \cite{FDJ1,FDJ2}. 
\textcolor{black}{A rich set of literature on anti-jamming techniques exists, for instance, reconfigurable intelligent surfaces (RIS) based techniques \cite{RIS_Anti_jam3,RIS_Anti_jam4,RIS_R}, where \cite{RIS_R} proposes an active RIS-assisted method using a Stackelberg game formulation, and \cite{R1} proposes a graph attention network-driven hierarchical learning framework for UAV communications.
Such techniques are applicable in scenarios where topology adaptation, deployment \textcolor{black}{optimization}, and beamforming-based mitigation can be effectively exploited, whereas the proposed countermeasures are designed for close-proximity omnidirectional jamming scenarios with fixed receiver positions.} A preliminary version of this work appeared in \cite{NCC}, which introduced a basic DTRTF variant without analysis.

\begin{figure}%
\vspace{-0.4cm}
\begin{center}
    \subfloat[]
{\includegraphics[scale=0.29]{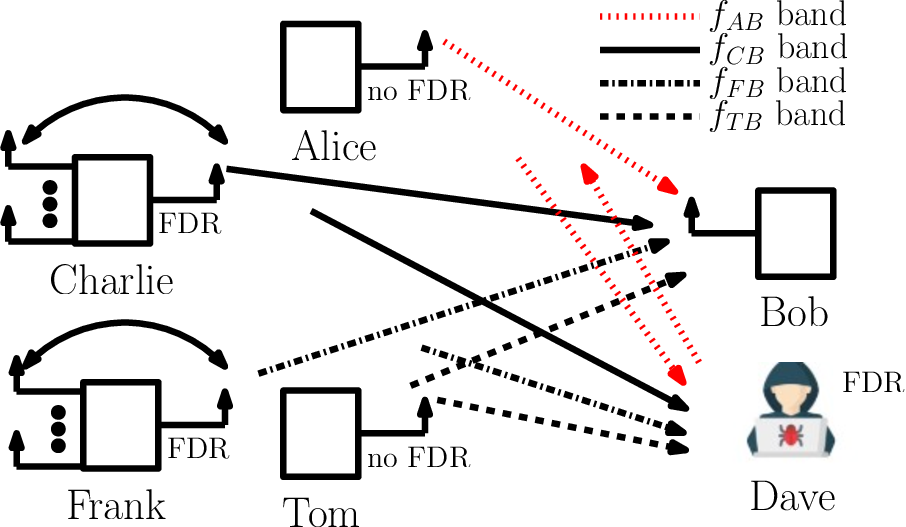}\label{SMdiag}
}
    \subfloat[]{
{\includegraphics[scale=0.29]{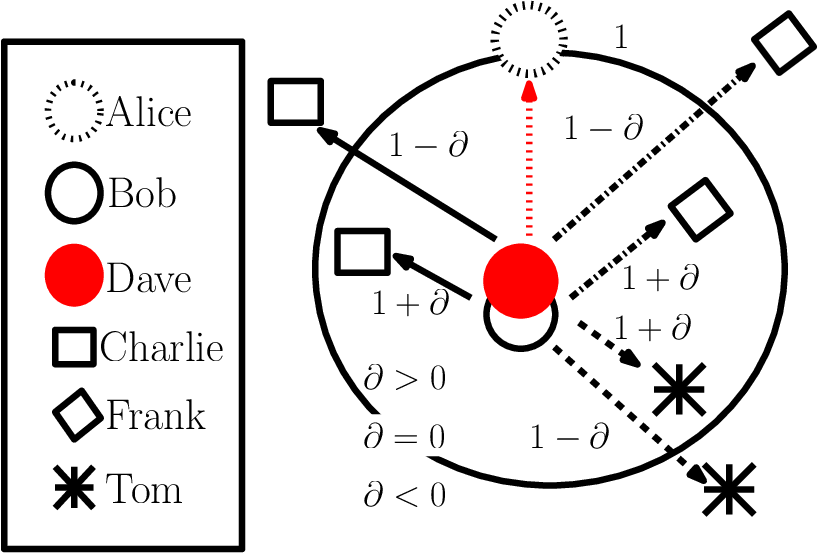}\label{Geometry}}
    \label{geometry}}%
    \vspace{-0.1cm}
    \caption{\textcolor{black}{(a)  Network model depicting uplink communication between the UEs and the base station. The reactive adversary, namely, Dave, is seen injecting jamming energy on the victim Alice, while monitoring all network frequencies.
    (b) Geometry capturing the relative positions of the UEs with respect to Bob, wherein, $\partial$ captures the variable for large-scale fading. }}
    \label{Systemmodel}
\end{center}
\vspace{-0.3cm}
\end{figure}

\color{black}

% \clearpage

\section{Network Model} \label{SM}

We consider a single-cell based wireless network, as shown in Fig. \ref{Systemmodel}, wherein a number of user equipments (UEs) communicate with the base station through their uplink channels on orthogonal frequency bands. We assume that the network is crowded and heterogeneous, i.e., all the uplink frequencies of the base station are allocated to the UEs, and the UEs have arbitrary rate requirements for their uplink channels. One such instantiation of the network consists of four UEs, namely, Alice, Charlie, Tom and Frank, that communicate with the base station, namely, Bob \textcolor{black}{over a bandwidth of $W$ Hertz} on the \textcolor{black}{center} frequencies $\!f_{AB}$, $\!f_{CB}$, $\!f_{TB}$ and $\!f_{FB}$, respectively. The \textcolor{black}{average energies of the UEs per channel use are respectively denoted by $E_{A}, E_{C}, E_{T},$ and $E_{F}$}. 
\textcolor{black}{
We assume that the uplink channels of Alice, Charlie, Tom and Frank are frequency-flat and are quasi-static, with a coherence-time of $\!\tau_A$, $\!\tau_C$ ,
$\!\tau_T$ and $\!\tau_F$ time-slots, respectively.} Henceforth, one time-slot refers to the time taken for one channel use, which is $\frac{1}{W}$ seconds. Among the four UEs, Alice has critical information to communicate with Bob, which may be delay-tolerant or have strict low-latency constraints to reach the base station. Therefore, she is a potential victim of a DoS attack from an adversary. Also, given that coherent communication for Alice might result in pilot contamination attacks from adversaries, we assume that she communicates with Bob using non-coherent On-Off keying (OOK) modulation. 
\textcolor{black}{This implies that Bob decodes Alice's information symbols without using the instantaneous channel state information (CSI).} 
Conversely, Charlie, Tom and Frank, which are not directly under adversarial attack, have strict high-rate requirements on their messages, and they communicate with Bob using coherent modulation schemes.
To support coherent demodulation on $f_{CB}$, $f_{TB}$ and  $f_{FB}$ bands, we assume that Charlie, Tom and Frank broadcast pilot symbols \cite[Section 6.4.1.1]{3gpp} followed by information symbols within the coherence-time, such that \textcolor{black}{$\tau_{C} = \tau_{C}^{p}+\tau_{C}^{d}$, $\tau_{T} = \tau_{T}^{p}+\tau_{T}^{d}$ and $\tau_{F} = \tau_{F}^{p}+\tau_{F}^{d}$, where $\tau_{C}^{p}$, $\tau_{T}^{p}$ and $\tau_{F}^{p}$, denote the number of time-slots for pilots, and $\tau_{C}^{d}$, $\tau_{T}^{d}$ and $\tau_{F}^{d}$, denote the number of time-slots for data, by Charlie, Tom and Frank, respectively. Here $\tau_{C}^{p}$, $\tau_{T}^{p}$ and $\tau_{F}^{p}$ are chosen such that the corresponding CSI are estimated with reasonable accuracy \cite{PCC1}.}   \textcolor{black}{Also, Charlie and Frank have  a requirement of high spectral-efficiency on their uplink frequencies, thus, they are equipped with practical FDRs with multiple receive antennas and single transmit antenna, however, there is no such requirement at Tom, therefore, he is not equipped with an FDR.}

\subsection{Threat Model} \label{AM}

\textcolor{black}{In the above network, we also assume the presence of a reactive adversary, namely Dave, with single transmit antenna, that intends to execute a DoS threat on the critical messages of Alice. We assume that Dave, who is equipped with an FDR, has the knowledge of the uplink frequency of Alice. As a result, he transmits jamming energy over the $f_{AB}$ band and also monitors the various frequency bands in the network for any possible countermeasures from Alice.}\footnote{\textcolor{black}{Dave is equipped with a perfect FDR, implying perfect SIC. This is to assume a worst-case scenario for covertness.}} \textcolor{black}{\textcolor{black}{To detect the presence of countermeasures, Dave employs a generalized energy detection framework that includes both statistical detectors and data-driven detectors. The statistical component of Dave’s generalized energy detector is as follows. First, on each frequency band, he compares the statistical distribution of the energies of received symbols before and after the attack using a KLD estimator. Second, he monitors the instantaneous energy of the  received symbols across different frequencies by using his knowledge of the modulation formats used by the UEs and the CSI that he obtains by listening to the pilots within each coherence block. \textcolor{black}{In addition to these statistical detectors, Dave also uses data-driven detectors that rely on supervised ML-classifiers, namely random forest, multi-layer perceptron, and convolutional neural network, along with an unsupervised ML-classifier, namely isolation forest, trained on observed energy sequences to learn discriminative patterns that may reveal the presence of a countermeasure.}\footnote{\textcolor{black}{The scope of the considered threat model is not limited to ML-based implementations, as the reactive jammer employs both statistical and data-driven detectors to detect potential countermeasures, and can thus be viewed as a broader class of reactive adversaries that make decisions based on the observed sequences.}}}}
\textcolor{black}{Subsequently, on detecting variations in the generalized energy statistics on the $f_{AB}$ band, Dave suspects a possible countermeasure and pours all his jamming energy randomly on one of the other bands. Conversely, if he detects variations in the energy statistics on any other band, he again suspects a countermeasure, and pours all his jamming energy on that band. Consequently, in both scenarios, the communication of other UEs with Bob is disrupted,  
thus, countermeasures should ensure that the probability of getting detected is minimum, and Dave is engaged only to the $f_{AB}$ band.}

\color{black}
\subsection{Overview of the Proposed Strategies}

To assist reliable and covert communication for Alice, we propose a family of cooperative mitigation strategies wherein Alice takes the assistance from a nearby helper node, \textcolor{black}{who is already using coherent-modulation on its band. First, we ask Alice to move to the frequency band of helper node and transmit her information symbols on the time-slots when the helper is communicating its information.} Subsequently, \textcolor{black}{we ask the incumbent user of that frequency band to listen to her information symbols using an FDR and forward the same to Bob in a different time-slot provided the victim's latency constraints can be met.} 
\textcolor{black}{We note that this cooperation between Alice and the helper takes place only in the data transmission phase, while the pilot transmission phase remains unaltered.} \textcolor{black}{In our network model, we can choose either Charlie or Frank as the helper node since both are equipped with FDRs and they are capable of listening to Alice's information symbols while transmitting their information symbols. Henceforth, \textcolor{black}{in this paper, we denote the frequency band of the helper, the average energy of the helper, the number of time-slots in a coherence-block, number of pilots per coherence-block, number of information symbols per coherence-block of the helper channel by $f_{HB}, E_{H}, \tau_{H}, \tau^{p}_{H}$ and $\tau^{d}_{H}$, respectively, where $f_{HB} \in \{f_{CB}, f_{FB}\}$, $E_H \in \{E_C, E_F \}$, $\tau_H \in \{\tau_C, \tau_F \}$, $\tau^{p}_H \in \{\tau^{p}_C, \tau^{p}_F \}$ and $\tau^{d}_H \in \{\tau^{d}_C, \tau^{d}_F \}$}.}

\textcolor{black}{In contrast to prior contributions against reactive adversaries with only statistical detectors,} our model considers practical FDRs at the helper node, which incur a delay of several time-slots, to achieve reasonable accuracy in its self-interference cancellation (SIC) modules. \textcolor{black}{To formally capture this delay metric, earlier referred to as \emph{reaction-time} of FDRs},\footnote{\textcolor{black}{We note that the reaction-time of the helper can be any positive real number, however, the particular time-slot at which the helper incorporates the victim's messages into its information symbols, in the form of energy and phase modifications, will be an integer.}} we define \textcolor{black}{$n_{fd}\! \in \!\mathbb{Z}_{+}$} as the maximum number of time-slots required by the helper node to keep the residual interference $\rho$ of his SIC block, below a certain threshold, denoted by $\rho_{th}$, \textcolor{black}{needed for reliably decoding Alice's information symbols.} \textcolor{black}{Given that pilot symbols are transmitted within a coherence-block, Alice's decoded information symbol cannot always be incorporated in the helper's frame right after $n_{fd}$ time-slots. With $\tau_{H}$ time-slots for a coherence-block, it is straightforward to verify that Alice's information symbol can be incorporated with a delay of $n_{fr}\! \triangleq \! (\lceil \frac{n_{fd}}{\tau_{H}}\rceil \!+\! 1) \tau_{H}$ time slots. Thus, $n_{fr}$ characterizes the effective delay offered by the FDR at the helper node.} Furthermore, given that Alice may have low-latency requirements on her messages, we define the parameter $m\! \in\! \mathbb{Z}_{+}$, to capture the maximum number of time-slots for which Bob can wait to decode Alice's symbol. Thus, to cater to diverse requirements on the delay offered by the helper node, and the low-latency constraints on Alice, we design mitigation strategies that are applicable for different combinations of $m\! \geq\! 0$ and $\textcolor{black}{n_{fr}}\! >\! 0$.
For $m \geq \textcolor{black}{n_{fr}}$ and $m \!\neq\! 0$, i.e., when the delay introduced by the helper node to decode Alice's bits is less than or equal to the maximum number of time-slots for which Bob can wait for Alice's bits, we propose the Delay-Tolerant Rate-Three-Fourth (DTRTF) scheme. On the other hand, for $m\! < \!\textcolor{black}{n_{fr}}$ and $m \!\geq\! 0$,\footnote{Note that $m=0$ represents the case when Bob wants to decode Alice's messages instantly upon observing the received symbols.} we propose the Low-Latency Constrained Rate-Three-Fourth (LLCRTF) scheme. 
\textcolor{black}{We note that these mitigation strategies are designed by preserving the instantaneous and distributional energy metrics required by statistical detectors at Dave, we later demonstrate that they cannot be detected with a high-probability under  data-driven detectors that employ ML-classifiers.}
\textcolor{black}{Also, these strategies are applicable for quasi-static channels with arbitrary coherence-time as long as $\tau_H^p$ is sufficient enough to achieve coherent communication, and in such a case, the minimum coherence-time is $\tau_H^p\! +\! 1$.} \textcolor{black}{Suppose $\tau_H^p=1$, which captures a quasi-static channel with coherence-time 2 time-slots, Bob uses one pilot symbol to obtain CSI, and in such a case, the obtained CSI would be inaccurate resulting in degraded error performance.}

\section{DTRTF Mitigation Strategy} \label{DTMS}
In this section, we propose the DTRTF scheme,  to cater to the case when $m\geq \textcolor{black}{n_{fr}}$. This strategy allows Bob to decode Alice's bits with a delay of $\textcolor{black}{n_{fr}}$ time-slots. 
\textcolor{black}{Owing to the fact that the pilot transmission phase remains unaltered and the data transmission phase undergoes changes in the proposed strategies, we only present the data transmission phase of the DTRTF scheme. As shown in Fig. \ref{nfcbdiag}, we  exclude the pilots and only focus on the time-slots for data transmission to aggregate a set of $2n$ symbols on the helper's band. Here $n =\big \lceil \frac{n_{fd}}{\tau_{H}} \big \rceil \tau^{d}_{H}$ refers to the number of helper's  symbols for which Alice's decoded symbols cannot be incorporated due to the effective delay $n_{fr}$ of the FDR.} Therefore, in the first $n$ time-slots, Alice and the helper form an uplink multiple access channel (MAC) to transmit their symbols using a portion of their energies over $f_{HB}$ band. Meanwhile, they cooperatively pour their remaining energies over $f_{AB}$ band to generate a pseudorandom OOK sequence using a preshared key. In the subsequent $n$ time-slots, the helper transmits a modified version of his symbols on $f_{HB}$ band by incorporating Alice's bits that were decoded $n$ time-slots ago using his FDR. Meanwhile, Alice takes complete charge of transmitting dummy OOK symbols over $f_{AB}$ band. 
Note that the helper transmits his symbols in all the $2n$ time-slots, whereas Alice transmits her symbols only in the first $n$ time-slots. As a result, the DTRTF scheme can be viewed as a helper-friendly scheme where the helper retains his transmission rate while Alice takes a hit on her transmission rate by 50\%. For the same reason, we refer to our strategy as the rate-three-fourth scheme since the total number of symbols transmitted over the two frequency bands across $2n$ time-slots is three-fourth of the number of transmitted symbols without any countermeasure. Also, we refer to our scheme as delay-tolerant as Bob needs to wait for at least $n$ time-slots to decode Alice's bits.

Next, we will explain the \textcolor{black}{signaling} methods over the $f_{HB}$ and $f_{AB}$ bands in detail. 
\begin{figure}[t]
   \begin{center}
       {\includegraphics[scale=0.30]{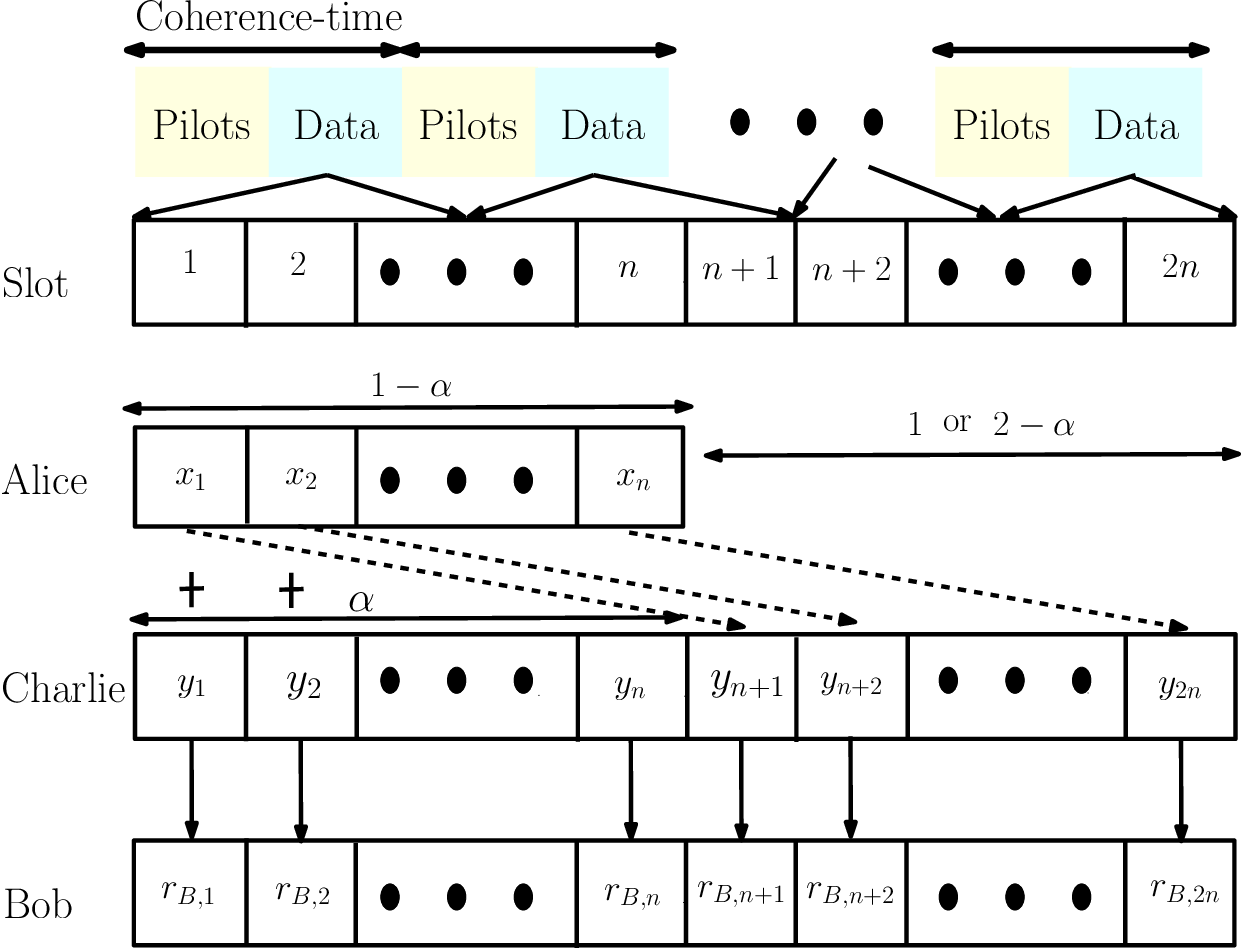}}
   \end{center}  
\caption{\textcolor{black}{Depiction of the frame structure and the corresponding energy levels over $f_{HB}$ in the DTRTF scheme. \textcolor{black}{Only the data transmission phase is captured for exposition.}}}
\label{nfcbdiag}
    \end{figure}

 \begin{figure}
 \vspace{-0.4cm}
 \begin{center}
     \subfloat[]
{
\includegraphics[scale=0.26]{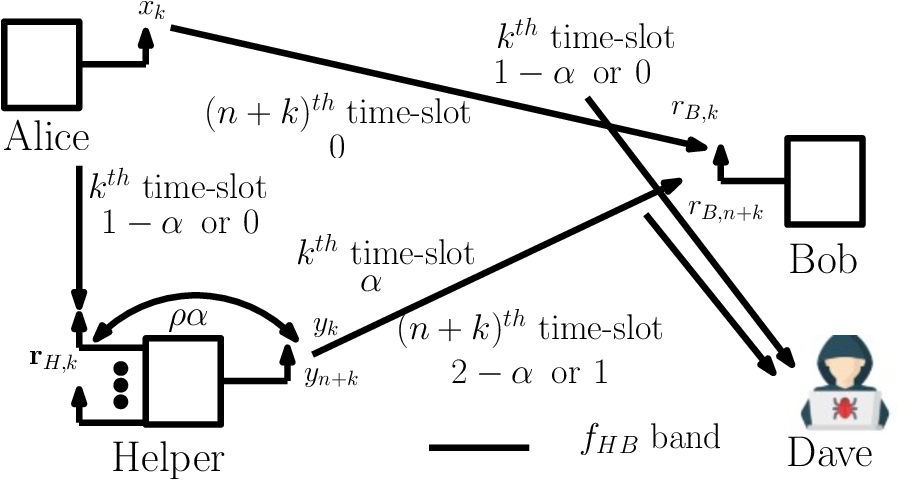}
    \label{fcbkn}}%
    \hfil
    \subfloat[]
{
\includegraphics[scale=0.26]{SM_1_new.eps}
    \label{fabntimeslots}}%
 \end{center}
    \vspace{-0.3cm}
    \caption{ 
    \textcolor{black}{Representation of the transmission during the $k^{th}$ and $(n+k)^{th}$ time-slot over (a) $f_{HB}$ band, and (b) $f_{AB}$ band.}
    }
    \label{fcb2diags}
\end{figure}
\subsection{\textcolor{black}{Signaling} over $f_{HB}$ Band}

As shown in Fig. \ref{fcbkn}, in $k^{th}$ time-slot, where $k \in [n]$, Alice communicates her bits modulated using OOK, which is scaled by $\sqrt{(1-\alpha)E_H}$, while the helper communicates his symbol scaled by $\sqrt{\alpha E_H}$. Here, $\alpha\in (0,1)$ denotes the energy-splitting factor, which is a design parameter. The received symbol at Bob in $k^{th}$ time-slot, denoted as $r_{B,k}$, is 
$r_{B,k}\!=\!\sqrt{(1-\alpha)E_H}h_{AB,k}x_k\!+\!\sqrt{\alpha E_H}y_{k} h_{HB,k}\!+\!w_{B,k},$  
where \textcolor{black}{$h_{AB,k}\!\sim \!\mathcal{CN}(0, 1)$ is the channel between Alice and Bob that captures small-scale fading,} $x_k\!\in\! \{0, 1\}$
denotes Alice's bits, 
\textcolor{black}{$y_k \!\in \! \mathcal{S}(j+1)$, $\mathcal{S}$ denotes the complex constellation used by the helper (this could be $M$-PSK, $M$-QAM), $j\! \in \! \{0,1, \ldots, M-1\}$ denotes the symbol transmitted by the helper using a coherent modulation scheme of order $M$},
$h_{HB,k}\!\sim\! \mathcal{CN}(0, 1 \!+\!\partial)$ is the channel between the helper and Bob,  $w_{B,k}\!\sim\! \mathcal{CN}(0, N)$ is the additive white Gaussian noise (AWGN) at Bob, and the subscript $k$ denotes the $k^{th}$
time-slot. 
Since the helper is equipped with an FDR, he receives Alice's bits transmitted over $f_{HB}$, while simultaneously transmitting his symbols. \textcolor{black}{The received vector at the helper in $k^{th}$ time-slot, denoted using $\mathbf{r}_{H,k}$, is 
${\mathbf{r}_{H,k}\!=\!\sqrt{(1-\alpha)E_H}\mathbf{h}_{AH,k}x_k\!+\! \mathbf{h}_{HH,k}\!+\!\mathbf{w}_{H,k},}$
where $\mathbf{h}_{AH,k}\!\sim\! \mathcal{CN}(\mathbf{0}_{N_H}, \sigma^{2}_{AH,k}\mathbf{I}_{N_H})$ is the $N_H \times 1$ channel between Alice and the helper that captures small-scale fading, $\mathbf{h}_{HH,k}\! \sim \!\mathcal{CN} (\mathbf{0}_{N_H},\alpha \rho \mathbf{I}_{N_H})$ is  $N_H \times 1$ loop interference (LI) channel \cite{LInew} at the helper, and $\mathbf{w}_{H,k}\!\sim\! \mathcal{CN}(\mathbf{0}_{N_H}, N \mathbf{I}_{N_H})$ is the AWGN at
the helper.} 
Given that Alice is the victim node, our priority is to ensure utmost reliability in transmitting her bits to Bob. 
To achieve this, the helper decodes Alice's bit transmitted in $k^{th}$ time-slot, denoted as $\bar{x}_k$, based on $\mathbf{r}_{H,k}$.
Subsequently, owing to $n$ time-slot delay offered by his practical FDR, the helper incorporates this decoded bit into his symbol through energy and phase modifications in the $(n\!+\!k)^{th}$ time-slot. If $\bar{x}_k\!=\!1$, the helper transmits his symbol without any modifications, and the received symbol at Bob in $(n\!+\!k)^{th}$ time-slot, denoted as $r_{B,n+k}$, is
${r_{B,n+k}\!=\!\sqrt{E_H}h_{HB,n+k}y_{n+k} \!+\! w_{B,n+k},}$
where $h_{HB,n+k}\!\sim\! \mathcal{CN}(0, 1\!+\!\partial)$ is the channel between the helper and Bob, 
\textcolor{black}{$y_{n+k} \!\in\! \mathcal{S}(j+1)$},
$w_{B,n+k}\!\sim\! \mathcal{CN}(0, N)$ is the AWGN at Bob,  and  subscript $(n+k)$ denotes the $(n+k)^{th}$ time-slot. 
If $\bar{x}_k\!=\!0$, the helper transmits  his symbol with an energy of $2\!-\!\alpha$ and incorporates an additional phase shift of $\theta$, such that 
${r_{B,n+k}\!=\!\sqrt{(2-\alpha)E_H}h_{HB,n+k}y_{n+k} e^{
\iota \theta}\!+\!w_{B,n+k}.}$
It should be noted that Alice remains inactive during the $(n+k)^{th}$ time-slot over $f_{CB}$ band, while her symbols are transmitted in all the $2n$ time-slots.

\subsection{\textcolor{black}{Signaling} over $f_{AB}$ Band}
As Dave is constantly monitoring the statistical distribution on the energies of the received symbols on the $f_{AB}$ band, it is crucial to maintain OOK \textcolor{black}{signaling} in each time-slot. To achieve this, as shown in Fig. \ref{fabntimeslots}, Alice and the helper combine their remaining energies, using a preshared key over  $f_{AB}$ band, creating pseudorandom bit sequences during the first $n$ time-slots. The received symbol at Dave during the $k^{th}$ time-slot, denoted as $r_{D,k}^{AB}$, can be represented as 
$r_{D,k}^{AB}=\sqrt{2E_{A} - (1 - \alpha)E_{H}}h_{AD,k}x_{d,k} + \sqrt{(1 - \alpha)E_{H}}h_{HD,k}x_{d,k} +w_{D,k}^{AB},$
where $h_{AD,k}\sim \mathcal{CN}(0, 1)$ is the channel between Alice and Dave \textcolor{black}{that captures small-scale fading}, $x_{d,k}\in \{0, 1\}$
denotes pseudorandom bits, $h_{HD,k}\sim \mathcal{CN}(0, 1+\partial)$ is the channel between the helper and Dave, and $w_{D,k}^{AB}\sim \mathcal{CN}(0, N)$ is the AWGN at Dave. During the $(n+k)^{th}$ time-slot, the helper remains silent over $f_{AB}$ band, whereas Alice transmits a pseudorandom OOK sequence from the alphabet $\{0, \sqrt{2E_A}\}$. \textcolor{black}{We highlight that the proposed strategy is feasible as long as the average energies of Alice and the helper satisfy the relation $E_A \geq \frac{E_H}{2}$, which is necessary to ensure that Alice transmits non-zero energy on  $f_{AB}$ band in $k^{th}$ time-slot.}

\subsection{Salient Features of the DTRTF Scheme}\label{SFsub}
In this subsection, we present the salient features of the DTRTF scheme with respect to the energy statistics maintained over the $f_{AB}$ and $f_{HB}$ bands, and then propose a suitable \textcolor{black}{optimization} problem to choose  $\alpha$. 

\begin{Proposition}\label{SF}
\textcolor{black}{For DTRTF, when $\partial = 0$},\footnote{\textcolor{black}{ Robustness analyses on the reliability and covertness for $\partial \neq 0$ are discussed in the later sections.}}, for any $\alpha \in (0, 1)$,
1) the statistical distribution of the energies of received symbols on $f_{AB}$ band is identical before and after countermeasure.
2) The average energy of the received symbols on $f_{CB}$ band  for $2n$ time-slots is unity, which is consistent with the before countermeasure case.
\end{Proposition}

\noindent Although the average energy of the received symbols is maintained over the $f_{HB}$ band for any $0 < \alpha < 1$, it is clear that the statistical distribution of the energies of the received symbols before and after the countermeasure is not identical because the information symbols are scaled as a function of $\alpha \in (0, 1)$.
Similarly, the instantaneous energy of the received symbols after the countermeasure is different from that before the countermeasure, and also depends on $\alpha$. 
Therefore, we must characterize the probability with which the DTRTF scheme can be detected at Dave using statistical detectors. On the reliability front, because the information symbols of Alice and the helper are scaled as a function of $\alpha$, the average error-rates of the two users is also a function of $\alpha$. \textcolor{black}{Thus, to ensure high reliability and \textcolor{black}{covertness}, we propose a problem, referred to as \probref{main_opt_problem_n_time_slots}, which aims to determine the optimal value of $\alpha$, denoted as $\alpha_{opt}^{\Omega}$, which jointly minimizes the weighted sum of $P_{Eavg}^{\Omega}$ and $P_{Davg}^{\Omega}$, where $P_{Eavg}^{\Omega}$ denotes the average probability of decoding error at Bob and $P_{Davg}^{\Omega}$ denotes the average probability of detection at Dave when using the statistical detectors. 
The weighting factors assigned to $P_{Eavg}^{\Omega}$ and  $P_{Davg}^{\Omega}$ are  denoted by $w$ and $1-w$, respectively, where $w \in (0,1)$, with $w=0.5$ serving as a neutral baseline as it provides a balanced operating point between the two metrics in the absence of any prior preference. The generalized weighted formulation enables adaptation to application-specific priorities, such that $0 \leq w < 0.5$ and $0.5 < w \leq 1$ prioritize covertness and reliability, respectively, while $w=0$ and $w=1$ correspond to no reliability and no covertness, respectively.}
 \footnote{\textcolor{black}{Unless stated otherwise, all discussions correspond to $w=0.5$. The variation of $\alpha_{opt}^{\Omega}$ with $w$ is discussed in Section \ref{weights_and_alpha}.}}

\begin{center}
\problem\label{main_opt_problem_n_time_slots}
\textcolor{black}{$\alpha_{opt}^{\Omega}= \arg \mathop {\min }\limits_{\alpha \in (0, 1)} w P_{Eavg}^{\Omega} + (1-w)P_{Davg}^{\Omega}.$}
\end{center}

To address \probref{main_opt_problem_n_time_slots}, we must derive the expressions for $P_{Eavg}^{\Omega}$ and $P_{Davg}^{\Omega}$ as a function of $\alpha$. \textcolor{black}{Additionally, these expressions also depend on the modulation scheme used by helper and the energies $E_A$ and $E_H$.  Henceforth, for the rest of the sections, we set $E_{A} = 0.5$ and $E_{H} = 1$, although all the results can be generalized for any $E_{A}, E_{H}$ with $E_{A} \geq \frac{E_{H}}{2}$. Furthermore, we pick Charlie as the helper, using 
$M$-ary Phase Shift Keying (PSK) to communicate with Bob, as it offers the worst-case covertness scenario.}\footnote{\textcolor{black}{For the impact of the proposed strategies when QAM is used, we refer the readers to Section \ref{Robustness_QAM}.}}
In the next section, we discuss the decoding strategies at Bob to derive $P_{Eavg}^{\Omega}$ for DTRTF with $\theta=\pi/M$.

\begin{figure*}
\vspace{-0.5cm}
\begin{footnotesize}
\begin{IEEEeqnarray}{rcl}\label{JMAP}
\hat{a}_k, \hat{b}_k, \hat{b}_{n+k} &=& \arg \mathop {\max }\limits_{a_k,b_k,b_{n+k}} f\big( r_{B,k},r_{B,n+k} \big| x_k=a_k, y_k=e^{-({\iota 2 \pi b_k}/{M})},  y_{n+k}=e^{-({\iota 2 \pi b_{n+k}}/{M})}, h_{CB,k},h_{CB,n+k} \big)
\end{IEEEeqnarray}
\vspace{-0.3cm}
\begin{IEEEeqnarray}{rcl}\label{JMAPre}
\hat{a}_k, \hat{b}_k, \hat{b}_{n+k}&=&\arg \mathop {\max }\limits_{a_k,b_k,b_{n+k}} \big[ f_1\big(r_{B,k} \big| x_k =a_k,y_k=e^{-({\iota 2 \pi b_k}/{M})}, h_{CB,k}  \big) \nonumber \\ & &
\textstyle\sum_{\hat{x}_k \in \{a_k, \bar{a}_k\}} P_{a_k\hat{x}_k} f_{2}\big(r_{B,n+k} \big| \bar{x}_k, y_{n+k} =e^{-({\iota 2 \pi b_{n+k}}/{M})}, h_{CB,n+k}  \big) \big]
\end{IEEEeqnarray}
\vspace{-0.3cm}
\begin{IEEEeqnarray}{rcl}\label{fx1}
f\big( r_{B,1},r_{B,2}\big| x_1 =1, y_1=e^{-({\iota 2 \pi b_1}/{M})}, y_2=e^{-({\iota 2 \pi b_2}/{M})}, h_{CB,1}, h_{CB,2} \big)=f_{11} (P_{11} f_{21} + P_{10} f_{20})
\end{IEEEeqnarray}
\vspace{-0.3cm}
\begin{IEEEeqnarray}{rcl}\label{fx0}
f\big( r_{B,1},r_{B,2}\big| x_1 =0, y_1=e^{-(c{\iota 2 \pi b_1}/{M})}, y_2=e^{-({\iota 2 \pi b_2}/{M})}, h_{CB,1}, h_{CB,2} \big)=f_{10} (P_{00}  f_{20} + P_{01} f_{21})
\end{IEEEeqnarray}  
\end{footnotesize}
\hrule
\end{figure*}
\section{Error Analysis of the DTRTF Scheme} \label{DecodingatBob}

\textcolor{black}{Recall that based on decoded bit $\bar{x}_k$, Charlie transmits his information symbol in $(n+k)^{th}$ time-slot to Bob. Thus, the probability of decoding error of Alice's bits at Charlie influences the probability of decoding error at Bob. In light of this, we first discuss the error analysis at Charlie.}

\textcolor{black}{Since Charlie is equipped with FDR, he listens to Alice's bit $x_k$ in the $k^{th}$ time-slot, while transmitting his $M$-PSK symbol. Using the received vector at Charlie ($\mathbf{r}_{H,k}$), SIC block of the FDR reduces the interference level from $\rho$ to $\rho_{th}$ thereby resulting in $\tilde{\mathbf{r}}_{H,k} = \sqrt{(1-\alpha)E_H}\mathbf{h}_{AH,k}x_k+ \tilde{\mathbf{h}}_{HH,k}+\mathbf{w}_{H,k},$ where $\tilde{\mathbf{h}}_{HH,k} \sim \mathcal{CN} (\mathbf{0}_{N_H},\alpha \rho_{th} \mathbf{I}_{N_H})$. Subsequently, Charlie sets an optimal threshold, denoted by $\tau_{opt}$, which is computed using the various energy levels used by Alice, the second order statistics of the AWGN, and the statistics of the residual interference offered by his FDR. Then, Charlie performs non-coherent energy detection by comparing $\tau_{opt}$ with the energy of the received vector, i.e., $\tilde{\mathbf{r}}_{H,k}^H\tilde{\mathbf{r}}_{H,k}$. The decision rule is as follows: (a) if $|\tilde{\mathbf{r}}_{H,k}|^2 > \tau_{opt}$, then $\bar{x}_k=1$, and  (b) if $|\tilde{\mathbf{r}}_{H,k}|^2 < \tau_{opt}$, then $\bar{x}_k=0$. For the mentioned non-coherent energy detector, the probability of decoding bit-0 as bit-1 and bit-1 as bit-0, are denoted using $P_{01}$ and $P_{10}$, respectively. Note that $P_{01}$ and $P_{10}$ inherently considers the residual interference of the FDR. Given that the probability of decoding error at Bob depends on the probability of decoding error at Charlie, we perform the error analysis at Bob by assuming that Bob has complete knowledge about $P_{01}$ and $P_{10}$.}

Now, we discuss a practical decoding strategy at Bob to retrieve both Alice's  and Charlie's symbols transmitted on $f_{CB}$ band.
Since Alice's information is transmitted in both $k^{th}$ and $(n\!+\!k)^{th}$ time-slots, Bob can use an optimal joint decoder to recover $x_k$, $y_k$, and $y_{n+k}$ from the received symbols $r_{B,k}$ and $r_{B,n+k}$. 
In such a case, the decoding metric of the Joint Maximum A Posteriori (JMAP) decoder is given by \eqref{JMAP}, 
where $f( r_{B,k},r_{B,n+k} | x_k, y_k,  y_{n+k}, h_{CB,k},h_{CB,n+k})$ is the conditional probability density function (CPDF) of $r_{B,k}$ and $r_{B,n+k}$, given $x_k$, $y_k$, $y_{n+k}$, $h_{CB,k}$, $h_{CB,n+k}$. Here, $a_k\!\in\! \{0,1\}$ and $b_k,b_{n+k}\!\in\!\{0, 1,..., M\!-\!1\}$, denote the search space of JMAP decoder.
Given the statistical independence of $r_{B,k}$ and $r_{B,n+k}$ conditioned on $\!x_k$, $\!\bar{x}_k$, $\!y_k$, $\!y_{n+k}$, $\!h_{CB,k}$, $\!h_{CB,n+k}$, the JMAP decoder reduces to \eqref{JMAPre},
where $f_1(r_{B,k} | x_k \!=\!a_k, y_k\!=\!e^{-({\iota 2 \pi b_k}/{M})}, h_{CB,k})\!$ is the CPDF of $r_{B,k}$ given $x_k$, $y_k$, $h_{CB,k}$,  and 
$f_2(r_{B,n+k} | \bar{x}_k, y_{n+k}\!=\!e^{-({\iota 2 \pi b_{n+k}}/{M})}, h_{CB,n+k})\!$ is the CPDF of $r_{B,n+k}$ given $\bar{x}_k$, $y_{n+k}$, $h_{CB,n+k}$, and $\bar{a}_k$ denotes the complement of $a_k$. 
Henceforth, for exposition, without the loss of generality, we assume $k\!\!=\!\!1$ and $n\!\!=\!\!1$ for the ease of notations in our error analysis.
Thus, ${r}_{B,1}$ and ${r}_{B,2}$ are 
received on time-slots 1 and 2, respectively.
The CPDF of ${r}_{B,1}$ given $x_1\! =\! 1$, denoted as $f_{11}({r}_{B,1}|x_1 = 1,y_1=e^{-({\iota 2 \pi b_1}/{M})}, h_{CB,1})$, and the CPDF of ${r}_{B,1}$ given $x_1 = 0$, denoted as $f_{10}({r}_{B,1}|x_1\! =\! 0,y_1\!=\!e^{-({\iota 2 \pi b _1}/{M})}, h_{CB,1})$, take the complex Gaussian structures with variance $N_{1b}\!=\!N\!+\!1\!-\!\alpha$ and $N_{0b}\!=\!N$, respectively. Similarly, the CPDF of $r_{B,2}$ given $\bar{x}_1$, $y_2$, and $h_{CB,2}$, denoted as $f_{21}(r_{B,2} | \bar{x}_1\!=\!1, y_2 \!=\!e^{-({\iota 2 \pi b_2}/{M})}, h_{CB,2})\!$ for $\bar{x}_1\!=\!1$, and $f_{20}(r_{B,2}| \bar{x}_1\!=\!0, y_2 \!=\!e^{-({\iota 2 \pi b_2}/{M})}, h_{CB,2})\!$ for $\bar{x}_1\!=\!0$, and take the complex Gaussian structure with variance $N_{0b}\!=\!N$, however, with different mean values based on $\bar{x}_1$. The overall CPDF for $x_1\!=\!1$ and $x_1\!=\!0$, 
\begin{comment}
  $f\left( r_{B,1},r_{B,2}\left| x_1 \right.=1, y_1=e^{-\frac{i 2 \pi b_1}{M}}, y_2=e^{-\frac{i 2 \pi b_2}{M}}, h_{CB,1},h_{CB,2} \right)$ and $f\left( r_{B,1},r_{B,2}\left| x_1 \right.=0, y_1=e^{-\frac{i 2 \pi b_1}{M}}, y_2=e^{-\frac{i 2 \pi b_2}{M}}, h_{CB,1},h_{CB,2} \right)$  
\end{comment} 
are given in \eqref{fx1} and \eqref{fx0}, respectively.
Here, $P_{ij}$ denotes the probability that Charlie decodes Alice's bit-$i$ as bit-$j$, where $i, j \in \{0, 1\}$ \cite{V3}.
As \eqref{fx1} and \eqref{fx0} are Gaussian mixtures scaled by Charlie’s decoding error probabilities, obtaining the overall error probability using JMAP decoder in \eqref{JMAPre} is challenging.
Thus, we propose a sub-optimal decoder called Sub-Optimal DTRTF (SODTRTF) decoder, where Bob first decodes Charlie’s symbol in time slot 1, and then jointly decodes Alice’s bit and Charlie’s second symbol in time slot 2. To decode Charlie's symbol $y_1$ in time-slot 1, Bob obtains $\hat{a}_1$, $\hat{b}_1$ given by 

\vspace{-0.3cm}
\begin{footnotesize}
\begin{IEEEeqnarray}{rcl}\label{Dtimeslot1}
\hspace{-5mm}\hat{a}_1,\hat{b}_1&=& \arg \mathop {\max }\limits_{a_1,b_1} f_1\big(r_{B,1} \left| {x}_1=a_1, y_1 \right.=e^{-({\iota 2 \pi b_1}/{M})}, h_{CB,1}  \big),
\end{IEEEeqnarray}     
\end{footnotesize}

\noindent and discards $\hat{a}_1$.
Subsequently, during time-slot 2, Bob jointly decodes Alice's bit $x_1$ and Charlie's symbol $y_2$ as 

\vspace{-0.3cm}
\begin{footnotesize}
\begin{IEEEeqnarray}{rcl}\label{D21}
\hspace{-5mm}\hat{a}_1, \hat{b}_2 &=& \arg \mathop {\max }\limits_{a_1,b_2} f_2\big(r_{B,2} \left| {x}_1=a_1, y_2 \right.=e^{-({\iota 2 \pi b_2}/{M})}, h_{CB,2}  \big).
\end{IEEEeqnarray}
\end{footnotesize}

\noindent Unlike the JMAP decoder, the decoding metrics in \eqref{Dtimeslot1} and \eqref{D21} are analytically tractable for error analysis.
For given $h_{CB,1}$ and $h_{CB,2}$, if $P_{E1}^{\Omega}$ and $P_{E2}^{\Omega}$ denote the probabilities of decoding error for time-slots 1 and 2, respectively, then the overall probability of decoding error of SODTRTF decoder, denoted as $P_{E}^{\Omega sub}$, can be upper bounded as

\begin{footnotesize}
\begin{IEEEeqnarray}{rcl}\label{PesubD}
P_{E}^{\Omega sub} \leq P_{E1}^{\Omega}+P_{E2}^{\Omega}.
\end{IEEEeqnarray}    
\end{footnotesize}

\noindent Next, we provide upper bounds on $P_{E1}^{\Omega}$ and $P_{E2}^{\Omega}$.

\vspace{-0.2cm}
\subsection{Upper Bound on $P_{E1}^{\Omega}$}
%Recall that the  received symbol at Bob during time-slot 1 ($r_{B,1}$), comprises of Alice's bit $x_1$ and Charlie's symbol $y_1$. 
%Given that Alice's bit is transmitted in both the two time-slots, Bob discards Alice's bit transmitted during time-slot 1, and only decodes Charlie's symbol $y_1$. 
With $a_1,b_1$ representing the indices jointly chosen by Alice and Charlie, the decoding metric in \eqref{Dtimeslot1} introduces an error event if $(b_1 \neq \hat{b}_1)$ and $(a_1=\hat{a}_1) \cup (a_1 \neq \hat{a}_1)$.
If the probability of this error event is denoted as $\Pr \left(\hat{b}_1\neq b_1\right)$, then the overall probability of decoding error for $y_1$, conditioned on $h_{CB,1}$, is given by

\begin{footnotesize}
\vspace{-2mm}
\begin{IEEEeqnarray}{rcl}\label{Pesumsum}
\hspace{-5mm}P_{E1}^{\Omega} = \frac{1}{2M} \textstyle\sum_{ b_1 \in \{0, 1, \ldots, M-1\}} \Pr(\hat{b}_1\neq b_1).
\end{IEEEeqnarray} 
\end{footnotesize}

Prior to calculating every term of the right-hand-side of \eqref{Pesumsum}, in Fig. \ref{Constellation_diagram_TS1}, we present a 2-dimensional  constellation diagram showing the superposition of Alice's and Charlie's information symbols in time-slot 1 for $M\!=\!4$. 
 The dots and circles correspond to the $M$-PSK constellation, for $x_1\!=\!0$ and $x_1\!=\!1$, respectively. In particular, symbols $(a_1, b_1)$ with $a_1=0$ and $a_1=1$ are shown as dots and circles, respectively. We refer to Alice’s bits and Charlie’s symbols interchangeably as $(a_1,b_1)\in\{0,1\}\times\{0,1,\ldots,M-1\}$ or as points in the constellation, since these representations are in one-to-one correspondence.
 Using this mapping, the next theorem provides an upper bound on $P_{E1}^{\Omega}$.

\begin{figure}[t!]%
\vspace{-5mm}
\begin{center}
        \subfloat[]
{\includegraphics[scale=0.27]{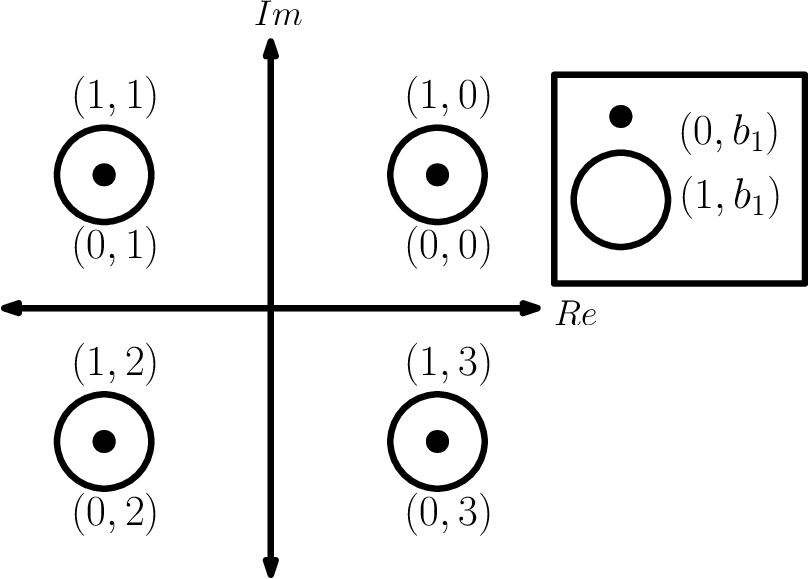}\label{Constellation_diagram_TS1}}
    \hfil
    \subfloat[]
{\includegraphics[scale=0.27]{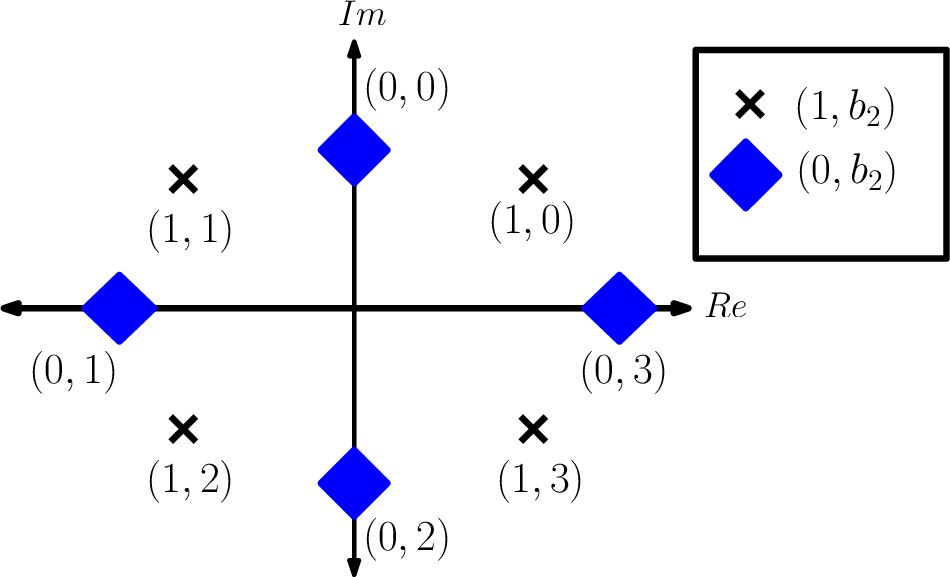}\label{Constellation_diagram_TS2}}
\end{center}
\vspace{-0.1cm}
    \caption{\textcolor{black}{(a) Constellation diagram illustrating the (a)
    symbols jointly communicated by Alice and Charlie during time-slot 1, presented in the form of $(x_1, y_1)$, (b) 
    symbols communicated by Charlie during time-slot 2, presented in the form of $(\bar{x}_1, y_2)$.}}
    \label{Constdiagram}
\end{figure}

\begin{theorem}\label{T1}
At high SNR i.e.,  $N \ll 1$, an upper bound on $P_{E1}^{\Omega}$ conditioned on $h_{CB,1}$, denoted as $P_{E1}^{\Omega ub}$, is 
$ P_{E1}^{\Omega ub}=\Pr[(1,1)\rightarrow(0,0)]+\Pr[(1,1)\rightarrow(1,0)] +\Pr[(0,1)\rightarrow(0,0)]+\Pr[(0,1)\rightarrow(1,0)]$
where $\Pr[(a_1,b_1)\rightarrow(\bar{a}_1,\bar{b}_1 )]$ denotes the probability that $(a_1,b_1)$ is decoded as $(\bar{a}_1,\bar{b}_1)$, when $(a_1,b_1)$ is transmitted by Alice and Charlie.

\end{theorem}

\begin{figure*}[ht!]
\vspace{-0.4cm}
\textcolor{black}{
\begin{footnotesize}
\begin{IEEEeqnarray}{rcl}\label{Prop2exp}
\hspace{-5mm}  P_{A1}=1-Q_{1}\big(\sqrt{N_{1b} r},\sqrt{Ns}\big),  
P_{A2}=Q\big(\sqrt{{\alpha}/{N_{1b}}}|h_{CB,1}|\big), 
P_{A3}=Q\big(\sqrt{{\alpha}/{N}}|h_{CB,1}|\big), 
P_{A4}=Q_{1}\left(\sqrt{Nr},\sqrt{N_{1b}s}\right) 
\end{IEEEeqnarray}
\vspace{-0.2cm}
   \begin{IEEEeqnarray}{rcl} \label{TH3exp1}
& & P_{A1}^{avg}=\left({N_{1b} \sqrt{\alpha}}/{(1-\alpha)^2}+1\right)^{(-1)},
P_{A2}^{avg}=\textstyle\sum_{i=1}^{3}{k_i}/\left({\frac{t_i \alpha}{N_{1b}}+1}\right),
P_{A3}^{avg} =\textstyle\sum_{i=1}^{3}{k_i}/\left({\frac{t_i \alpha}{N}+1}\right),
P_{A4}^{avg}=\left({N}/{N_{1b}}\right)^{({N_{1b}}/{1-\alpha})}P_{A1}^{avg}
\end{IEEEeqnarray} 
\end{footnotesize}
}
\hrule
\end{figure*}
 
Henceforth, for the ease of notations, we denote
 $\Pr[(1, 1)\! \rightarrow\!(0, 0)]$, $\Pr[(1,1)\! \rightarrow\! (1,0)]$, $\Pr[(0,1)\! \rightarrow \! (0,0)]$ and $\Pr[(0,1)\! \rightarrow \! (1,0)]$ by $P_{A1}$, $P_{A2}$, $P_{A3}$ and $P_{A4}$, respectively. 
In the next proposition, we provide their expressions, derived from their definitions using first principles.

\begin{Proposition} \label{PA}
For a given $\alpha$, $N$, and $h_{CB,1}$, the expressions of $P_{A1}$, $P_{A2}$, $P_{A3}$ and $P_{A4}$ are given in \eqref{Prop2exp}, where $r\!=\!{2\sqrt{\alpha} |h_{CB,1}|^2}/{(1-\alpha)^2}$,  $s\!=\!\frac{2}{1-\alpha}\!\big(ln \frac{N_{1b}}{N}\!+\! \frac{\sqrt{\alpha} |h_{CB,1}|^2}{1-\alpha} \big)\!$, and $Q_1(.,.)$ and $Q(.)$ denote Marcum-Q function of first order and Q-function, respectively.

\end{Proposition}

\subsection{Upper Bound on $P_{E2}^{\Omega}$}
With $a_1,b_2$ representing the indices jointly chosen by Alice and Charlie, the decoding metric in \eqref{D21} raises an error event, denoted by $\Delta_{(a_1,b_2)\rightarrow (\hat{a}_1,\hat{b}_2)}$, is given by, 

\begin{footnotesize}
\vspace{-2mm}
   \begin{IEEEeqnarray}{rcl} \label{delta2}
\Delta_{(a_1,b_2)\rightarrow (\hat{a}_1,\hat{b}_2)} \!\triangleq \! \frac{f_{2}\big(r_{B,2} \left | {x_1}\!=\!a_1, y_2 \right.\!=\!e^{-({\iota 2 \pi b_2}/{M})}, h_{CB,2}  \big)}{f_{2}\big(r_{B,2} \left | {x_1}\!=\!\hat{a}_1, y_2 \right.\!=\!e^{-({\iota 2 \pi \hat{b}_2}/{M})}, h_{CB,2}  \big)} \leq 1,\nonumber
\end{IEEEeqnarray}   
\end{footnotesize}

\noindent when $(\hat{a}_1 \!\neq\! a_1)$ or $(\hat{b}_2 \!\neq \!b_2)$.
Using $\Delta_{(a_1,b_2)\rightarrow (\hat{a}_1,\hat{b}_2)}$, the overall probability of decoding error for $x_1$ and $y_2$, conditioned on $h_{CB,2}$ is 

\begin{footnotesize}
\vspace{-3mm}
    \begin{IEEEeqnarray}{rcl}\label{TS2sum}
 P_{E2}^{\Omega} = \frac{1}{2M} \textstyle\sum_{a_1=0}^{1} \textstyle\sum_{b_2=0}^{M-1} \Pr\big(\Delta_{(a_1,b_2)\rightarrow (\hat{a}_1,\hat{b}_2)}\big).
\end{IEEEeqnarray} 
\end{footnotesize}

\noindent In Fig. \ref{Constellation_diagram_TS2}, for $M=4$, we visualize a 2-dimensional constellation diagram representing Charlie's symbols transmitted in time-slot 2.
%During time-slot 2, Charlie transmit symbols that belong to one of the two groups of $M$-PSK symbols: regular symbols (with unit energy) and rotated symbols (with $2-\alpha$ energy). 
Here, the black crosses and the blue diamonds denote $M$-PSK constellation, for   $\bar{x}_1=1$ and $\bar{x}_1=0$, respectively.
As we proceed, we refer to Alice's and Charlie's symbols as $(a_1, b_2) \in \{0, 1\} \times \{0, 1, \ldots, M-1\}$, or as points within the constellation, as these two share a one-to-one mapping. 
In the next theorem, we present an upper bound on $
P_{E2}^{\Omega}$ by using this one-to-one mapping.

\begin{theorem}\label{Pe2th}
At high SNR i.e., $N \ll 1$, an upper bound on $P_{E2}^{\Omega}$ conditioned on $h_{CB,2}$, denoted as $P_{E2}^{\Omega ub}$, can be expressed as 
$
P_{E2}^{\Omega ub}=[ 2 P_{11} (P_{B1}+P_{B2})+ P_{01} (1-P_{B1}) +  2 P_{00} P_{B3}+ P_{10} (1-P_{B3})]/2, 
$
\noindent where $P_{B1}=\Pr (\Delta_{(1, 1)\rightarrow (0, 0)})$, $P_{B2}=\Pr (\Delta_{(1, 1)\rightarrow (1, 2)})$ and $P_{B3}=\Pr (\Delta_{(0, 0)\rightarrow (1, 1)})$.
\end{theorem}

In the following proposition, we present expressions for $P_{B1}$, $P_{B2}$ and $P_{B3}$, which can be derived from their definitions using first principles.

\begin{Proposition} \label{PB0PB1}
For a given $\alpha$, $N$ and $h_{CB,2}$, 
$P_{B1}\!=\!P_{B3}=Q\big(\frac{|h_{CB,2}|d}{\sqrt{2 N}}\big),
P_{B2}\!=\!Q\big(\frac{|h_{CB,2}|}{\sqrt{2 N}}\big),
$
where $d=\sqrt{3-\alpha-2\sqrt{2-\alpha} cos (\pi/M)}$.
\end{Proposition}

Given that $P_{E1}^{\Omega ub}$ and $P_{E2}^{\Omega ub}$ are functions of $h_{CB,1}$ and $h_{CB,2}$, respectively, the next task is to find expectation $\mathbb{E}_{|h_{CB,1}|^2,|h_{CB,2}|^2}[P_{E}^{\Omega sub}]$. Using the linearity property of the expectation operator, we have Theorem \ref{PEtotalDTRTFtheorem}, wherein, the upper bound is derived by using \cite{Marcum,q_approx}.
\begin{theorem} \label{PEtotalDTRTFtheorem}
\textcolor{black}{
    An upper bound on average probability of decoding error at Bob using SODTRTF decoder is given by}
\begin{IEEEeqnarray}{rcl}\label{PEtotalDTRTF}
  \hspace{-4mm} \textcolor{black}{ \mathbb{E}_{|h_{CB,1}|^2,|h_{CB,2}|^2}[P_{E}^{\Omega sub}] \leq
    P_{UEavg}^{\Omega sub}\triangleq P_{avg1}^{\Omega ube}+P_{avg2}^{\Omega ube}, }
\end{IEEEeqnarray}
where $P_{avg1}^{\Omega ube}\!\!=\!\!P_{A1}^{avg}\!\!+\!\!P_{A2}^{avg}\!\!+\!\!P_{A3}^{avg}\!\!+\!\!P_{A4}^{avg},$ and
${P_{avg2}^{\Omega ube}}\textcolor{black}{=}\textcolor{black}{[ 2 P_{11} (P_{B}^{avg}+P_{B2}^{avg})+ P_{01} (1-P_{B}^{avg})}+2 P_{00} P_{B}^{avg}+ P_{10} (1-P_{B}^{avg})]/2 .$
Here $P_{B}^{avg}=\textstyle\sum_{i=1}^3 {k_{i}}/\left({({t_{i}d}/{2N})+1}\right)$
and
$P_{B2}^{avg}=\textstyle\sum_{i=1}^3 {k_{i}}/ \left({({t_{i}}/{2N})+1}\right),$ and $P_{A1}^{avg}$-$P_{A4}^{avg}$ are given in \eqref{TH3exp1}. The constants are $k_{1}\! = \! 0.168$, $k_{2} \!= \! 0.144$, $k_{3}\! = \!0.002$, $t_{1}\! =\! 0.876$, $t_{2}\! =\! 0.525$, and $t_{3}\! =\! 0.603$.
Since $P_{B1}$ and $P_{B3}$ in Proposition \ref{PB0PB1} are identical, averaging over $|h_{CB,2}|$ yields the same expression, denoted as $P_{B}^{avg}$.

% The expression of $P_{A1}^{avg}$,  $P_{A2}^{avg}$, $P_{A3}^{avg}$, $P_{A4}^{avg}$ are given in \eqref{TH3exp1}, and that of $P_{B}^{avg}$ and $P_{B2}^{avg}$  are
% \begin{IEEEeqnarray}{rcl} \label{TH3exp2}
% P_{B}^{avg}=\textstyle\sum_{i=1}^3 {k_{i}}/\left({\frac{t_{i}d}{2N}+1}\right), \quad \text{and} \quad
% P_{B2}^{avg}=\textstyle\sum_{i=1}^3 {k_{i}}/ \left({\frac{t_{i}}{2N}+1}\right),
% \end{IEEEeqnarray} 
% respectively. Also, $k_{1}\! = \! 0.168$, $k_{2} \!= \! 0.144$, $k_{3}\! = \!0.002$, $t_{1}\! =\! 0.876$, $t_{2}\! =\! 0.525$, and $t_{3}\! =\! 0.603$.
% As the expressions of $P_{B1}$ and $P_{B3}$ given in Proposition \ref{PB0PB1} are identical, thus averaging them over $|h_{CB,2}|$ yields the same result, denoted as $P_{B}^{avg}$.
\end{theorem}

To validate the derived expressions, in Fig. \ref{DTRTF_SNR}, we use Monte-\textcolor{black}{Carlo} simulations to plot the average probability of decoding error 
using the JMAP decoder (defined in \eqref{JMAP}, with probability of error denoted as $P_{Eavg}^{\Omega}$) and SODTRTF decoder (defined in \eqref{Dtimeslot1} and \eqref{D21}, with probability of error denoted as $P_{E}^{\Omega sub}$) for $N_C=2$.
Also, we plot analytical upper bound on $P_{E}^{\Omega sub}$, denoted as $P_{UEavg}^{\Omega sub}$ (given in Theorem \ref{PEtotalDTRTFtheorem}), denoted by $P_{UEavg}^{\Omega sub}$.
The curves of $P_{Eavg}^{\Omega}$ and $P_{E}^{\Omega sub}$ nearly overlap, demonstrating the near-optimality of the SODTRTF decoder.
Also, the curve of  $P_{UEavg}^{\Omega sub}$ lies above the other two curves, confirming it as a valid upper bound. All three plots exhibit the same trend, with their global minima occurring at similar values of $\alpha$. 
\begin{figure}[t]
\begin{center}
         \subfloat[]
{
\includegraphics[scale=0.11]{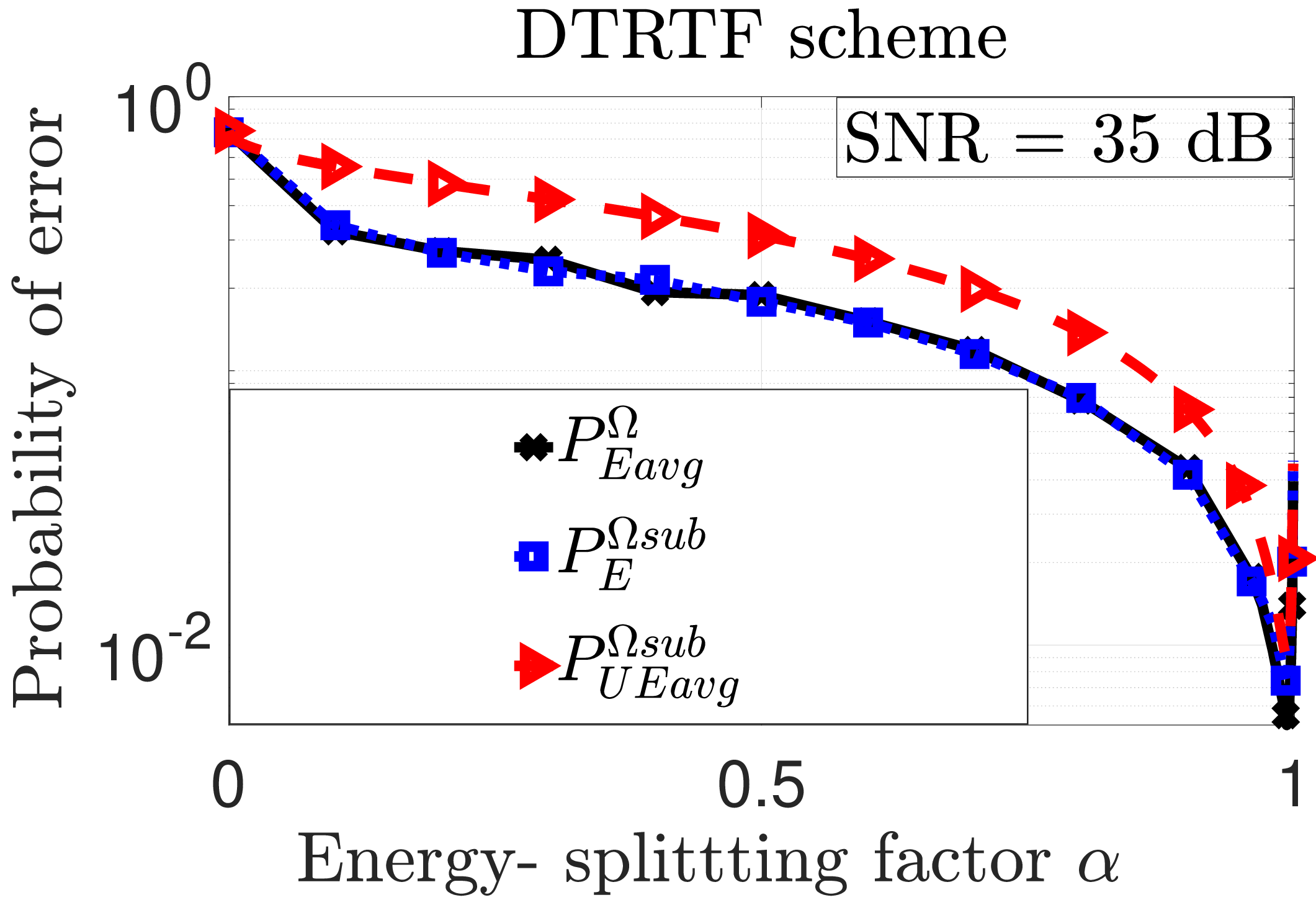}
    \label{DTRTF_SNR}}%
         \subfloat[]
     {
\includegraphics[scale=0.11]{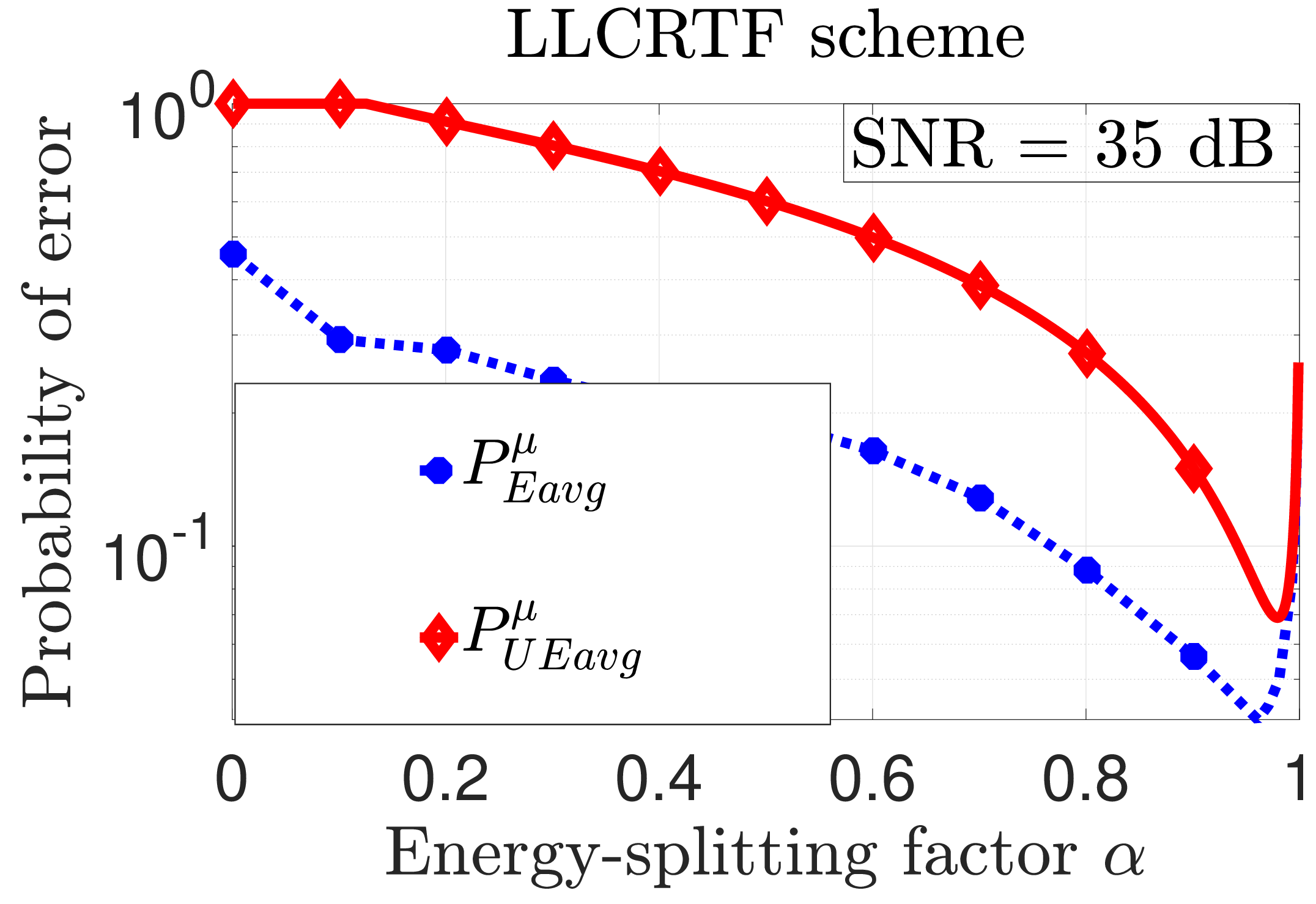}
    \label{LLCRTF_SNR}}%
    \vspace{-0.2cm}
    \caption {(a) Error rates  of optimal decoder ($P_{Eavg}^{\Omega}$) and SODTRTF decoder ($P_{E}^{\Omega sub}$), versus energy-splitting factor $\alpha$, along with the derived upper bound ($P_{UEavg}^{\Omega sub}$). (b) Error rates of LLCRTF ($P_{Eavg}^{\mu}$), with its derived upper bound ($P_{UEavg}^{\mu}$).}
    \end{center}
\end{figure}

\textcolor{black}{Although we have proposed the DTRTF scheme for $m \geq n_{fr}$ and $m \neq 0$, it can also be used for $m < n_{fr}$
and $m \neq 0$. However, in such a case, the residual interference would be higher than $\rho_{th}$, leading to increased values of $P_{10}$, $P_{01}$, thereby degrading the overall error probability at Bob. Despite the benefits of the DTRTF scheme for $m  \geq n_{fr}$, it is not applicable when $m=0$. Therefore, we present the LLCRTF scheme, which is applicable for $m<n_{fr}$ in general, and particularly relevant when $m=0$.}

\section{LLCRTF Mitigation Strategy} \label{LLCA}
This section presents the LLCRTF scheme, which is applicable when $m<n_{fr}$.
The LLCRTF scheme follows the same structure as the DTRTF scheme, however, Charlie does not embed Alice's bits into his transmitted symbols in $(n+k)^{th}$ time-slot because the latency-constraint is less than the delay offered by Charlie's FDR. In that time-slot, Charlie uniformly selects one of two $M$-PSK constellations to preserve the average energy over $f_{CB}$.
This randomness is captured using a binary random variable $x_k^d \in \{0, 1\}$: if $x_k^d=1$, Charlie transmits a unit-energy $M$-PSK symbol, otherwise, he
transmits $M$-PSK symbol with $(2-\alpha)$ energy and phase-shift of $\pi/M$.
As no embedding is performed, $n\geq 1$ can be predetermined by all nodes.\footnote{\textcolor{black}{Since LLCRTF does not require FDRs, this countermeasure is an appealing choice for low-powered internet-of-things networks.}} Except for the above mentioned changes on $f_{CB}$ band, the proposed schemes have the same \textcolor{black}{signaling} method on  $f_{AB}$ band, hence, Proposition \ref{SF} also applies to LLCRTF scheme.

Compared to DTRTF, LLCRTF reduces complexity at Charlie by eliminating FDR-based decoding of Alice's bits.\footnote{\textcolor{black}{As the LLCRTF scheme does not require a helper with an FDR, any of the three UEs, i.e., Tom, Frank and Charlie can be selected as the helper. For consistency with the DTRTF scheme, we assume Charlie acts as the helper without using his FDR capability.} } 
As no embedding is performed, Bob decodes Alice's and Charlie's symbols in  $k^{th}$ time-slot, and only Charlie's symbols in  $(n+k)^{th}$ time-slot.
Similar to DTRTF, the average probability of error depends on $\alpha$, thus 
 we present \probref{main_opt_problem_n_time_slots_latency} that determines the optimal $\alpha$, denoted by $\alpha ^{\mu}_{opt}$, which jointly minimizes the average probability of decoding error at Bob  associated with decoding $x_k$, $y_k$, and $y_{n+k}$, (denoted by $P_{Eavg}^{\mu}$) and the average probability of detection at Dave under statistical detectors (denoted by $P_{Davg}^{\mu}$). 

\begin{center}
\problem\label{main_opt_problem_n_time_slots_latency}
$\alpha ^{\mu}_{opt}= \arg \mathop {\min }\limits_{\alpha \in (0, 1)} P_{Eavg}^{\mu} +P_{Davg}^{\mu}.$
\end{center}

We note that using $r_{B,k}$, Bob jointly decodes Alice's and Charlie's symbols transmitted in $k^{th}$ time-slot, and  using $r_{B,2}$, Bob decodes Charlie's symbols transmitted in $(n+k)^{th}$ time-slot. We follow the footsteps of Theorem \ref{PEtotalDTRTFtheorem}, and present an upper bound on $P_{Eavg}^{\mu}$ in Theorem \ref{PEtotalLLCRTFtheorem}.

\begin{theorem} \label{PEtotalLLCRTFtheorem}
     An upper bound on average probability of decoding error at Bob for the LLCRTF scheme is given by
$P_{UEavg}^{\mu}\triangleq P_{avg1}^{\mu ube}+P_{avg2}^{\mu ube}$,
where $P_{avg2}^{\mu ube}=P_{B}^{avg}+P_{B2}^{avg}$, and
$P_{avg1}^{\mu ube}= ({2(P_{A1}^{avg}+P_{A2}^{avg}+P_{A3}^{avg}+P_{A4}^{avg})+P_{A5}+P_{A6}})/{2}$.
\noindent The expressions of $P_{A1}^{avg}$, $P_{A2}^{avg}$, $P_{A3}^{avg}$, $P_{A4}^{avg}$, $P_{B}^{avg}$ and $P_{B2}^{avg}$ are given in Theorem \ref{PEtotalDTRTFtheorem}, and
$P_{A5} \triangleq  \Pr[(1,1)\rightarrow (0,1)] =1-\big(\frac{N_{1b}}{N_{0b}}\big)^{\frac{N_{0b}}{\alpha-1}}$ and
$P_{A6} \triangleq \Pr[(0,1)\rightarrow (1,1)]=\big(\frac{N_{0b}}{N_{1b}}\big)^{\frac{N_{1b}}{1-\alpha}}.$

\end{theorem}

Now, to prove the validity of our results, in Fig. \ref{LLCRTF_SNR}, we plot $P_{Eavg}^{\mu}$ and $P_{UEavg}^{\mu}$, as a function of $\alpha$.  
From the plots, we observe that both curves exhibit similar trend, and their minima are around the same value of $\alpha$.

In the following section, we will discuss the accuracy with which our proposed schemes are detected by Dave.

\section{\textcolor{black}{Covertness} Analysis at Dave} \label{CD}
In this work, a countermeasure is said to be covert against a particular detector at Dave if it fails to detect the countermeasure with a high probability. 
\textcolor{black}{Based on the threat model in Section \ref{SM}, we \textcolor{black}{analyze} the covertness of our countermeasures against both statistical detectors and data-driven detectors.
Henceforth, we only discuss the covertness on  $f_{AB}$ and  $f_{CB}$ bands since the other frequency bands are not altered implicitly. 
Since the countermeasures are laid to evade statistical detectors, we restrict this section to the covertness analysis at Dave for statistical detectors, while the covertness against data-driven detectors is discussed separately in Section \ref{Hardwar_and_ML}.}
With respect to $f_{AB}$, while the first result of Proposition \ref{SF} shows the effectiveness of the proposed schemes against the KLD-estimator based detector, the instantaneous energy detector is ineffective due to the non-coherent Alice-Bob channel. Thus, we focus on the \textcolor{black}{covertness} analysis of our countermeasures when Dave uses the two detectors on $f_{CB}$ band. However,  in solving \probref{main_opt_problem_n_time_slots} and \probref{main_opt_problem_n_time_slots_latency},  deriving $P_{Davg}^{\Omega}$ and $P_{Davg}^{\mu}$ is challenging as the probability of detection of KLD-estimator based detector is analytically intractable. Hence, in the next section, we derive the expressions for average probability of detection of instantaneous energy detector, and then use it in place of $P_{Davg}^{\Omega}$ and $P_{Davg}^{\mu}$ for optimizing $\alpha$. Subsequently, in Section \ref{sec:kld_fcb}, we show that the optimized $\alpha$ yields KLD estimates that remain close to those obtained without the countermeasure.

\subsection{Instantaneous Energy Detector on $f_{CB}$ Band}
Recall that Dave possesses complete knowledge of the modulation schemes used by different network users. As a result, Dave uses the knowledge of the $M$-PSK constellation at Charlie, to monitor the envelope of the symbols on the $f_{CB}$ band using an instantaneous energy detector.
Furthermore, under a worst-case scenario for Alice wherein Dave estimates $|h_{CD,k}|$, i.e., instantaneous magnitude of the channel between himself and Charlie, to monitor the envelope. Given that unit energy $M$-PSK symbols are transmitted before implementing our proposed countermeasures, Dave raises a detection event on $f_{CB}$ band when the received instantaneous energy, after normalized by $|h_{CD,k}|$, is either below $1-\delta$ or above $1+\delta$, for some chosen $\delta >0$ 
\cite[Proposition 3]{ISIT}. Here, $\delta$ is chosen such that the probability of false alarm is bounded by a small number of Dave's choice.
When the above detector is employed against the proposed schemes, there is a non-zero probability of detection as $M$-PSK symbols transmitted by Charlie are scaled by either $\sqrt{\alpha}$ or $\sqrt{2-\alpha}$, for some $0< \alpha < 1$. Thus, we characterize the average probability of detection as a function of $\alpha$ in the following proposition.

\begin{Proposition} \label{PD}
For the instantaneous energy detector on the $f_{CB}$ band, 
(1) the average probability of detection of the DTRTF scheme is same as that of the Rate-Half Scheme (RHS) in \cite[Theorem 2]{ISIT}.
(2) the average probability of detection of the LLCRTF scheme can be derived from that of the DTRTF, as LLCRTF is a special case of DTRTF.
\end{Proposition}

From Proposition \ref{PD}, the average probability of detection for the instantaneous energy detector depends on $\alpha$. However, the choice of $\alpha$ also dictates the average probability of decoding error at Bob. Therefore, as formulated in \probref{main_opt_problem_n_time_slots}, the optimal $\alpha$ for the DTRTF scheme is the one that \textcolor{black}{minimizes} the sum of average probability of decoding error at Bob and average probability of detection at Dave. Given that we have upper bounds on the above terms, we propose to solve \probref{main_opt_problem_n_time_slots_minima}, instead of \probref{main_opt_problem_n_time_slots}, where $P_{UEavg}^{\Omega sub}$ is in \eqref{PEtotalDTRTF} and $P_{UD}^{avg}$ denotes the upper bound on the average probability of detection when using the instantaneous energy detector.

\begin{center}
\problem\label{main_opt_problem_n_time_slots_minima}
$\alpha_{opt}^{\Omega min}
= \arg\min_{\alpha \in (0,1)}
\big( P^{\Omega\mathrm{sub}}_{UE\mathrm{avg}} + P^{\mathrm{avg}}_{UD} \big)$. 
\end{center}

Given the complexities of the expressions, assessing the \textcolor{black}{behavior} of $P_{UEavg}^{\Omega sub}$ and $P_{UD}^{avg}$ with respect to $\alpha$ is non-trivial. Thus, we use simulation results, shown in the left figure of Fig. \ref{min_int_DTRTF_LLCRTF}, which reveal that the minima of $P_{UEavg}^{\Omega sub}+P_{UD}^{avg}$ occur near the intersection of $P_{avg \uparrow}^{\Omega}$ and $P_{avg \downarrow}^{\Omega}$. Here, $P_{avg \uparrow}^{\Omega}\!\triangleq \!P_{avg1}^{\Omega ube}$ and $P_{avg \downarrow}^{\Omega}\!\triangleq\! P_{avg2}^{\Omega ube}+P_{UD}^{avg}$,
are the increasing and decreasing functions of $\alpha$, respectively, where $P_{avg1}^{\Omega ube}$ and $P_{avg2}^{\Omega ube}$ are given in Theorem \ref{PEtotalDTRTFtheorem} and $P_{UD}^{avg}$ follows \cite[Theorem 2]{ISIT}. Hence, we solve \probref{main_opt_problem_n_time_slots_interesect} instead of \probref{main_opt_problem_n_time_slots_minima}.

\begin{center}
   \problem\label{main_opt_problem_n_time_slots_interesect}
$\alpha_{opt}^{\Omega in}
\ \text{such that}\
P^{\Omega}_{\mathrm{avg}\uparrow}(\alpha_{opt}^{\Omega in})
-
P^{\Omega}_{\mathrm{avg}\downarrow}(\alpha_{opt}^{\Omega in})
= 0$. 
\end{center}

 \begin{figure}
\vspace{-0.3cm}
\includegraphics[scale=0.19]{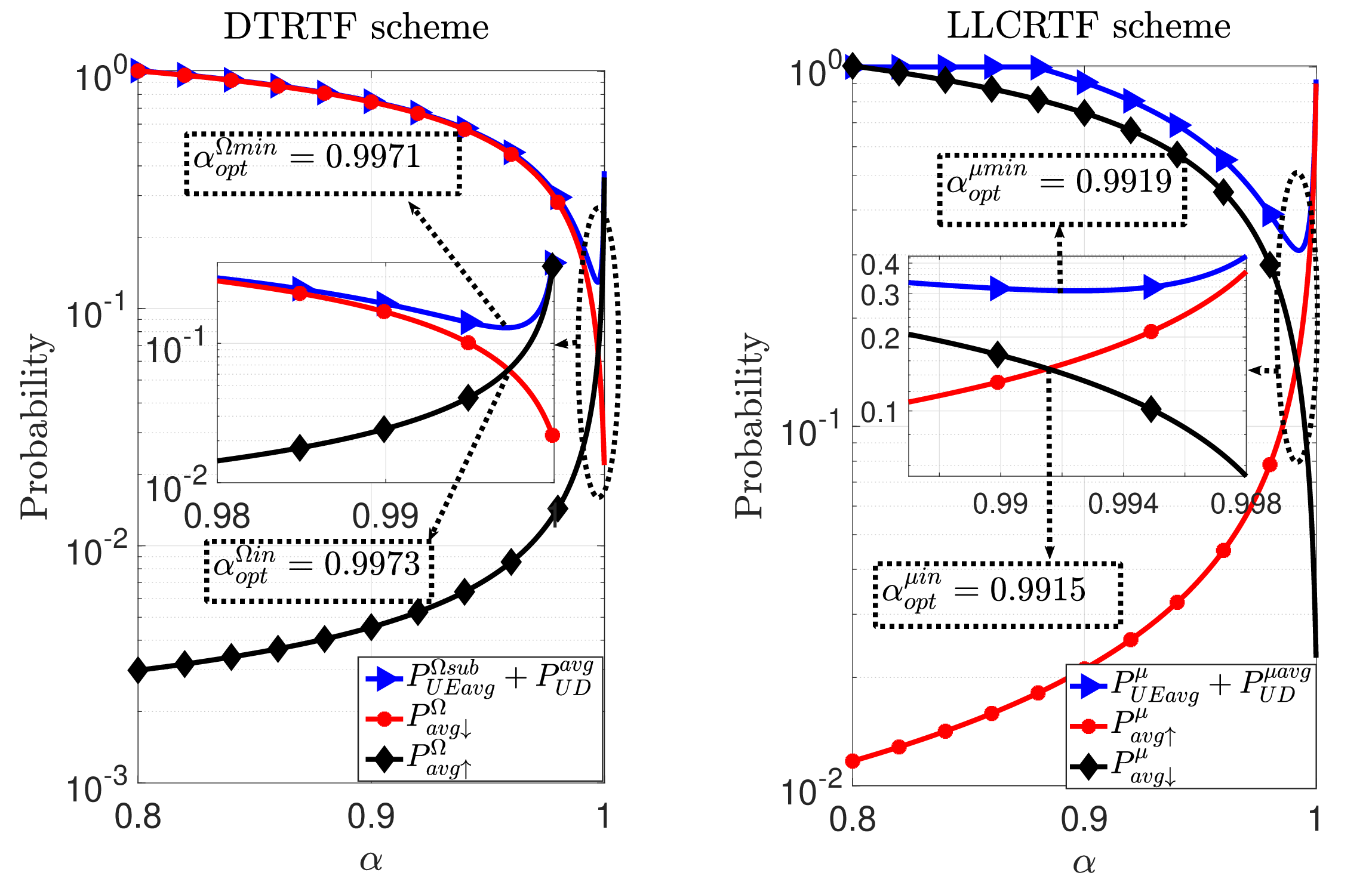}
\vspace{-0.1cm}
    \caption{\textcolor{black}{The intersection of $P_{avg \uparrow}^{\Omega}$ and $P_{avg \downarrow}^{\Omega}$ closely approximates the minimum of $P_{UEavg}^{\Omega sub}\! +\! P_{UD}^{avg}$ for DTRTF (left figure). The intersection of $P_{avg \uparrow}^{\mu}$ and $P_{avg \downarrow}^{\mu}$ closely approximates the minimum of of $P_{UEavg}^{\mu}+P_{UD}^{\mu avg}$ for LLCRTF (right figure).}}
\label{min_int_DTRTF_LLCRTF}
\vspace{-0.3cm}
\end{figure}

Similar to DTRTF scheme, as given in \probref{main_opt_problem_n_time_slots_latency}, the optimal $\alpha$ for LLCRTF scheme is the one that \textcolor{black}{minimizes} the sum of average probability of decoding error at Bob and average probability of detection at Dave. Hence, we solve \probref{main_opt_problem_n_tslatency_minima}.

\begin{center}
\problem\label{main_opt_problem_n_tslatency_minima}
$\alpha_{opt}^{\mu min}
= \arg\min_{\alpha \in (0,1)}
\big( P_{UEavg}^{\mu} +P_{UD}^{\mu avg} \big)$. 
\end{center}

Similar to DTRTF scheme, the simulation results in the right subfigure of Fig.~\ref{min_int_DTRTF_LLCRTF} show that the point of intersection of $P_{avg \uparrow}^{\mu}$ and $P_{avg \downarrow}^{\mu}$ closely approximates \probref{main_opt_problem_n_tslatency_minima}, where 
$P_{avg \uparrow}^{\mu}\!\triangleq\! P_{A5}\!+\!P_{A6}$ and $P_{avg \downarrow}^{\mu} \!\triangleq \!P_{A1}^{avg}\!+\!P_{A2}^{avg}\!+\!P_{A3}^{avg}\!+\!P_{A4}^{avg}\!+\!P_{UD}^{\mu avg}.$
Here, $P_{A1}^{avg}$, $P_{A2}^{avg}$, $P_{A3}^{avg}$ and $P_{A4}^{avg}$ are given in Theorem \ref{PEtotalDTRTFtheorem}, $P_{A5}$ and $P_{A6}$ are given in Theorem \ref{PEtotalLLCRTFtheorem}, and $P_{UD}^{\mu avg}$ follows from Proposition \ref{PD}. Hence, we solve \probref{main_opt_problem_n_latencyts_interesect} instead of \probref{main_opt_problem_n_tslatency_minima}.

\begin{center}
   \problem\label{main_opt_problem_n_latencyts_interesect}
$\alpha_{opt}^{\mu in}
\ \text{such that}\
P_{avg \uparrow}^{\mu}(\alpha_{opt}^{\mu in})
-
P_{avg \downarrow}^{\mu}(\alpha_{opt}^{\mu in})
= 0$. 
\vspace{-0.2cm}
\end{center}

\begin{tiny}
   \begin{center}
\begin{table}[h!]\caption{Solutions to \probref{main_opt_problem_n_time_slots_minima}, \probref{main_opt_problem_n_time_slots_interesect}, \probref{main_opt_problem_n_tslatency_minima}
and \probref{main_opt_problem_n_latencyts_interesect} for $N_C=1$.}\label{diff_alpha}
{%
\begin{center}
\begin{tabular}{|l|l|l|l|l|l|l|}
\hline
\multicolumn{2}{|c|}{}   & \multicolumn{2}{c|}{DTRTF} & \multicolumn{2}{c|}{LLCRTF}  \\ 
\hline
$$\text{SNR}(\text{dB})$$ & $P_{UF}^{avg}$  & $\alpha^{\Omega min}_{opt}$ & $\alpha^{\Omega in}_{opt}$ &   $\alpha^{\mu min}_{opt}$& $\alpha^{\mu in}_{opt}$  \\ \hline
\textcolor{black}{$20$ }          & \textcolor{black}{$10^{-1}$}                  & \textcolor{black}{$0.9819$}      & \textcolor{black}{$0.9903$}           & \textcolor{black}{$0.9467$}     & \textcolor{black}{$0.9560     $}  \\ \hline
$25$           & $10^{-1}$                  & $0.9933$      & $0.9970$           & $0.9796$     & $0.9842     $  \\ \hline
$30$           & $10^{-1}$                  & $0.9977$      & $0.9991$           & $0.9929$     & $0.9947     $  \\ \hline
$30$           & $10^{-2}$                  & $0.9919$      & $0.9920$           & $0.9768$     & $0.9757     $  \\ \hline
$35$          & $10^{-2}$                   & $0.9971$      & $0.9973$           & $0.9919 $ & $0.9915     $ 
\\ \hline
\end{tabular}
\end{center}
}
\end{table}
\end{center} 
\end{tiny}

\begin{figure}[ht!]
\vspace{-0.5cm}
\begin{center}
         \subfloat[]
{
\includegraphics[scale=0.11]{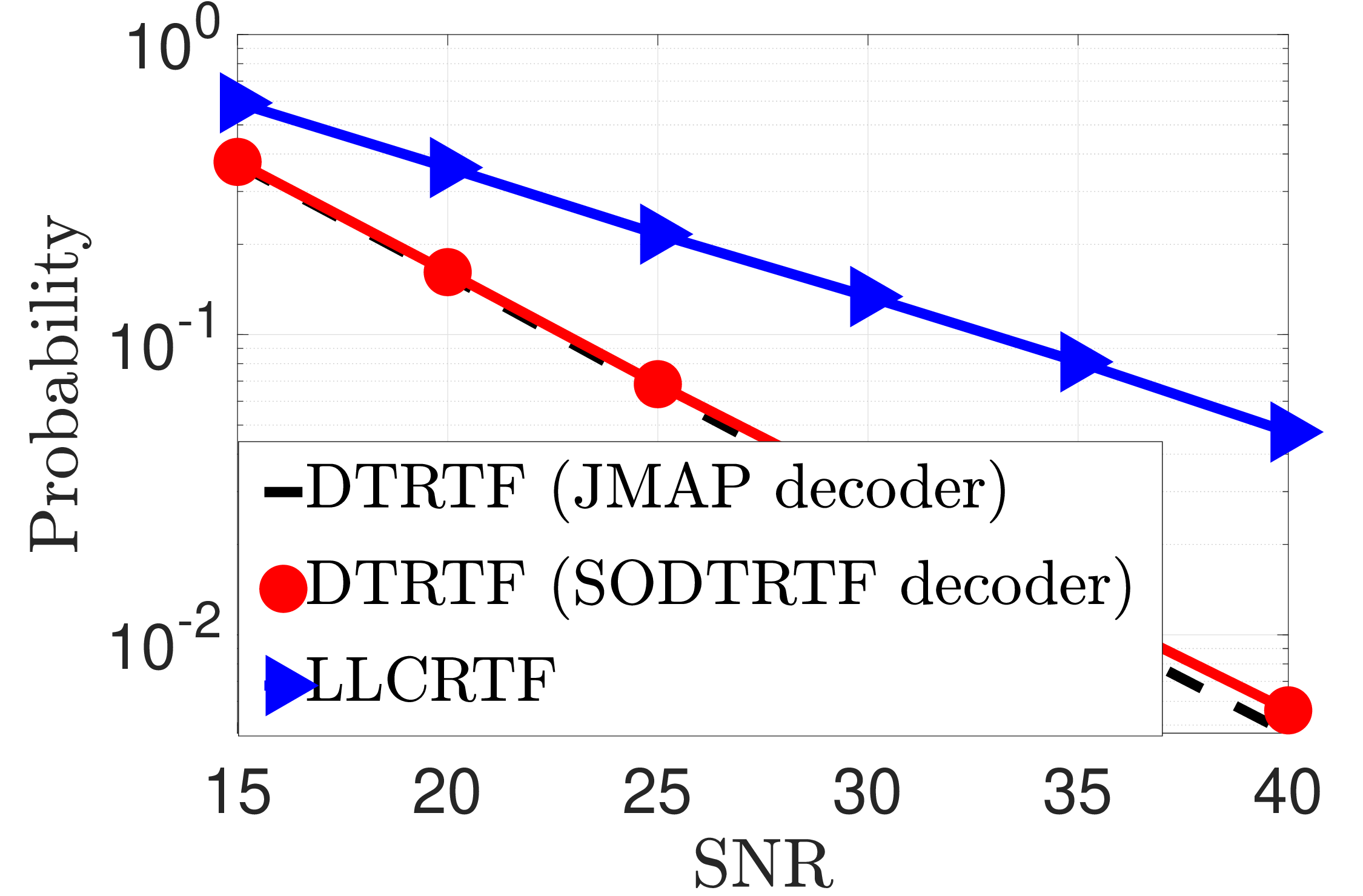}
    \label{SNR_15_40}}%
         \subfloat[]
     {
\includegraphics[scale=0.11]{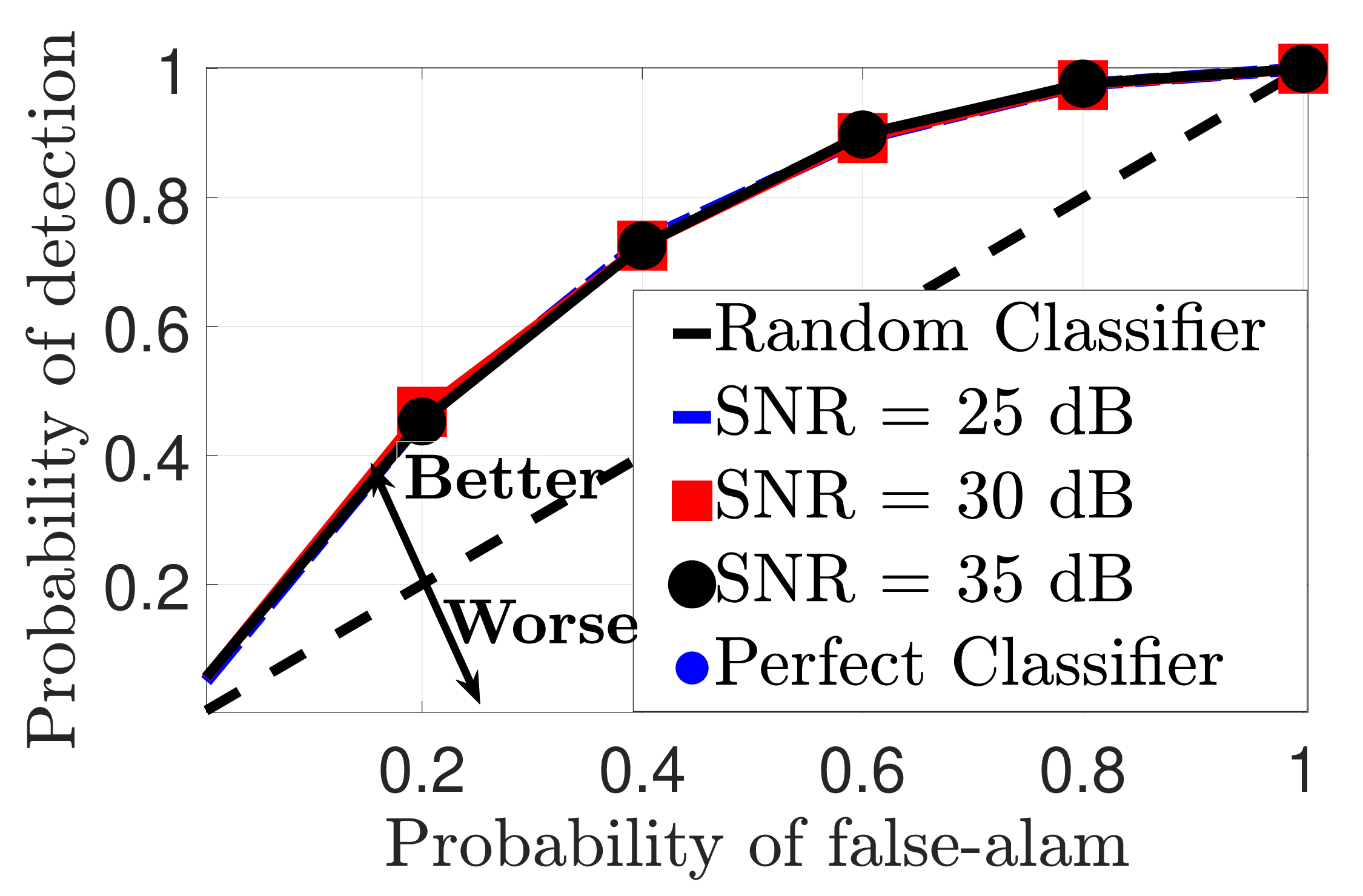}
    \label{ROC}}%
    \vspace{-0.2cm}
    \caption {(a) Sum of probability of decoding error at the destination and  probability of detection at the adversary, as a function of SNR, for the proposed schemes. (b) ROC curves for the instantaneous energy detector at Dave, for different SNRs.}
    \end{center}
    \vspace{-0.4cm}
\end{figure}

\begin{figure}[ht!]
\begin{center}
\includegraphics[scale=0.20]{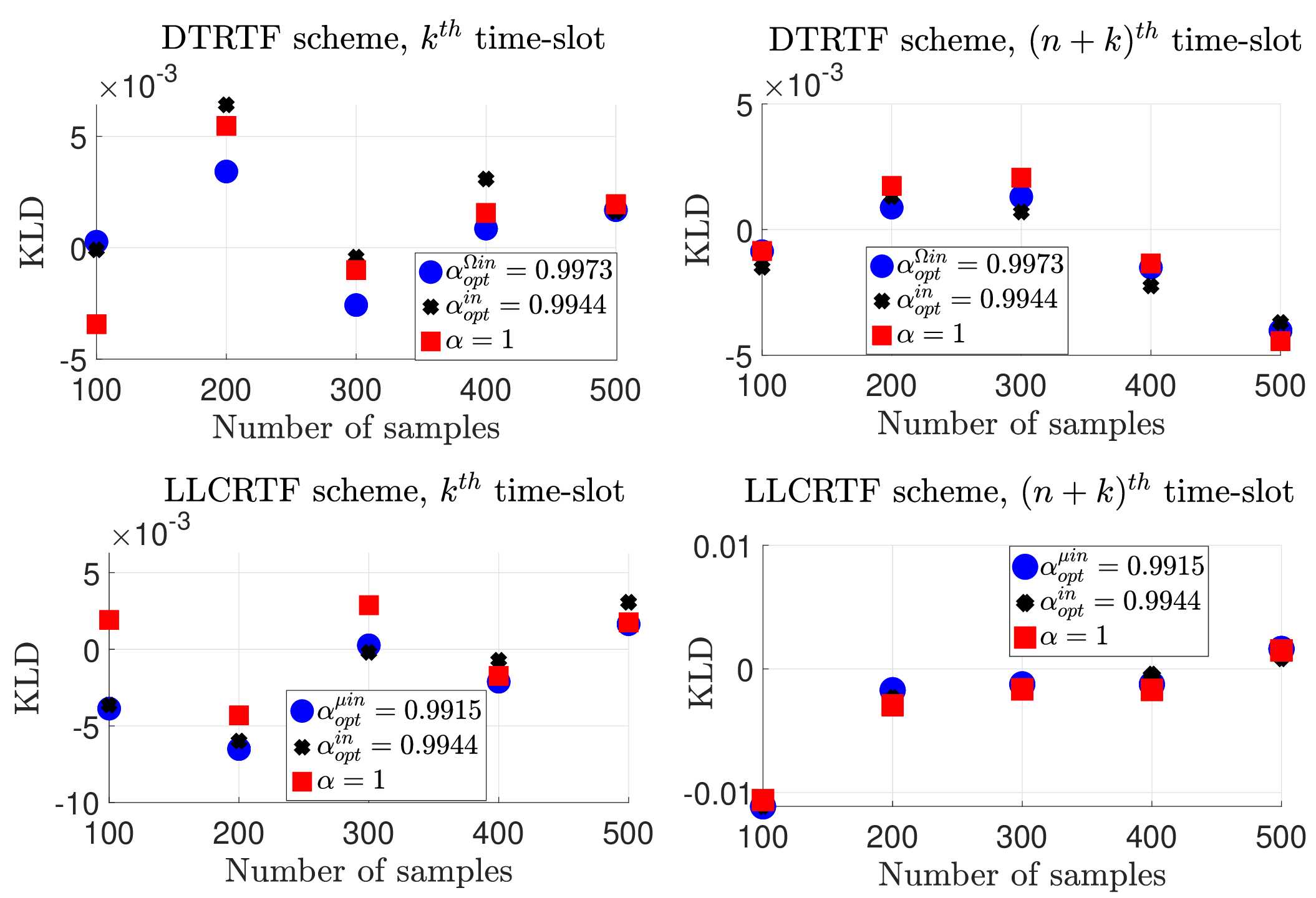}
\vspace{-0.1cm}
\caption{\textcolor{black}{Average KLD estimates on $f_{CB}$ for LLCRTF, DTRTF and RHS, for $N_{C}=1$, and SNR  $=35$ dB.}}
\label{KLDgraphfcb}
\end{center}
\vspace{-0.4cm}
\end{figure}

To validate our method of choosing $\alpha$, we numerically solve \probref{main_opt_problem_n_time_slots_minima}, \probref{main_opt_problem_n_time_slots_interesect}, \probref{main_opt_problem_n_tslatency_minima}, and \probref{main_opt_problem_n_latencyts_interesect}, and list the solutions in Table \ref{diff_alpha} as a function of SNR and the target false-positive rate, denoted by $P_{UF}^{avg}$. 
To generate these results, we use $N_C\!=\!1$, $\sigma_{AC}^2\!=\!4$, and an $\alpha$ step size of $0.0001$. 
The values in Table \ref{diff_alpha} confirm that \probref{main_opt_problem_n_time_slots_interesect} and \probref{main_opt_problem_n_latencyts_interesect} provide appropriate $\alpha$ for the proposed schemes. 
\textcolor{black}{Lastly, as our objective is to minimize the sum of average probability of decoding  error at Bob and average probability of detection at Dave for both the schemes, we examine how $\alpha$ affects this combined metric. 
\textcolor{black}{In this direction, we solve \probref{main_opt_problem_n_time_slots_minima} for SNR values between 15 dB and 40 dB, in 5 dB steps, for $N_C\!=\!3$, $\delta\!=\!0.842$, $\rho_{th}\! =\!10^{-5}$.
Then, in Fig. \ref{SNR_15_40}, we use Monte-Carlo simulations to plot the sum of the above mentioned probabilities for the optimal decoder \eqref{JMAP}, and the sub-optimal decoder \eqref{PesubD}, for DTRTF scheme.} 
Furthermore, we solve \probref{main_opt_problem_n_tslatency_minima} for various SNRs and $\delta\!=\!0.842$, and use these solutions to plot the sum of the probabilities for LLCRTF scheme.
These plots show that this sum decreases with SNR, demonstrating the efficacy of our proposed methods. We further observe that DTRTF  superior to LLCRTF, because in DTRTF, Charlie incorporates Alice's bits into his symbols  via phase and energy modification, providing an added layer of reliability to Alice's messages, which is impractical in LLCRTF due to its strict latency feature.} \textcolor{black}{Finally, to show the effectiveness of the DTRTF scheme against various values of probability of false-alarm of the instantaneous energy detector, in Fig. \ref{ROC}, we plot the Receiver Operating Characteristic (ROC) curves for different SNRs. To generate these plots, for a given average probability of false-alarm, we solve \probref{main_opt_problem_n_time_slots} for various SNRs, yielding $\alpha_{opt}^{\Omega}$, and then compute the corresponding average probability of detection for $\alpha_{opt}^{\Omega}$.
The resulting ROC curves of the countermeasure do not appear bulged outwards away from the random-classifier, thereby not providing high probability of detection for low probability of false alarm. This confirms the effectiveness of the proposed countermeasures against the instantaneous energy detector.
}

\subsection{KLD-Estimator Based Detector}
\label{sec:kld_fcb}
First, we solve \probref{main_opt_problem_n_time_slots_interesect} for the DTRTF scheme and \probref{main_opt_problem_n_latencyts_interesect} for the LLCRTF scheme with $N_C=1$ and SNR $=35$ dB, and obtain the value of $\alpha$ already optimized against the instantaneous-energy detector. 
Subsequently, we compare the statistical distribution of the received symbols at Dave before and after the countermeasure for these solutions using KLD-estimator \cite{KLD}, and present their KLD estimates in Fig. \ref{KLDgraphfcb}, along with the plots for $\alpha=1$.
In the same figure, we also plot the KLD estimates of RHS \cite{ISIT,TVT1} using its optimal value $\alpha_{opt}^{in}$. 
From Fig. \ref{KLDgraphfcb}, we infer that the KLD estimates for all the schemes are close to zero, which signifies that the statistical distribution of the received symbols on $f_{CB}$ are almost identical before and after the countermeasure for the solutions provided by \probref{main_opt_problem_n_time_slots_interesect} and \probref{main_opt_problem_n_latencyts_interesect}. In practice, Dave sets a threshold on the KLD estimate based on a tolerable false-positive rate and declares detection if the estimate exceeds it. Our results show that even with such a threshold,  the probability of detection of the proposed schemes remain small, as the KLD estimates for all schemes are close to those for $\alpha = 1$.

\color{black}

\begin{figure*}[!t]
\vspace{-0.4cm}
\subfloat[]
{\includegraphics[height=3.5cm,width=0.50\columnwidth]{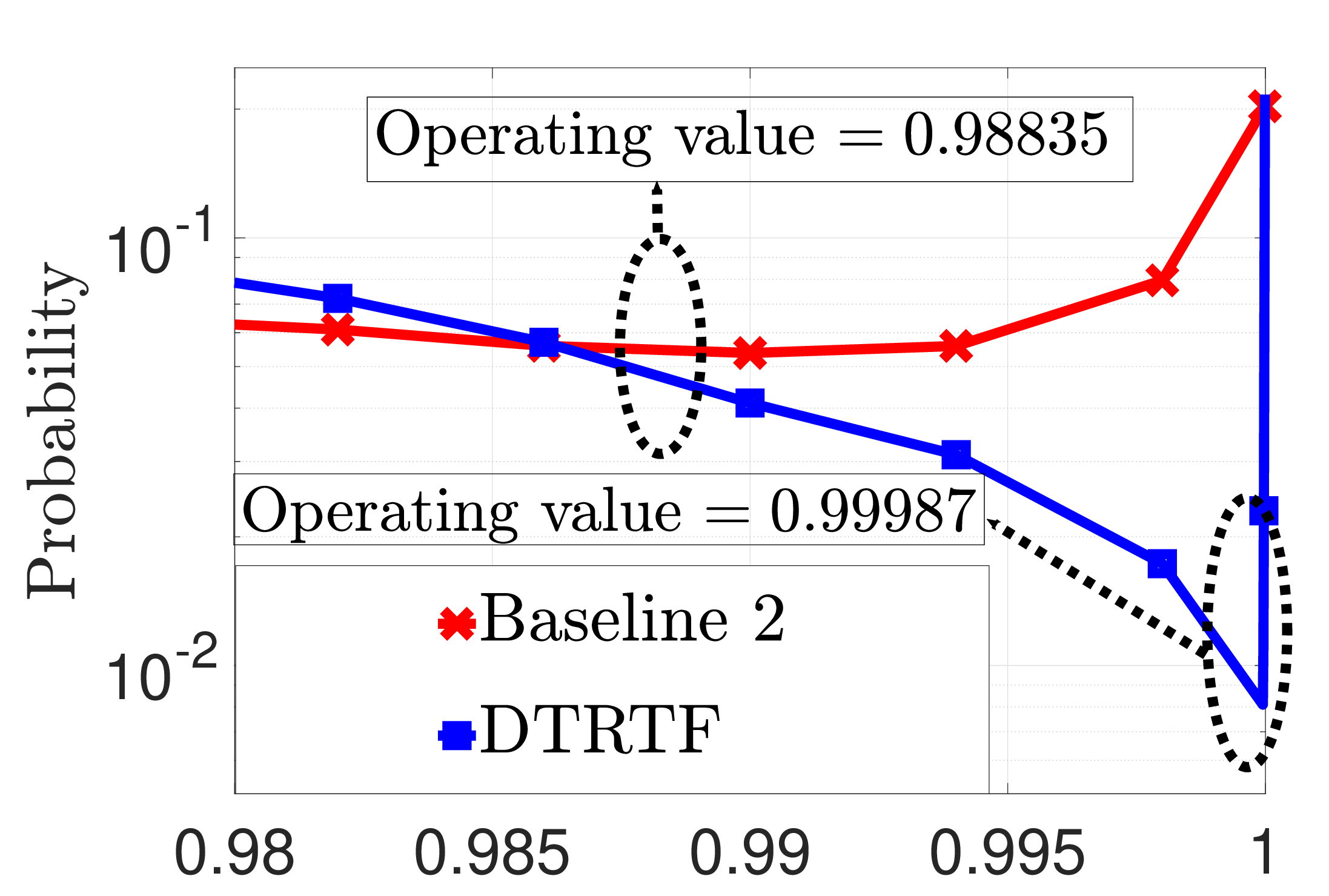}\label{PePdref3DTRTF}}
\hspace{5mm}
\subfloat[]
{\includegraphics[height=3.5cm,width=0.68\columnwidth]{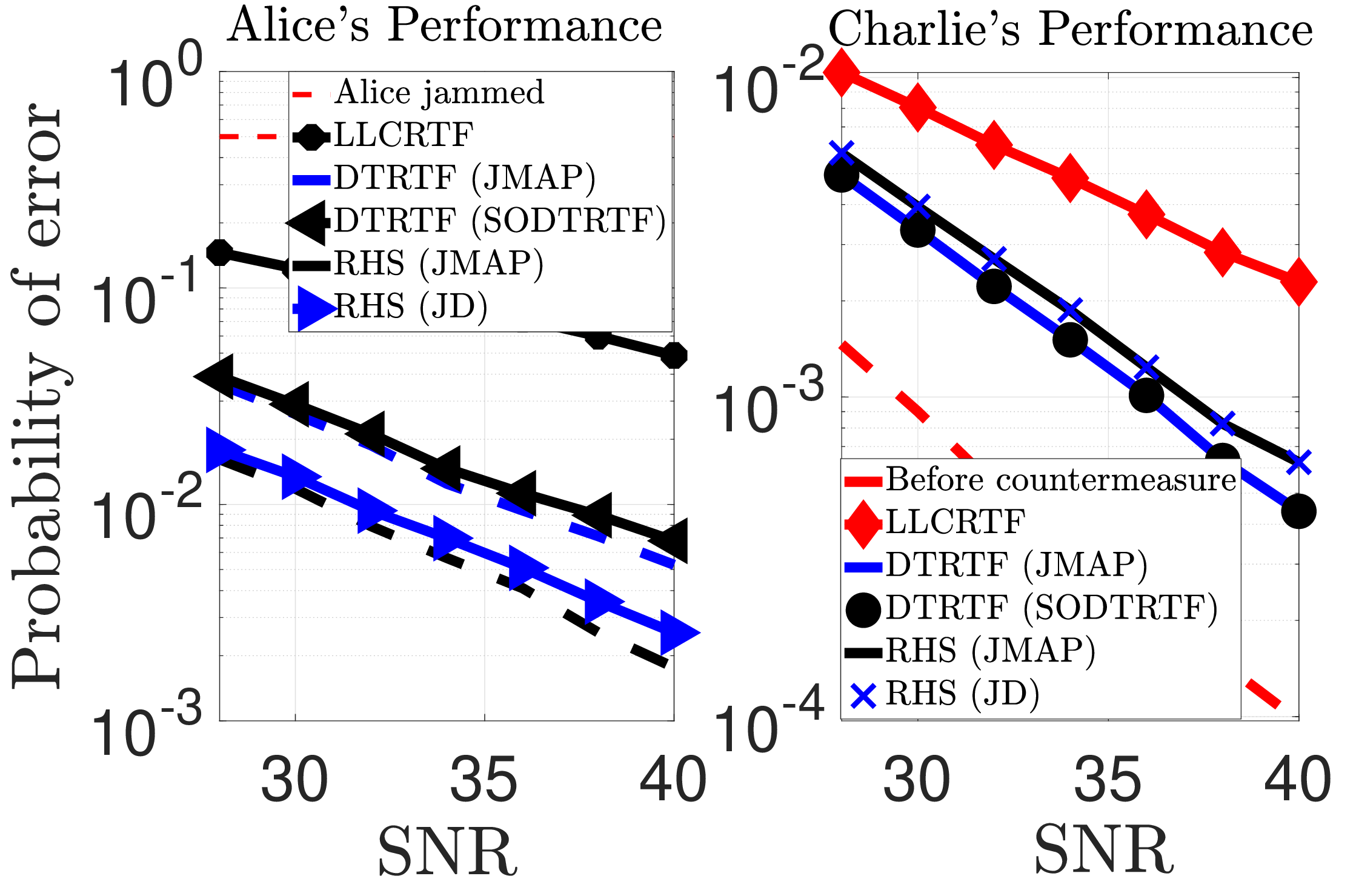}\label{Performance comparison}}
\hspace{5mm}
\subfloat[]
{\includegraphics[height=3.5cm,width=0.68\columnwidth]{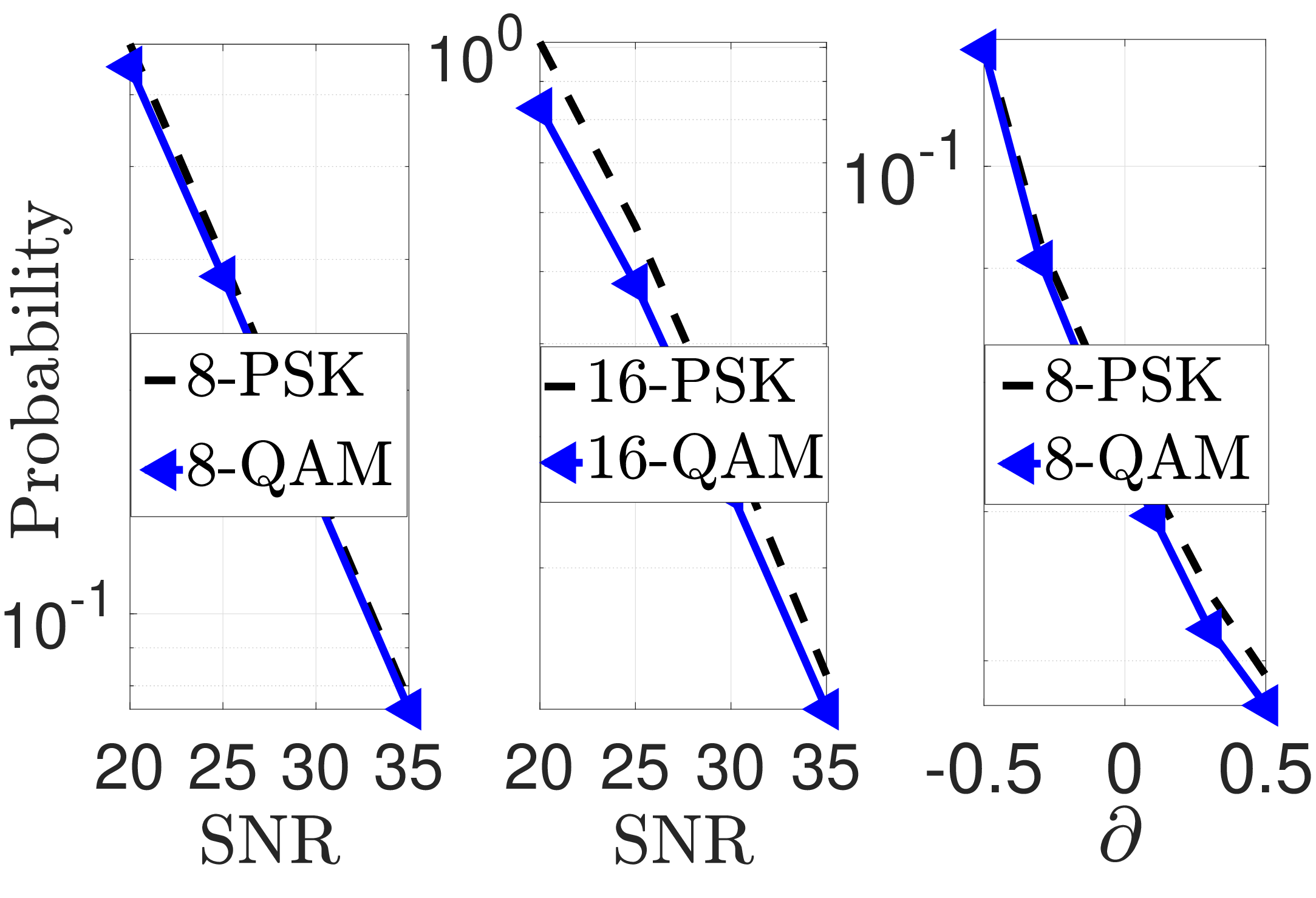}\label{diff_variance}}
\vspace{-0.2cm}
\caption{\textcolor{black}{(a) Sum of probability of decoding error and probability of detection for DTRTF and countermeasure in \cite{lowlatency}.  
(b) Average decoding error probability of DTRTF, LLCRTF and RHS.
(c) DTRTF performance against QAM and helper-node displacement.}}
\label{all_figs_page15}
\vspace{-4mm}
\end{figure*}

\section{Comparison with Baselines} \label{SR}
\textcolor{black}{
\cite{V2,lowlatency} consider threat models where the reactive adversary uses an average-energy detector, while \cite{ISIT,TVT1} consider threat models wherein the adversary uses statistical detectors.
In contrast, we consider a stronger threat model, wherein the adversary employs generalized energy detectors that include both statistical detectors and ML-based data-driven detectors.}
Consequently, the countermeasures in \cite{V2} and \cite{lowlatency} are not tailor-made for our threat model, and that in \cite{V2} assumes instantaneous decoding of  victim's bit, which is difficult under practical FDRs. Thus, we first compare the countermeasure in \cite{lowlatency} with our schemes.
In this direction, in Fig. \ref{PePdref3DTRTF}, we use Monte-Carlo simulations to plot the sum of probability of decoding error at Bob and the probability of detection at Dave as a function of $\alpha$, for $Eb/No$ of $34$ dB (owing to difference in spectral efficiencies), $N_C\!=\!190$, $\delta\!=\!0.49$, for DTRTF. Recall that the operating value of $\alpha$ is the value that minimizes the sum of the probabilities. 
We observe at this operating value, the sum of the probabilities for the countermeasure in \cite{lowlatency} is strictly higher than that for DTRTF, justifying the need for designing our countermeasure despite \cite{lowlatency}.} 

As our threat model and that in \cite{ISIT,TVT1} both consider statistical detectors at the adversary, we now compare our proposed strategies with that of RHS \cite{ISIT,TVT1}.
If we directly use the framework of RHS, the comparison is not fair as the rate of RHS is half whereas the rates of DTRTF and LLCRTF are three-fourth. 
Thus, for fairness, we use a modified RHS where the dummy symbol is replaced by a legitimate $M$-PSK symbol to equalize rates. 
In this modified RHS, in the first time-slot, Bob treats Alice’s symbol as interference and decodes Charlie’s symbol, then subtracts it before applying the JMAP and Joint Dominant (JD) decoders in \cite{ISIT,TVT1}.
Using this setup, together with the $\alpha$ values obtained by solving \probref{main_opt_problem_n_time_slots_minima} for DTRTF and \probref{main_opt_problem_n_tslatency_minima} for LLCRTF, for SNR from $28$ dB to $40$ dB, in $1$ dB steps, for $N_C\!=\!3$, $\delta\!=\!0.483$ and $\rho_{th}\!=\!10^{-5}$, we present the individual error performance of the two users for all three schemes in Fig.~\ref{Performance comparison}. Here, the left and right subfigures capture the average symbol error probabilities of Alice and Charlie, respectively.

From the left subfigure of Fig. \ref{Performance comparison}, we infer that Alice's error performance under DTRTF \textcolor{black}{is poorer} than under RHS. 
This is because, for a given SNR, the solution of \probref{main_opt_problem_n_time_slots_minima} for DTRTF is closer to unity  compared to that of RHS. 
As a result, in DTRTF scheme, the energy difference between bit-1 and bit-0 in $k^{th}$ time-slot decreases, and the distance between the corresponding concentric $M$-PSK circles in $(n+k)^{th}$ time-slot also reduces. 
Therefore, the individual error performance for Alice's bits in the DTRTF scheme is poorer compared to that in RHS. 
From the right figure of Fig. \ref{Performance comparison}, we infer that Charlie's error performance under DTRTF is better than under RHS. This is because, the DTRTF solution of \probref{main_opt_problem_n_time_slots_minima} is closer to unity compared to that of RHS, which shrinks the envelope of $M$-PSK symbols in first time-slot. Secondly, as Alice's bits are considered as interference in RHS, the interference at Bob increases. 
These effects reduce the SINR of the first-time-slot $M$-PSK symbol, leading to poorer error performance for Charlie under RHS compared to DTRTF. \textcolor{black}{Thus, the DTRTF scheme is a helper-friendly scheme, and is applicable when the helper node neither wants to compromise on their rate nor their error performance.}

\begin{figure}[ht!]
\vspace{-0.3cm}
\begin{center}
         \subfloat[]
{
\includegraphics[height=2.5cm, width=0.35\columnwidth]{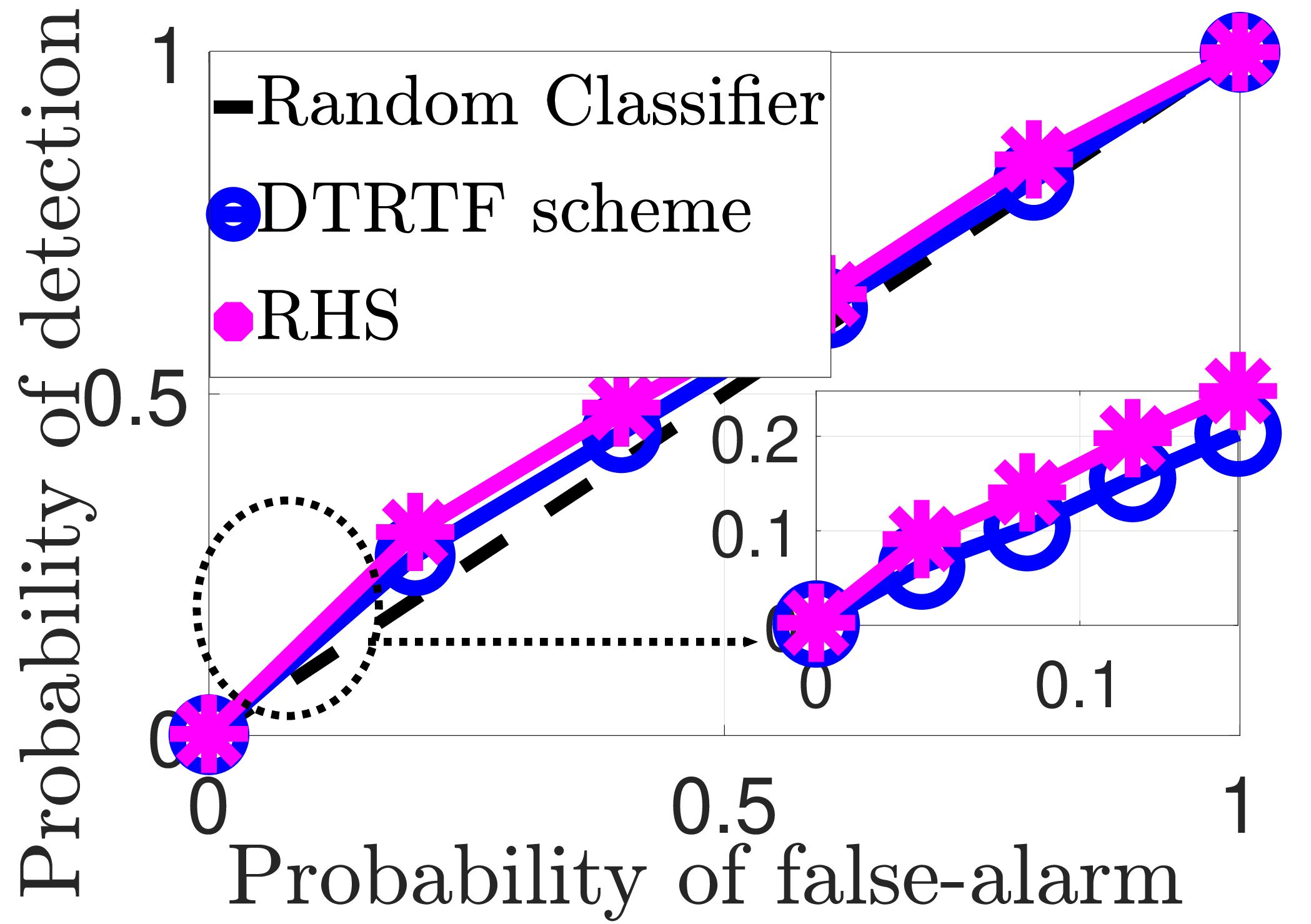}\label{RHS_DTRTF_ROC}%
}
         \subfloat[]
     {
\includegraphics[height=2.5cm, width=0.50\columnwidth]{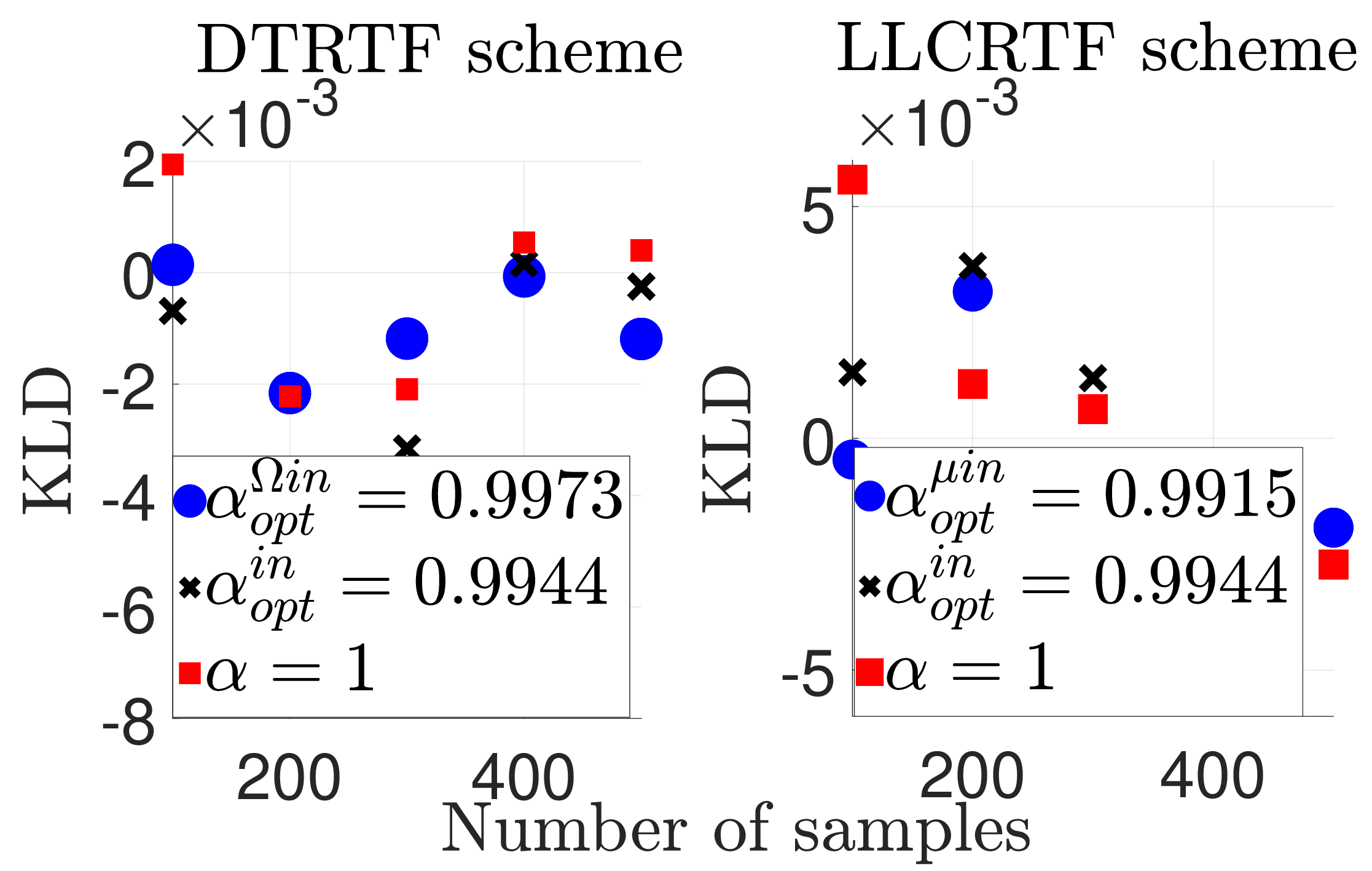}
    \label{KLDgraphfab2}}%
    \vspace{-0.2cm}
    \caption {\textcolor{black}{(a) ROC curves of the DTRTF scheme and the RHS. (b) Average KLD estimates on the  $f_{AB}$ for $\partial=0.5$.}}
    \end{center}
    \vspace{-0.4cm}
\end{figure}

\textcolor{black}{To compare the detection performance of the DTRTF scheme and that of RHS, in Fig. \ref{RHS_DTRTF_ROC}, for the same probability of error, we plot the ROC curves of both schemes for ML-classifiers at SNR $=25$ dB, $N_C=1$, and $\rho_{th}=10^{-5}$. From the plots, we observe that, for a fixed probability of false-alarm, the probability of detection for RHS is strictly higher than that of DTRTF. This is because the DTRTF solution of \probref{main_opt_problem_n_time_slots_minima} lies closer to unity compared to that of the RHS, thereby making the RHS more vulnerable to ML-classifiers than the DTRTF scheme.}

\color{black}
\section{Generalized Formulation of Strategies} \label{Robustness}
\textcolor{black}{In this section, we present the generalized formulation of the proposed strategies. First, we analyze their robustness by considering QAM at the helper node and arbitrary helper positioning relative to Alice and Bob. Subsequently, we examine the effect of the weighting factor $w$ on the corresponding optimal value of $\alpha$.}

\subsection{Robustness of the Proposed Strategies}\label{Robustness_QAM}
In this subsection, we analyze the robustness of the proposed strategies by considering QAM at the helper node and an arbitrary helper position relative to Alice and Bob.\footnote{\textcolor{black}{Alice continues to transmit her information modulated using OOK, and her relative position with respect to Bob is fixed.}} Here, we assume Frank acts as the helper and uses $M$-QAM to communicate with Bob. As posed in \probref{main_opt_problem_n_time_slots}, the performance with QAM is jointly determined by the probability of decoding error at Bob and the probability of detection at Dave. 
 Thus, for a fixed SNR, we vary $\alpha$ and use Monte-Carlo simulations to obtain the minimum of their sum.\footnote{\textcolor{black}{As the DTRTF scheme outperforms the LLCRTF scheme, we compare the performance of the former scheme with PSK and QAM.}}

To present the performance of DTRTF scheme with $M$-QAM, the leftmost and middle subfigures of Fig. \ref{diff_variance} plot the sum of probability of decoding error at Bob and probability of detection at Dave as a function of SNR for $M$-QAM, alongside the corresponding $M$-PSK plots.
From these plots, we observe that for a given SNR, the DTRTF scheme performs better with $M$-QAM than with $M$-PSK.
This is because, for a given SNR, the individual probabilities for $M$-QAM are lower than those for $M$-PSK, and therefore their sum is also smaller for $M$-QAM.

When the channel variances of Alice-Bob and Helper-Bob links are identical, i.e., $\partial=0$, the proposed strategies maintain the generalized energy statistics on both bands.
However, in practice, these channels may experience different large-scale fading, i.e., $\partial\! \neq \! 0$.
In the rightmost subfigure of Fig. \ref{diff_variance}, we plot the sum of probabilities versus $\partial$ for $M\!=\!8$, at SNR of $35$ dB. 
We observe that as $\partial$ increases, i.e., the helpers move closer to \textcolor{black}{center}, the sum decreases because the reliability of the Charlie-Bob and Frank-Bob links improves,  reducing the probability of error. Following Fig.~\ref{KLDgraphfcb}, in Fig. \ref{KLDgraphfab2}, we plot the KLD estimates for $\partial\! =\! 0.5$ for $f_{AB}$ band, and observe that they remain close to zero, indicating that the statistical distribution of energies of received symbols is nearly identical before and after countermeasure, even when $\alpha$ is \textcolor{black}{optimized} for $\partial\!=\!0$.

\subsection{\textcolor{black}{Effect of Weighting Factor on the Optimal \texorpdfstring{$\alpha$}{alpha}}}\label{weights_and_alpha}
\begin{figure}[ht!]
\vspace{-0.5cm}
\begin{center}
         \subfloat[]
{
\includegraphics[scale=0.11]{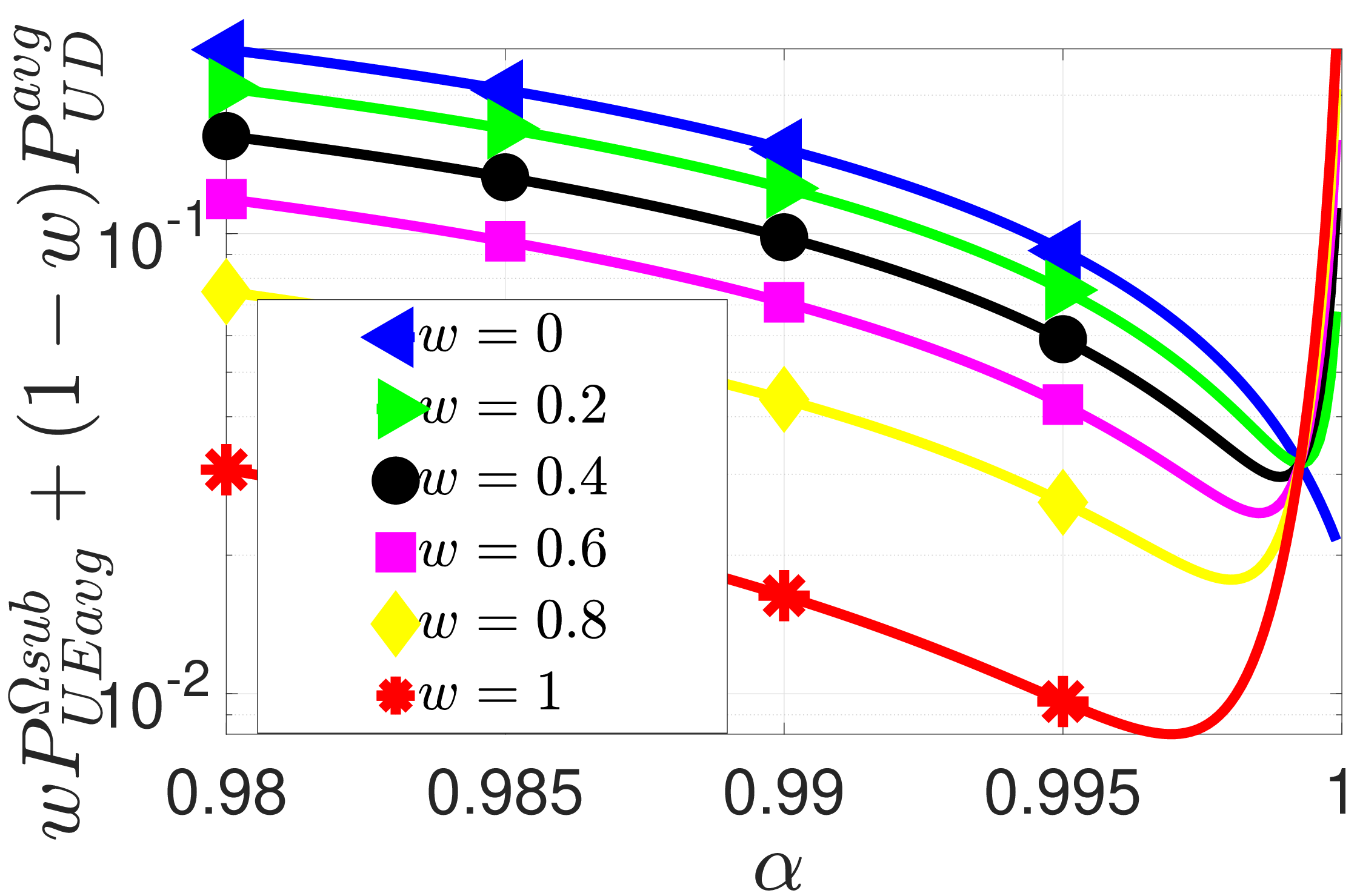}
    \label{PePdw}}%
         \subfloat[]
     {
\includegraphics[scale=0.11]{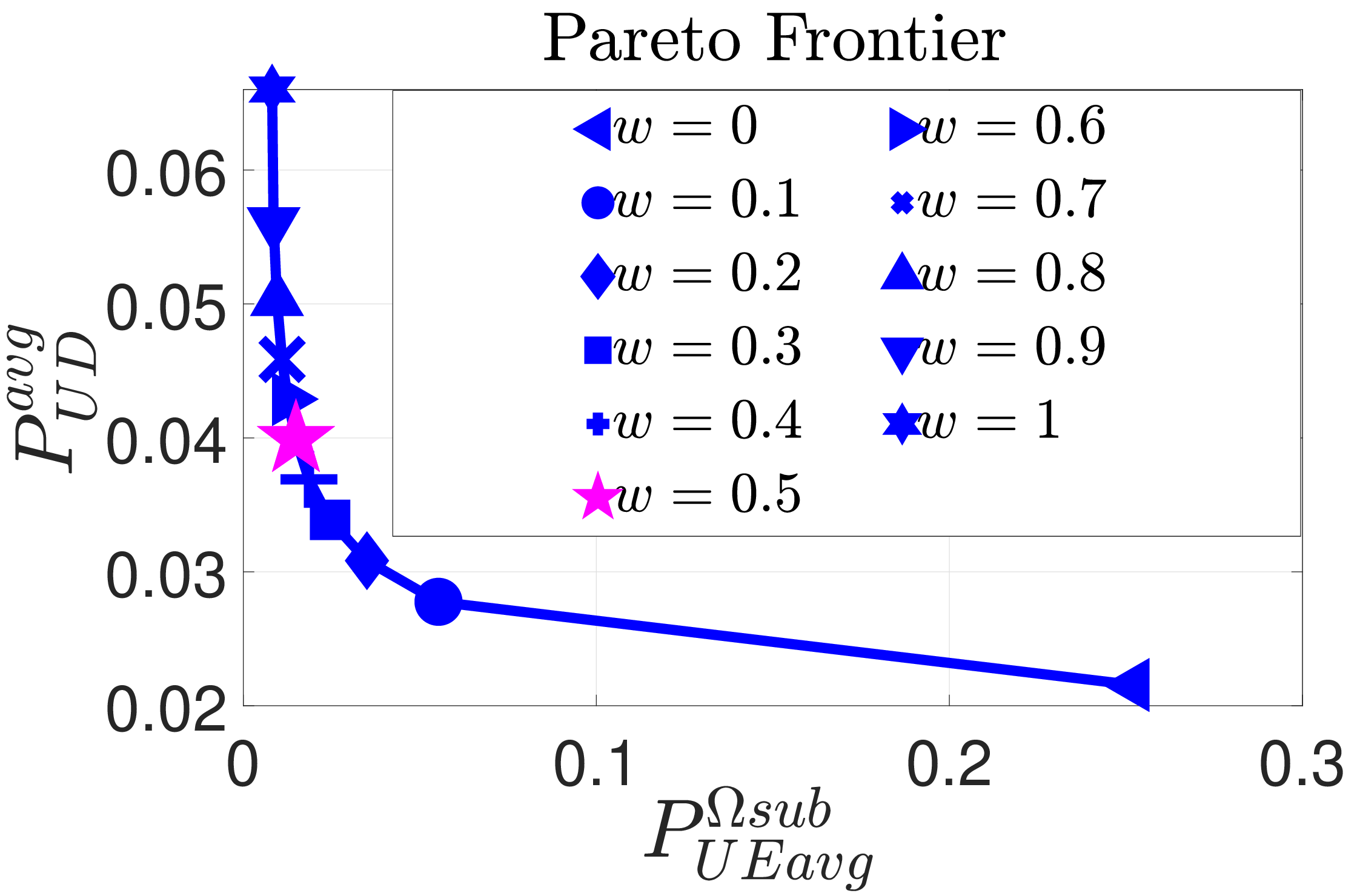}
    \label{PF}}%
    \vspace{-0.2cm}
    \caption {\textcolor{black}{(a) Weighted sum of the upper bounds on probabilities as a function of $\alpha$ for different values of $w$. (b) Corresponding upper bounds on probabilities for different values of $w$.}}
    \end{center}
    \vspace{-0.4cm}
\end{figure}

\textcolor{black}{In this subsection, we discuss the variation of optimal $\alpha$ with $w$ for DTRTF.
Given that we have upper bounds on $P_{Eavg}^{\Omega}$ and $P_{Davg}^{\Omega}$, i.e., $P_{UEavg}^{\Omega sub}$ and $P_{UD}^{avg}$, respectively, we obtain the optimal $\alpha$, denoted by $\alpha_{opt}^{\Omega w}$, by minimizing the weighted sum of these upper bounds rather than the weighted sum of the true expressions as given in \probref{main_opt_problem_n_time_slots}.
Subsequently, in Fig. \ref{PePdw}, we use  $N_C=3$,  $\rho_{th}=10^{-5}$, and SNR = $35$  dB to plot this sum as a function of $\alpha$ for different values of $w$.
From this figure, we observe that as  $w$ increases, the value of $\alpha_{opt}^{\Omega w}$ moves away from unity.
Also, we observe that $\alpha_{opt}^{\Omega w}$ varies within a narrow range of $\alpha$.
To observe the effect of $\alpha_{opt}^{\Omega w}$ on the individual upper bounds on the probabilities, in Fig. \ref{PF}, we plot the corresponding values of $P_{UEavg}^{\Omega sub}$ and $P_{UD}^{avg}$ for different values of $w$.
From this figure, we have the following observations.
First, even small variations in  $\alpha_{opt}^{\Omega w}$ result in significant changes in both upper bounds on the probabilities, thereby demonstrating the sensitivity of the system to $w$.
Second, when $w=0.5$, the corresponding upper bounds remain close to the origin, with the origin corresponding to the best-case scenario where both upper bounds are zero. Further, no other choice of weights results in upper bounds on the probabilities that are closer to the origin.
We observe similar results for the proposed LLCRTF scheme.}

\section{Hardware Implementation and ML-Based Detection Framework} \label{Hardwar_and_ML}

In this section, we discuss the covertness of the proposed countermeasures against a data-driven adversary equipped with ML-classifiers.
Towards this direction, we deploy the before-countermeasure and after-countermeasure setups on an SDR-based hardware testbed, collect dataset for training and testing purposes, and finally use the trained models to validate the covertness of our countermeasures.

\begin{figure}[ht!]
 \begin{center}
       {\includegraphics[scale=0.04]{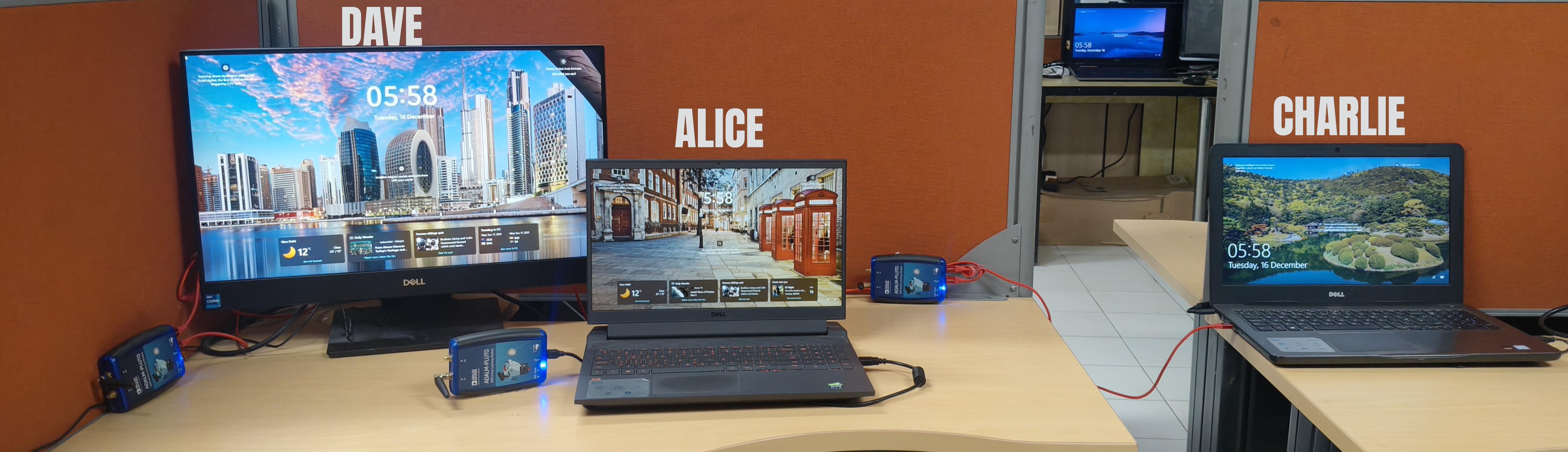}}
   \end{center} 
   \vspace{-0.1cm}
\caption{SDR testbed with Adalm-Pluto devices and Windows-11 MATLAB hosts implementing Alice, Charlie, and Dave.}
\label{Exp_setup}
\vspace{-0.4cm}
\end{figure}

\subsection{Hardware Testbed Description}
We employ three SDRs, operated via Adalm-Pluto devices, as shown in Fig. \ref{Exp_setup}, representing Alice, Charlie, and Dave.
Each Adalm-Pluto is connected to a Windows-11 machine running MATLAB-2024b, where all baseband signal processing is performed, while the Adalm-Plutos perform the RF transmission and reception. 
The transmitter and receiver setups follow the standard Adalm-Pluto QPSK examples in \cite{MathWorks_QPSK_Tx,MathWorks_QPSK_Rx}, with appropriate modifications incorporated to implement countermeasures.
\textcolor{black}{These experiments are conducted at a carrier frequency of 2.1 GHz with a bandwidth of 300 kHz, maximum transmit power of 7 dBm, and antenna gain of 2 dBi (Zenith), in an indoor quasi-static environment with both LoS and NLoS components, where the Alice-Bob and Alice-Charlie distances are 0.78 m and 0.47 m, respectively.}

Recall that the theoretical covertness analysis assumed Dave uses an instantaneous energy detector on $f_{CB}$ band. Hence, we implement the \textcolor{black}{signaling} scheme of $f_{CB}$ band on hardware testbed for both schemes, and exclude that of $f_{AB}$ band.
\textcolor{black}{Also, from Dave’s covertness perspective, the proposed schemes have similar \textcolor{black}{signaling} structures, and hence, the corresponding covertness analyses at Dave using ML-classifiers are also similar. Thus, to avoid repetitive discussions, we implement the \textcolor{black}{signaling} scheme of the LLCRTF scheme as a representative case and discuss the corresponding covertness analysis. Subsequently, we present a cost-effective SDR-based implementation of the DTRTF scheme to demonstrate the practical feasibility of the FDR-assisted \textcolor{black}{signaling} framework.}

\subsection{ML-classifiers and Dataset Setup}
To evaluate the detectability of the LLCRTF scheme under data-driven adversaries, we employ both supervised and unsupervised ML-classifiers. 
Given that the rule-based detector uses the instantaneous energy of the received symbols for detection, the energy of the received IQ samples is used to train and test the ML-classifiers.
Unless otherwise stated, all symbol energies hereafter refer to the energies of the received IQ samples.
The supervised models include Random Forest (RF) \cite{Random_Forest}, a neural-network-based Multi-Layer Perceptron (MLP) \cite{MLP_paper} with two hidden layers, and and \textcolor{black}{a Conv1D-based Convolutional Neural Network (CNN) with 16 filters \cite{CNN_paper}.}
Dave uses these models to detect the presence of countermeasure based on energy patterns.
Here, Class-0 contains the energies of unit-energy QPSK symbols, corresponding to before-countermeasure setup, and Class-1 contains the energies of symbols that deviate from unit-energy QPSK, representing the countermeasure effect.
\textcolor{black}{Since after-countermeasure samples are unavailable during training, Dave must detect the countermeasure based on unseen signal characteristics.
Thus, assuming the countermeasure alters scaling or modulation formats, we construct a surrogate Class-1.}\footnote{\textcolor{black}{Section \ref{Section_on_LLCRTF_as_Training} considers an adaptive adversary with labeled LLCRTF samples for training.}}
For training, Class-0 consists of the energies of unit-energy QPSK symbols, and Class-1 consists of an equal proportion of the energies of symbols of $16$-QAM, $64$-QAM, and QPSK with average energies $0.1$ and $0.01$.
For testing, we use the energies of unit-energy QPSK symbols, along with after-countermeasure symbols from both time-slots for $\alpha \!=\! 0.9877, 0.9901, 0.9964$.\footnote{\textcolor{black}{The values of $\alpha$ are chosen close to unity, as the optimal $\alpha$ that \textcolor{black}{minimizes} the sum of probability of decoding error at Bob and probability of detection at Dave using the rule-based detector lies near unity. Since the proposed countermeasures are intended to be deployed using this optimal $\alpha$, the ML-classifiers are evaluated using the corresponding datasets.}}
To avoid relying solely on the surrogate assumption, we also consider an unsupervised model, namely Isolation Forest (IF) \cite{Isolation_Forest}, trained only on energies of unit-energy QPSK symbols and tested on energies of after-countermeasure symbols.

\subsection{Data Collection and SNR Estimation} \label{data_Collection}

\begin{figure}[!t]
\vspace{-0.4cm}
\subfloat[]
{
\includegraphics[scale=0.09]{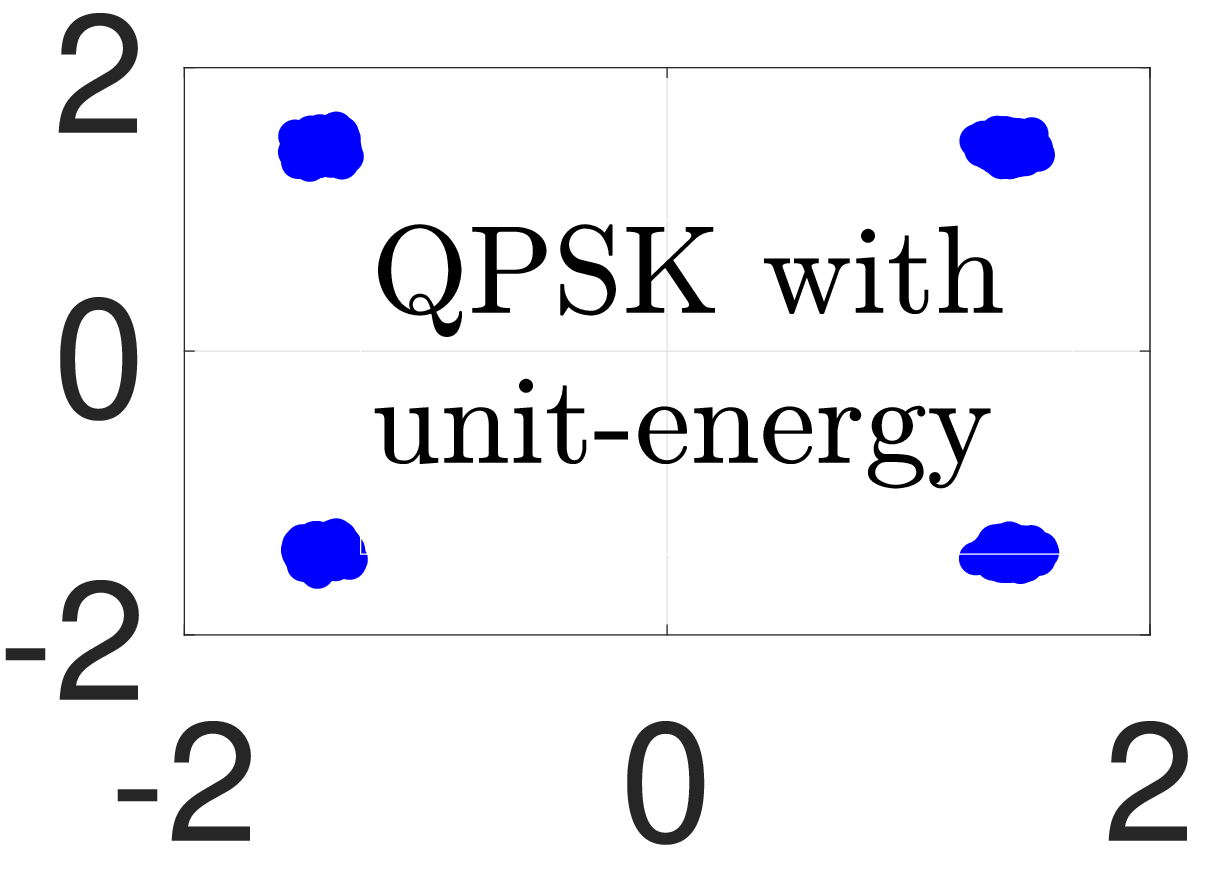}
    \label{QPSK_1}}%
    \hspace{-1mm}
         \subfloat[]
     {
\includegraphics[scale=0.088]{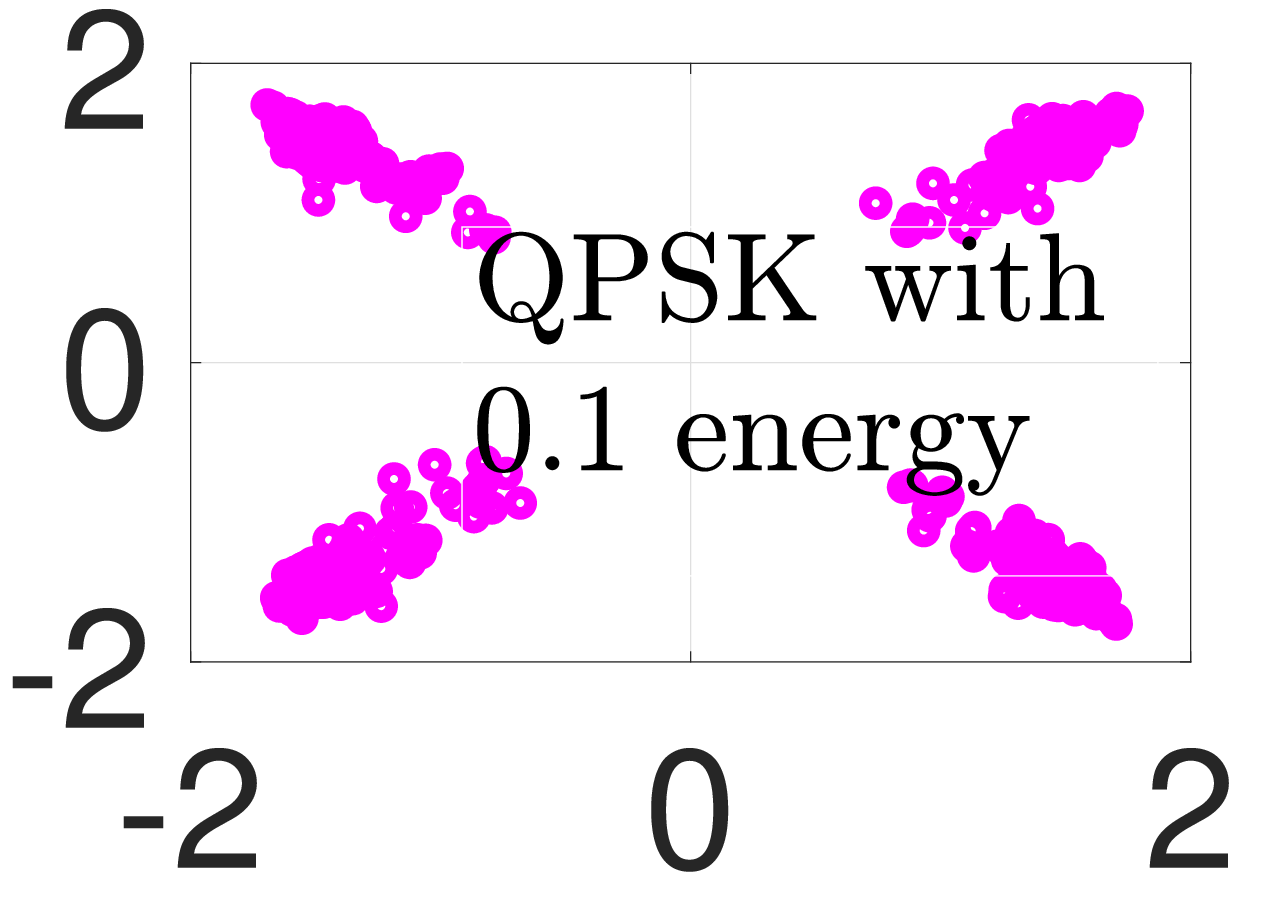}
    \label{QPSK_point1}}%
     \hspace{-1mm}
         \subfloat[]
     {
\includegraphics[scale=0.088]{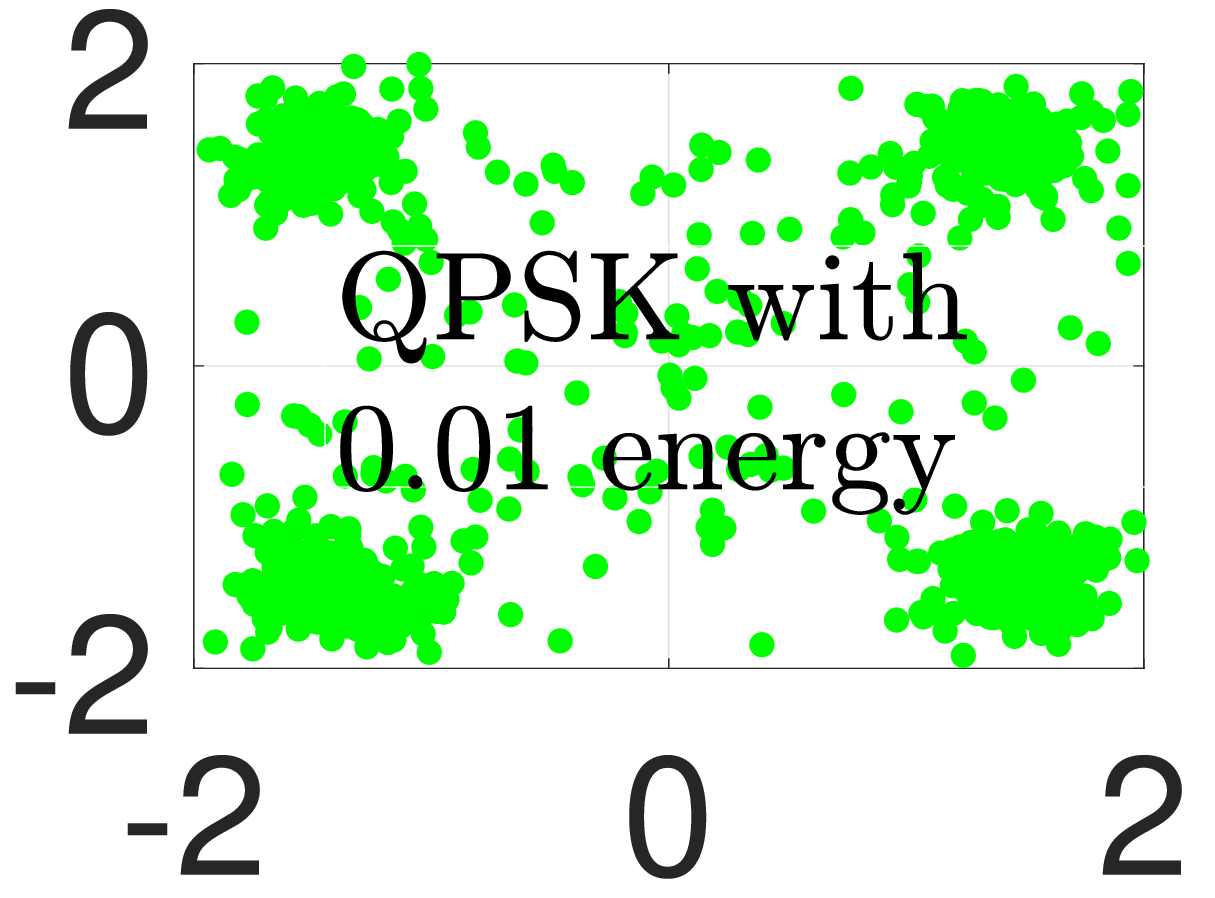}
    \label{QPSK_point_zero1}}%
   \hspace{-1mm}
    \subfloat[]
{
\includegraphics[scale=0.09]{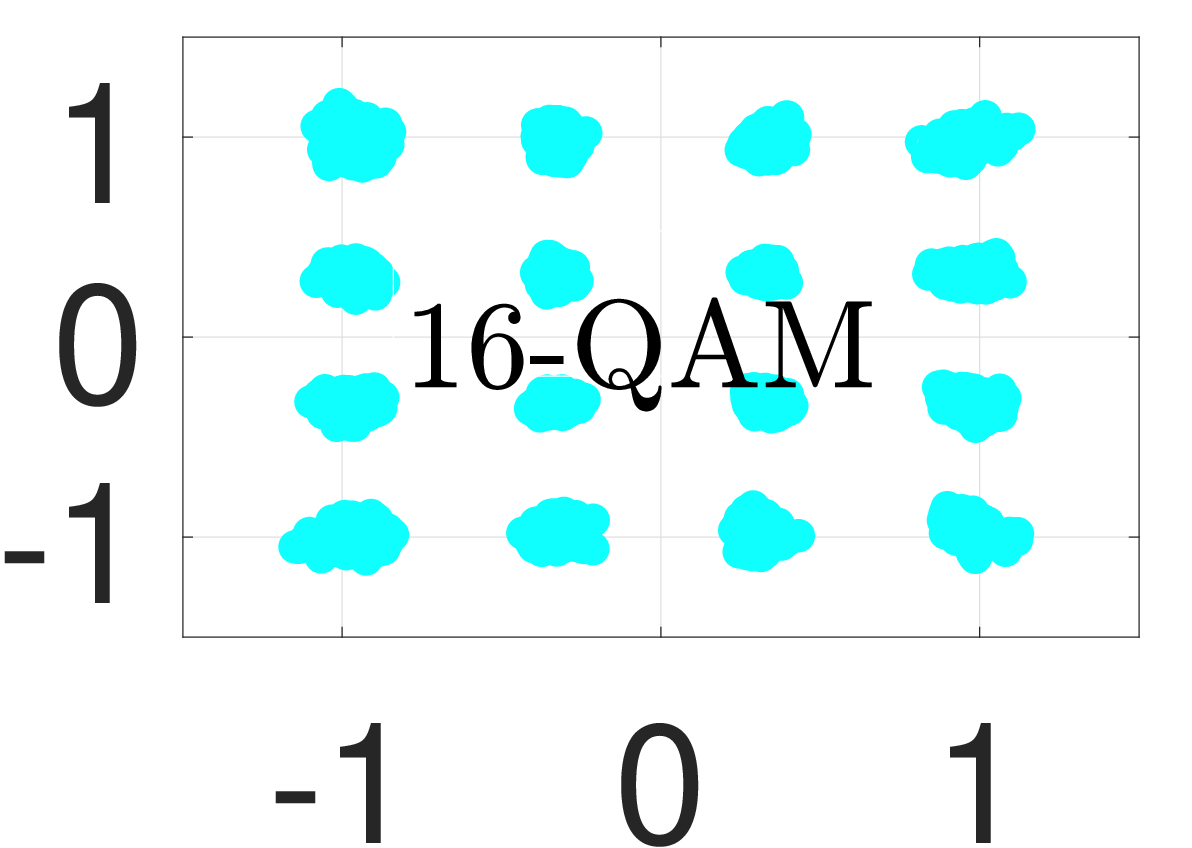}
    \label{QAM_16}}

    \vspace{-3mm}
     \subfloat[]
     {
\includegraphics[scale=0.09]{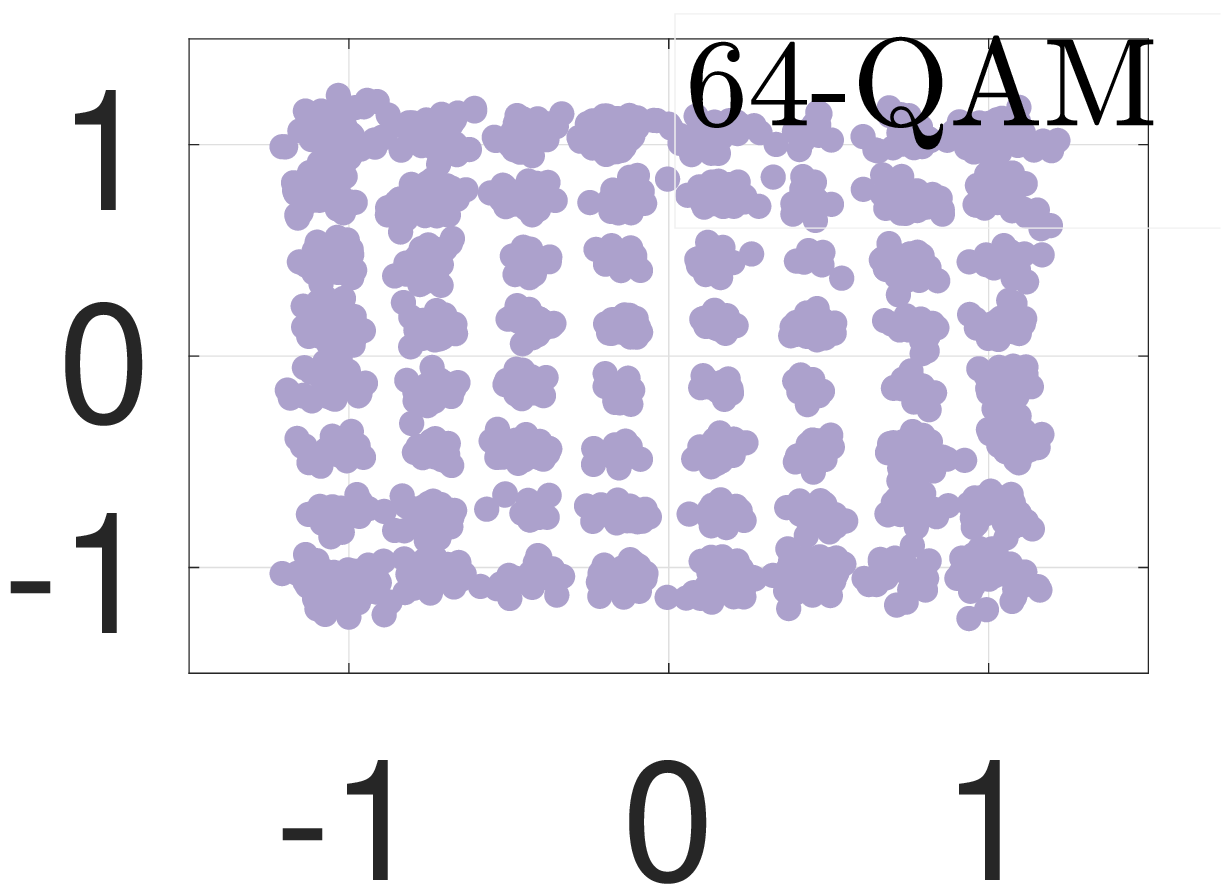}
    \label{QAM_64}}%}
    \hspace{-1mm}
                 \subfloat[]
{
\includegraphics[scale=0.09]{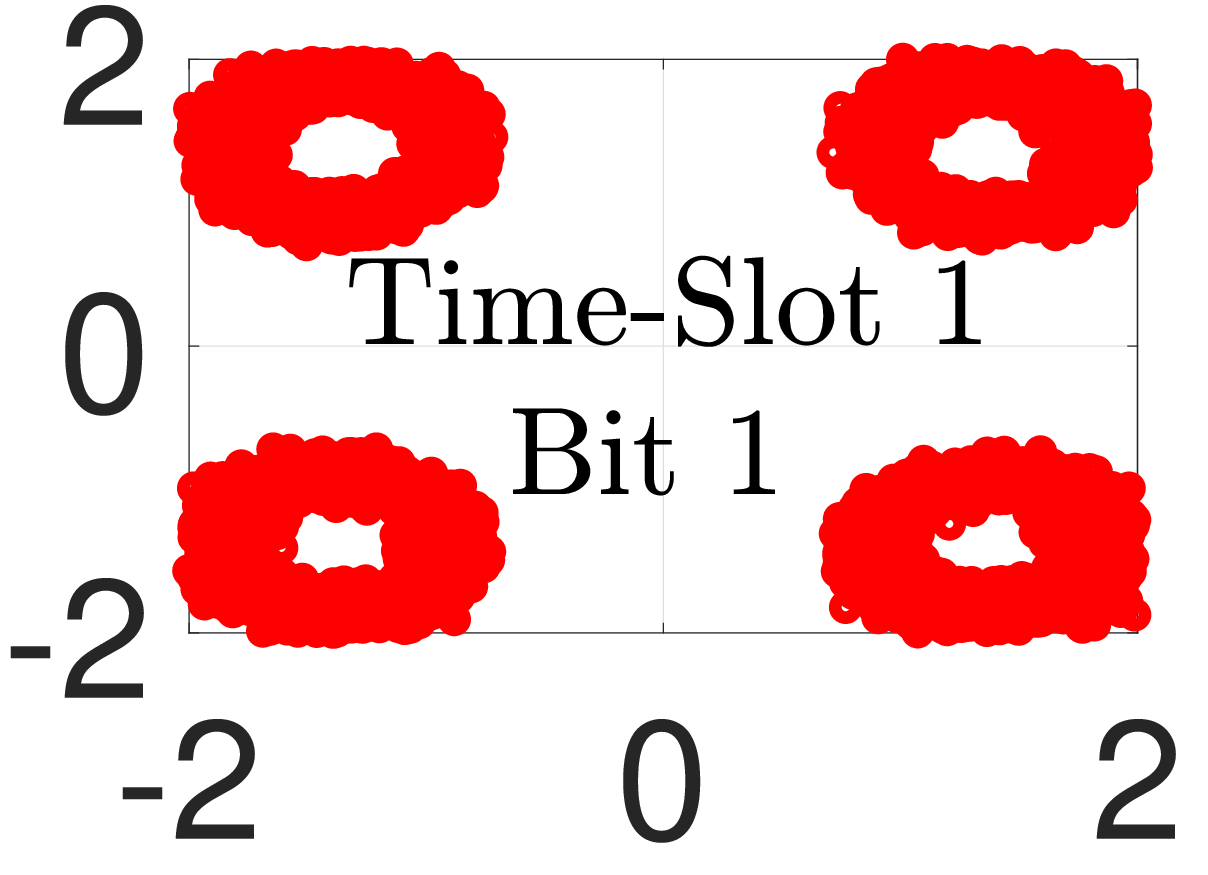}\label{TS_1_bit_1}
    }% 
    \hspace{-1mm}
    \subfloat[]
{
\includegraphics[scale=0.09]{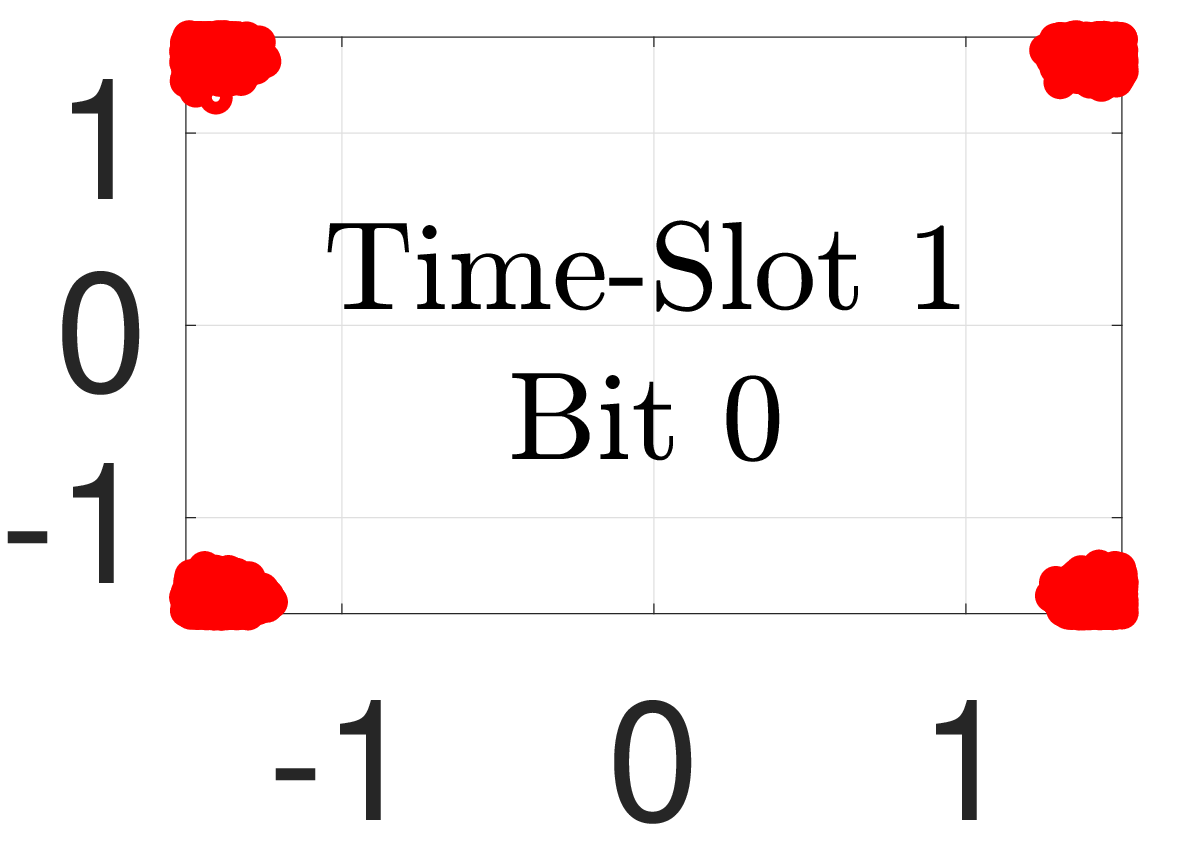}\label{TS_1_bit_0}
    }
    \hspace{-1mm}
     \subfloat[]
     {
\includegraphics[scale=0.09]{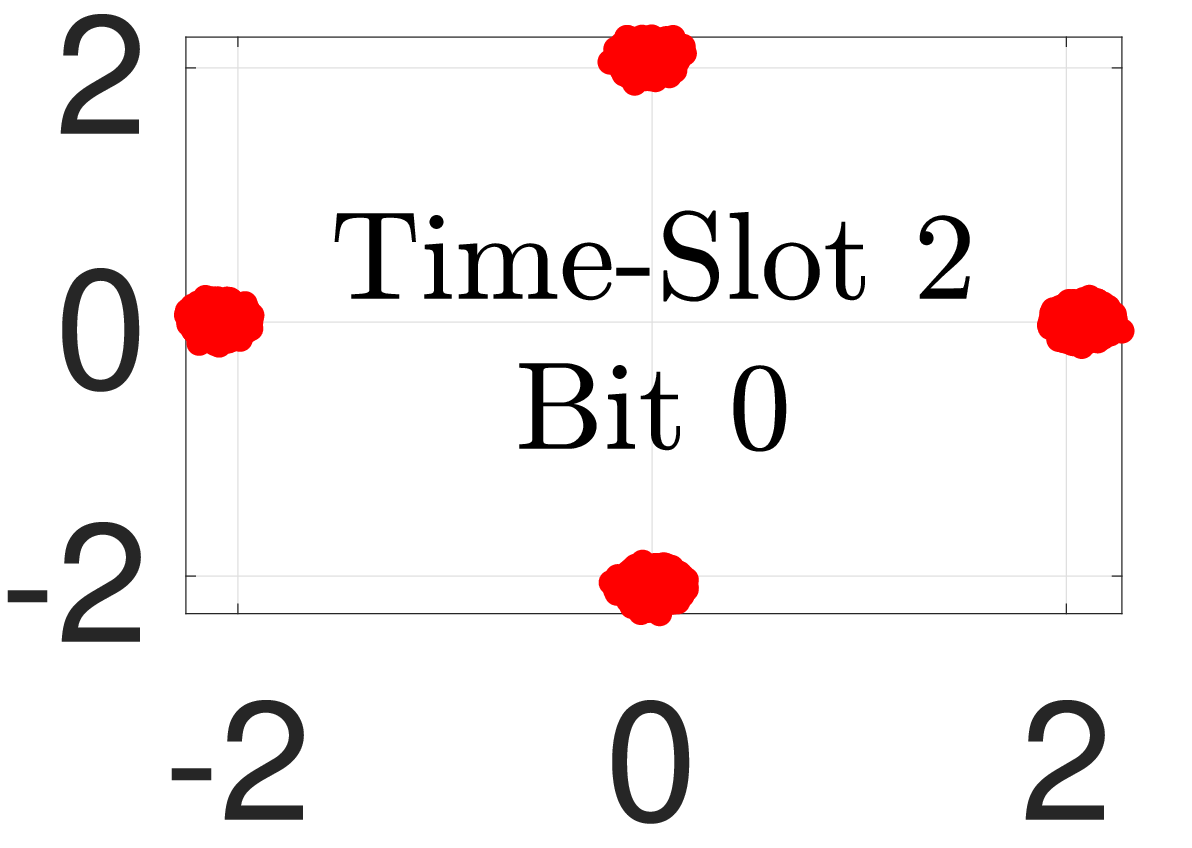}\label{TS_2_bit_0}
    }%
\vspace{-0.1cm}
\caption{\textcolor{black}{ Constellation diagrams of training dataset: (a) Class-0 and (b)–(e) Class-1, and of  testing dataset for after-countermeasure in (f)–(h), obtained after implementing the LLCRTF scheme  for $\alpha = 0.9901$ on SDR-based testbed.}}
\end{figure}

To generate training dataset for supervised ML-classifiers, we use the hardware setup given in Fig. \ref{Exp_setup}, and establish communication between Charlie and Dave.
Subsequently, we collect the IQ samples of received symbols
obtained after the carrier-synchroniser block in the receiver-chain for both classes.
All modulation references pertain to data symbols, while the preamble is consistently modulated using QPSK \cite{US9313065B2}.
In the before-countermeasure setup, Charlie transmits unit-energy QPSK symbols, and the corresponding energies form Class-0, while additional energy levels and modulation formats are implemented solely to construct the surrogate Class-1 for supervised learning. For IF, only energies of unit-energy QPSK symbols are used for training.
In Figs. \ref{QPSK_1}-\ref{QAM_64}, we show constellation diagrams of the collected IQ samples used to form the training dataset. 
We observe that unit-energy QPSK results in well-defined constellation clusters (Fig. \ref{QPSK_1}), whereas QPSK symbols transmitted with reduced average energy (Figs. \ref{QPSK_point1}, \ref{QPSK_point_zero1}), exhibit increased dispersion due to the corresponding reduction in effective SNR.
This is because, for low-average-energy transmissions, the Automatic Gain Control (AGC) at Dave boosts the power level of the received frame to its desired operating level, as a result, the noise also gets amplified. 
Also, higher-order QAM constellations (Figs. \ref{QAM_16},  \ref{QAM_64}), appear more scattered for a given noise level owing to their denser structure and smaller minimum Euclidean distance.

To generate testing dataset for after-countermeasure, we use the hardware setup given in Fig. \ref{Exp_setup} to implement the LLCRTF scheme.
For time-slot 1, when Alice and Charlie simultaneously transmit on $f_{CB}$ band, we establish communication from both to Dave, thereby creating a MAC, while for the other cases, communication only between Charlie and Dave.
Subsequently, we collect the IQ samples for $\alpha=0.9877, 0.9901, 0.9964$, and present in Figs. \ref{TS_1_bit_1}-\ref{TS_2_bit_0},  representative constellation diagrams for $\alpha \!=\! 0.9901$ of both time-slots.
We observe that during time-slot 1, when Alice transmits bit-1 on the $f_{CB}$ band and Charlie is simultaneously transmitting on the same band, their symbols superimpose, and as shown in Fig. \ref{TS_1_bit_1}, the constellation is visibly distorted.
In this MAC case, Alice and Charlie transmit with energies 
$1\!-\! \alpha$ and $\alpha$, respectively, and both preamble and data symbols are transmitted at these respective energy levels.
Since $\alpha$ is close to unity, Charlie’s preamble dominates the received signal, forcing synchronization onto Charlie, and Alice’s low-energy transmission appears as noise-like interference, leading to the distorted constellation in Fig.~\ref{TS_1_bit_1}.
In contrast, QPSK symbols transmitted by Charlie with energies $\alpha$ (Fig. \ref{TS_1_bit_0}) and $2\!-\!\alpha$ (Fig. \ref{TS_1_bit_1}), respectively,\footnote{We omit the constellation of time-slot 2, bit 1, as it is unit-energy QSPK symbol.} remain well clustered for $\alpha$ near unity.
From the above discussions based on the constellation dispersion, we conclude that 
symbols with varying transmit levels and modulation formats are reliably conveyed using the hardware testbed.
Both models use the same collected IQ samples, where the unit-energy QPSK symbols act as Class-0 and the LLCRTF symbols act as Class-1, while for IF, these LLCRTF symbols are treated as anomalous data.

Now we verify the correctness of the hardware testbed using SNR estimates at Dave. 
The first 13 received symbols in every frame $\!f$, denoted by 
$\mathbf{r}_x$, are the 13-chip Barker sequence modulated using QPSK \cite{Hardware_paper}.
The correlation metric $\mathbf{t_f}\!\! =\!\! \mathbf{r}_p^{\ast} \mathbf{r}_x$, where $\!\mathbf{r}_p$ denotes the reference vector.
The corresponding channel estimate $\hat{h}_{\text{est}}$, is obtained by averaging  $\mathbf{t_f}$, and the SNR at Dave, averaged over $F$ frames, is 
$\text{SNR}_{\text{dB}}\!\!\! = \!\!\! 10 \log_{10} \!\bigg(\!
\frac{\mathbb{E}\left[|\hat{h}_{\text{est}}|^2\right]}
{\mathbb{E}\left[|\mathbf{r}_x|^2\right]\! - \!\mathbb{E}\left[|\hat{h}_{\text{est}}|^2\right]}
\bigg)\!,$ where the numerator and the denominator denote the average signal power and average noise power, respectively.
From the experimental measurements,  with $F\!\! =\!\! 893$ frames per runs, the SNRs for QPSK with unit-energy, energies $\!0.1$, and $\!0.01$ are $\!30.44$~dB, $20.37$~dB, and $8.17$~dB, respectively.
The observed $\!\approx\!\! 10$~dB reduction per tenfold decrease in transmit energy validates the channel and SNR estimation, confirming the correctness of hardware testbed.

\begin{center}
    \begin{figure}[t]
\vspace{-0.4cm}
         \subfloat[]
{
\hspace{-2mm}\includegraphics[scale=0.11]{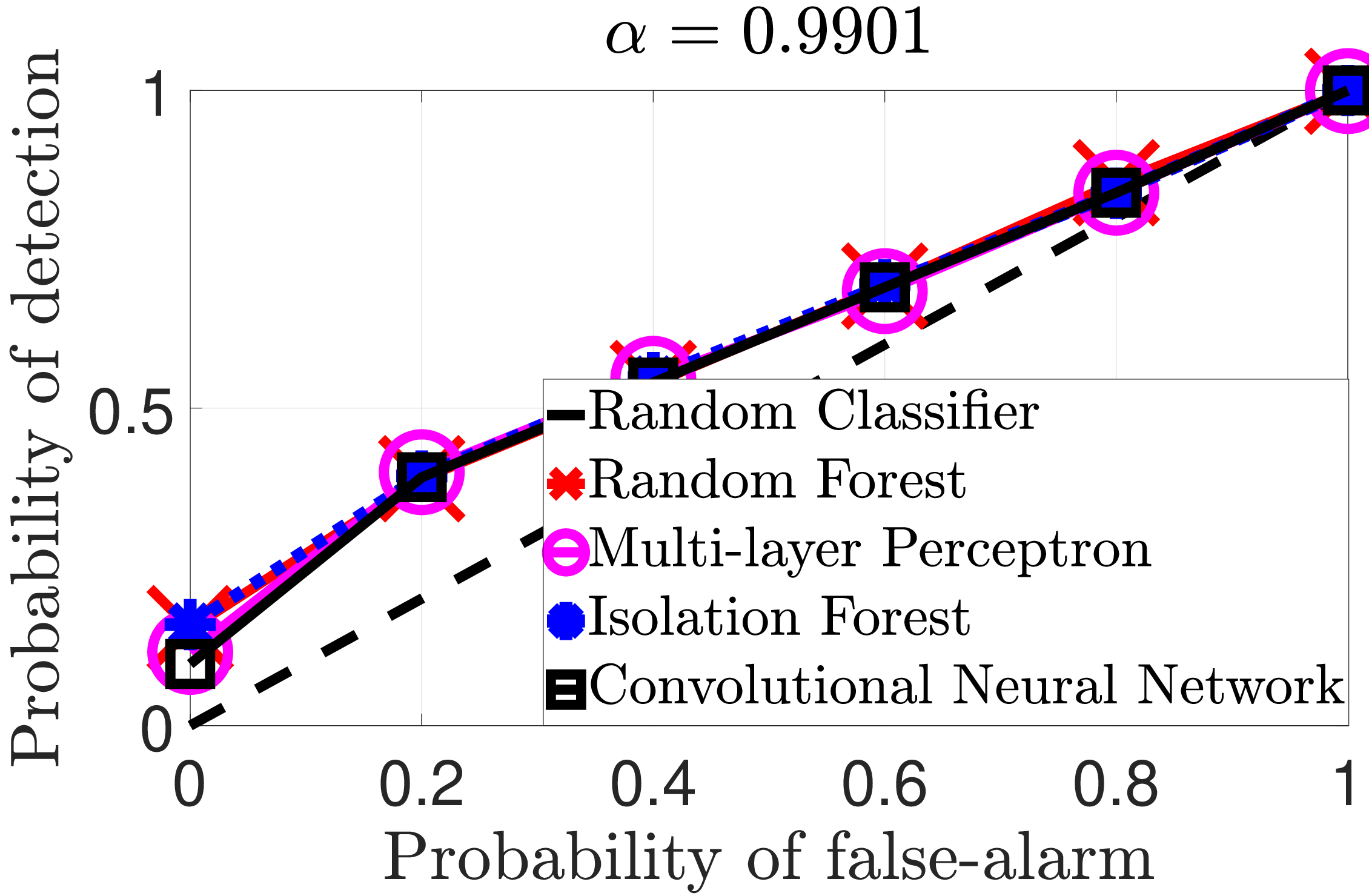}
    \label{alpha_9901}}%
    \subfloat[]
{
\includegraphics[scale=0.11]{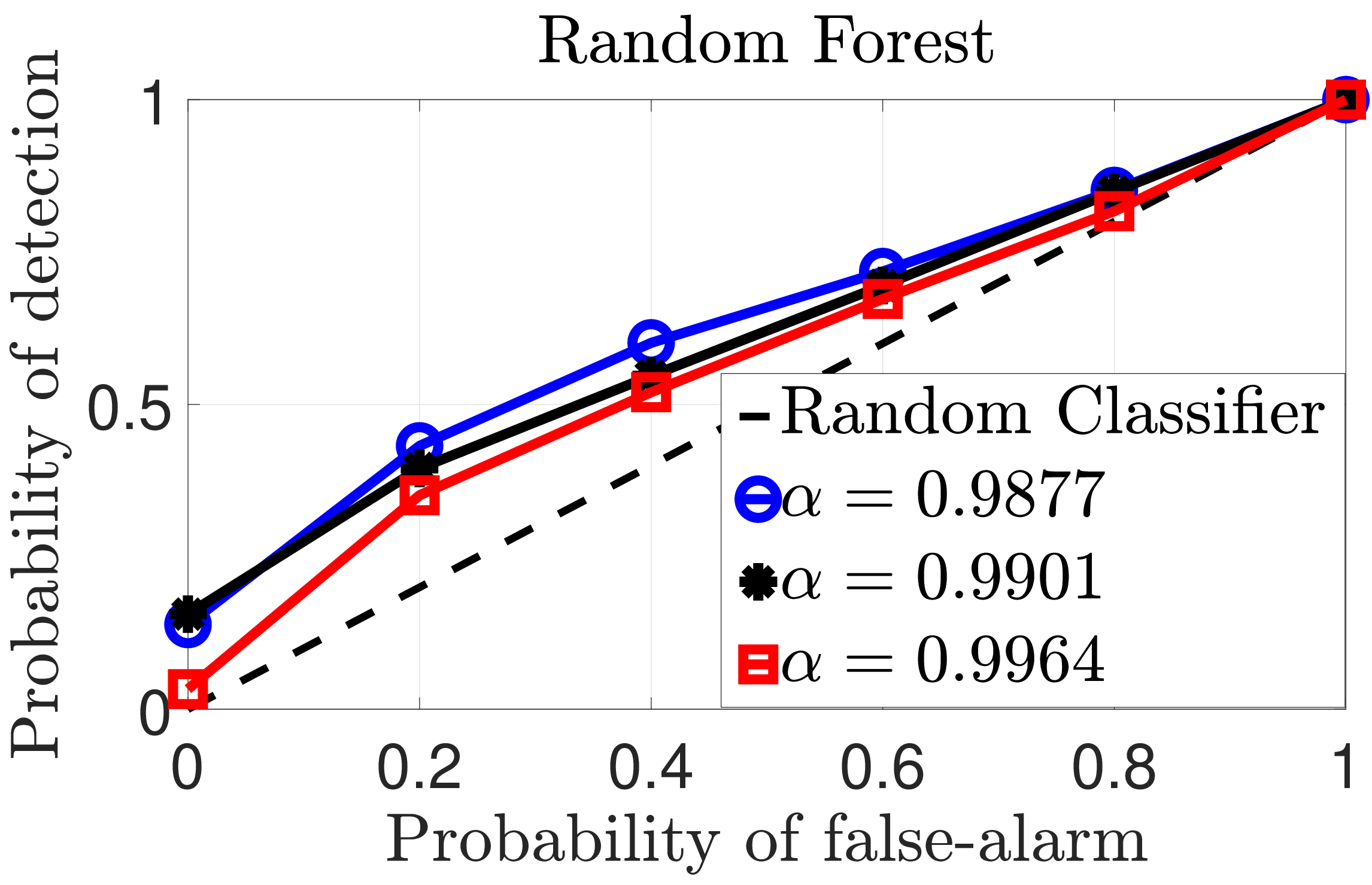}
    \label{RF_alpha}}
    \vspace{-0.1cm}
    \caption {\textcolor{black}{ROC curves for (a) $\alpha\!=\! 0.9901$,}  (b)  Random Forest. }
    \label{ROC_ML}
\end{figure}
\vspace{-0.3cm}
\end{center}

\begin{tiny}
\vspace{-0.5cm}
  \begin{table}[h!]
\centering
\caption{Performance metrics at $\alpha = 0.9901$.}
\label{tab:metrics_compact_extended}
\begin{tabular}{|c|c|c|c|}
\hline
\textbf{$P_{FA}$} & \textbf{Classifier} & \textbf{Accuracy} & \textbf{F1-score} \\ \hline
\multirow{3}{*}{0.2}
& RF & 0.575 & 0.452 \\ \cline{2-4}
& MLP & 0.580 & 0.461 \\ \cline{2-4}
& \textcolor{black}{CNN} & \textcolor{black}{0.5990} & \textcolor{black}{0.4981} \\ \cline{2-4}
& IF & 0.578 & 0.458 \\ \hline
\multirow{3}{*}{0.4}
& RF & 0.560 & 0.542 \\ \cline{2-4}
& MLP & 0.571 & 0.558 \\ \cline{2-4}
& \textcolor{black}{CNN} & \textcolor{black}{0.5760} & \textcolor{black}{0.5656} \\ \cline{2-4}
& IF & 0.573 & 0.561 \\ \hline
\end{tabular}
\end{table}  
\vspace{-0.5cm}    
\end{tiny}

\subsection{Detection Performance Analysis}\label{Surrogate_Class_1_for_training}
\textcolor{black}{To \textcolor{black}{analyze} the detection capability at Dave, we use 5000 samples per class for training and 1500 samples per class for testing, and implement the aforementioned ML-classifiers on the collected dataset.
For a target probability of false alarm, denoted $P_{\mathrm{FA}}$, the detection threshold is set to the $(1-P_{\mathrm{FA}})$-quantile of the classifier scores obtained from test samples corresponding to unit-energy QPSK symbols. The probability of detection is then computed as the fraction of after-countermeasure test samples exceeding this threshold.
By varying the threshold and computing the corresponding $P_{FA}$ and probability of detection, in Fig. \ref{alpha_9901}, we plot the ROC curves of ML-classifiers for $\alpha\! =\! 0.9901$, along with that of the random classifier.
The ROC curves of ML-classifiers do not exhibit a pronounced bulge away from that of the random classifier.
This observation confirms that at low $P_{\mathrm{FA}}$, high probabilities of detection cannot be achieved, thereby demonstrating that the  LLCRTF scheme cannot be detected with a high-probability  by a data-driven adversary equipped with ML-classifiers. We observe similar \textcolor{black}{behavior} for other values of $\alpha$.
To \textcolor{black}{analyze} detection capability with $\alpha$, in Fig. \ref{RF_alpha},  we plot the ROC curves of RF classifier for different values of $\alpha$.
For a given $P_{\mathrm{FA}}$, the probability of detection increases as 
$\alpha$ moves away from unity owing to the increased energy difference between the unit-energy QPSK symbols and the countermeasure symbols.
We observe similar \textcolor{black}{behavior} for other ML-classifiers.
Finally, Table~\ref{tab:metrics_compact_extended} reports accuracy and F1-scores for different $P_{\mathrm{FA}}$ at $\alpha = 0.9901$. The near-$0.5$ accuracies and low F1-scores align with the ROC results, further confirming that LLCRTF remains covert. We observe a similar trend for other  $\alpha$ and $P_{FA}$.}

\begin{figure}[ht!]
\vspace{-0.1cm}
\begin{center}
\includegraphics[scale=0.12]{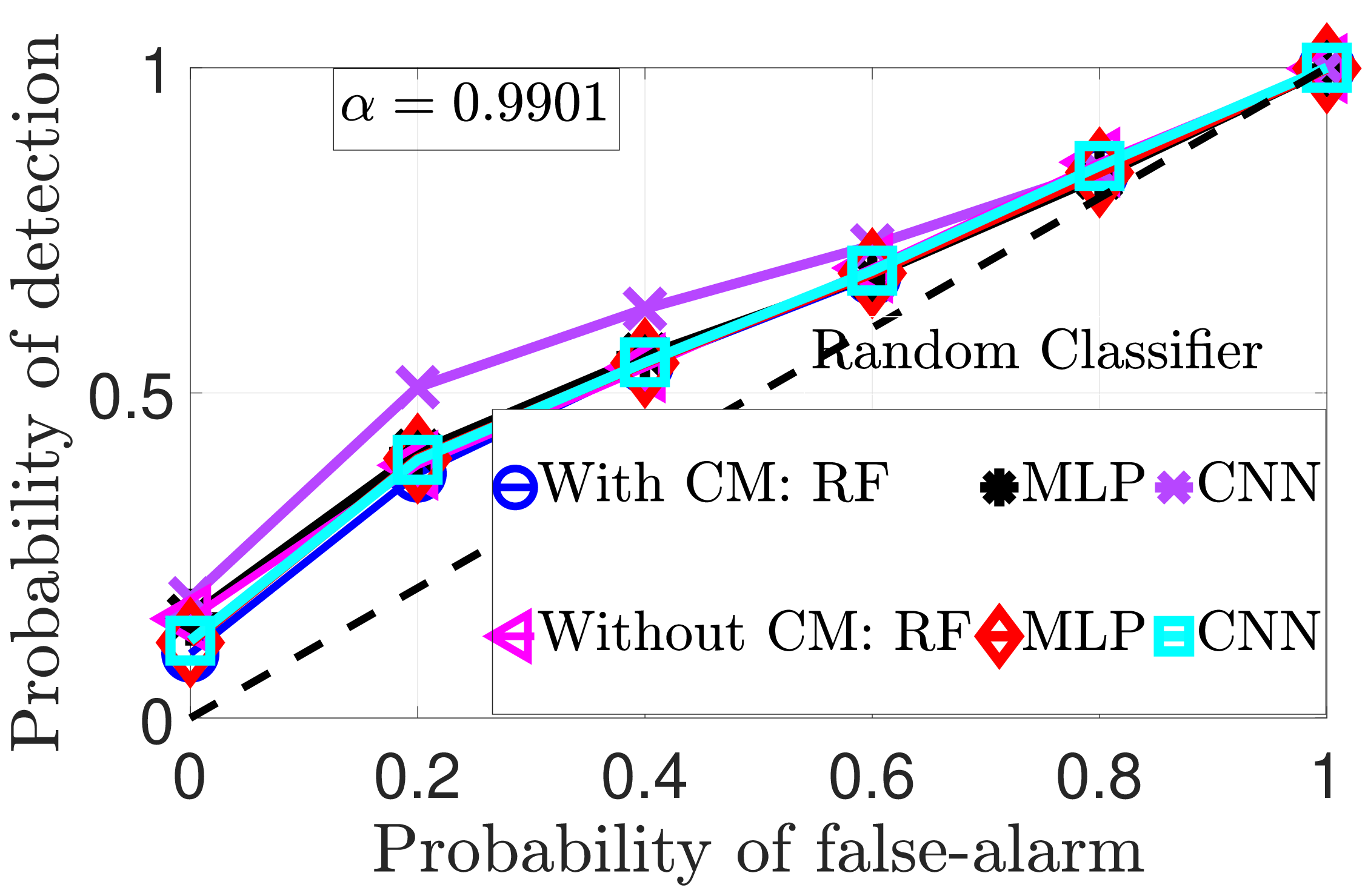}
\vspace{-0.1cm}
\caption{\textcolor{black}{{ROC curves of LLCRTF, under two training settings: with and without countermeasure (CM) samples in training.}}}
\label{CM_as_Training}
\end{center}
\vspace{-0.4cm}
\end{figure}

\subsection{\textcolor{black}{Detection with LLCRTF Samples for Training Class-1}}\label{Section_on_LLCRTF_as_Training}

\textcolor{black}{
Given that the surrogate training approach is assumption-driven and may not accurately capture the signal characteristics of the IQ samples of LLCRTF symbols, we consider an adaptive adversary that observes the post-countermeasure \textcolor{black}{behavior} and collects the IQ samples of LLCRTF for training. Since labeled LLCRTF  symbols are available, the adaptive adversary uses their energies to construct Class-1 and retrain the supervised models, thereby eliminating the need for surrogate Class-1.
For this case, in Fig. \ref{CM_as_Training}, for $\alpha=0.9901$, we plot the ROC curves for different ML-classifiers. In the same figure, for comparison, we also plot the ROC curves for the case when the adversary uses surrogate Class-1 samples for training.
From the figure, we observe that the ROC curves of all ML-classifiers do not bulge away from that of the random classifier for both cases.
Although the CNN exhibits slightly improved detection performance when LLCRTF samples are available for training, high probabilities of detection still cannot be achieved at low $P_{FA}$,
thereby  demonstrating that the LLCRTF scheme cannot be detected with a high-probability  by a data-driven adversary equipped with ML-classifiers, even when their labeled samples are available for training.}

\begin{figure}[t]
\vspace{-0.5cm}
\begin{center}
 \subfloat[]
{
\includegraphics[height=2cm, width=0.32\columnwidth]{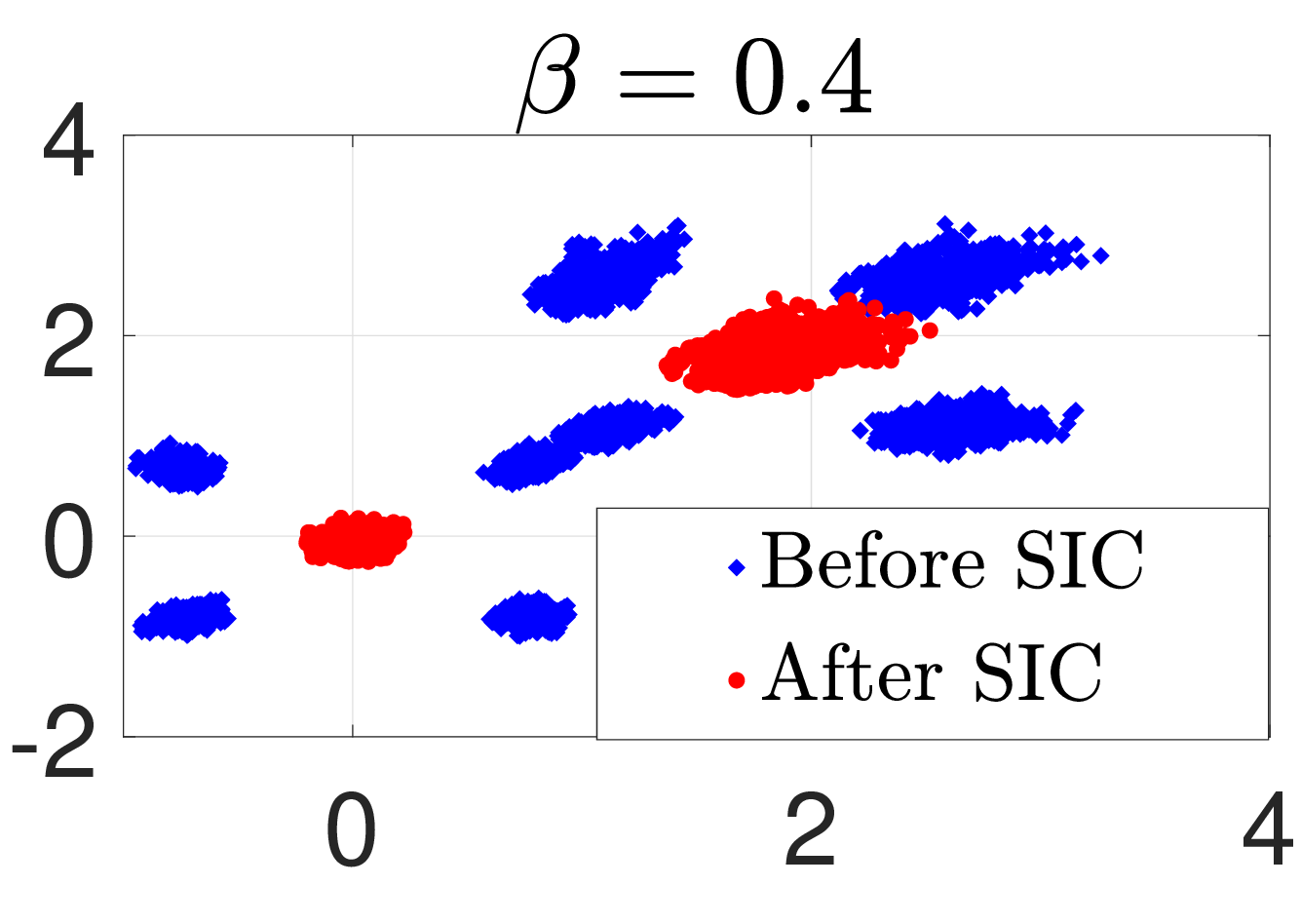}
    \label{b4}}%
         \subfloat[]
     {
\includegraphics[height=2cm, width=0.32\columnwidth]{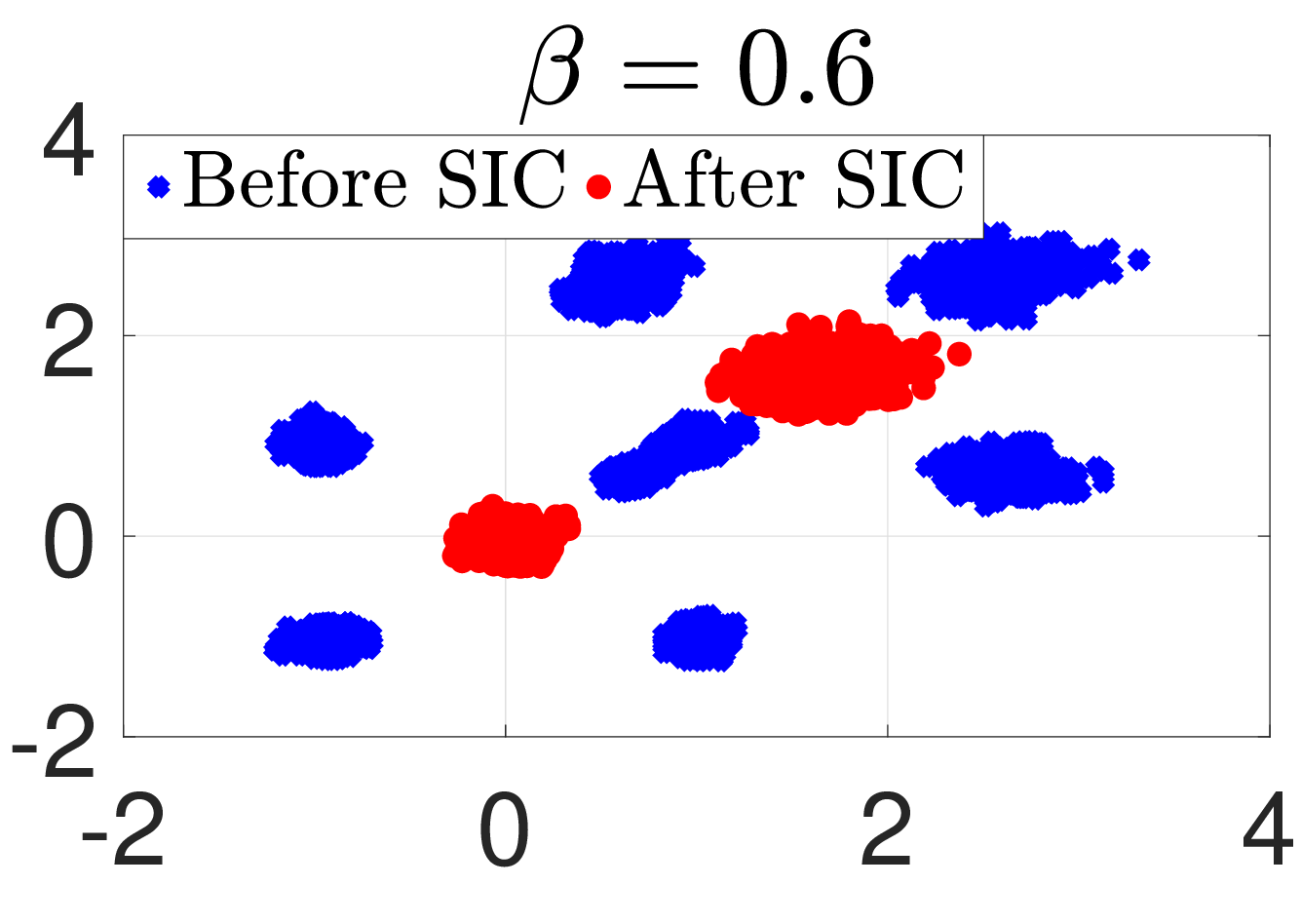}
    \label{b6}}%   
\end{center}
\vspace{-0.7cm}
   \begin{center}
         \subfloat[]
     {
\includegraphics[height=2cm, width=0.32\columnwidth]{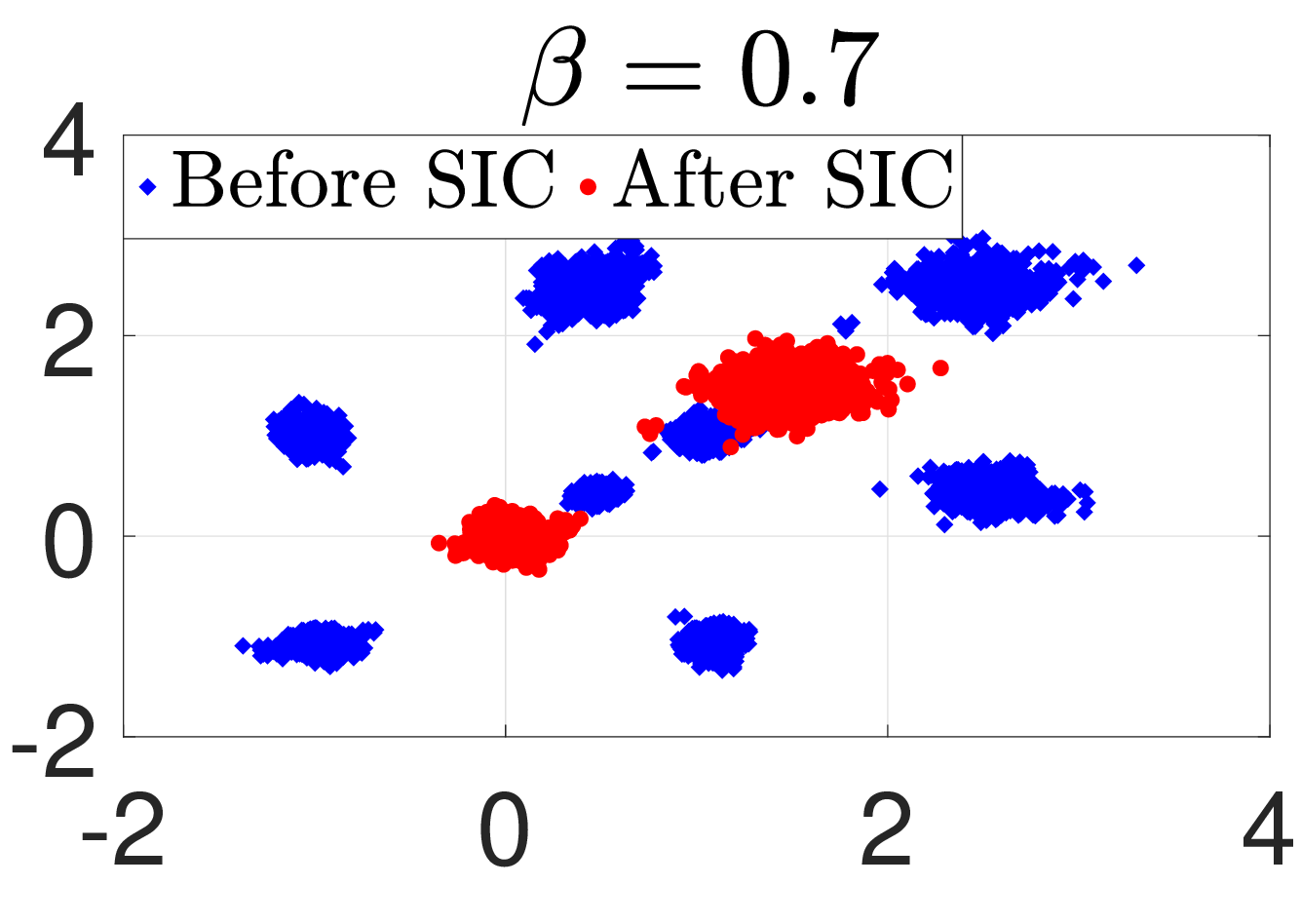}
    \label{b7}}%
    \subfloat[]
{
\includegraphics[height=2cm, width=0.32\columnwidth]{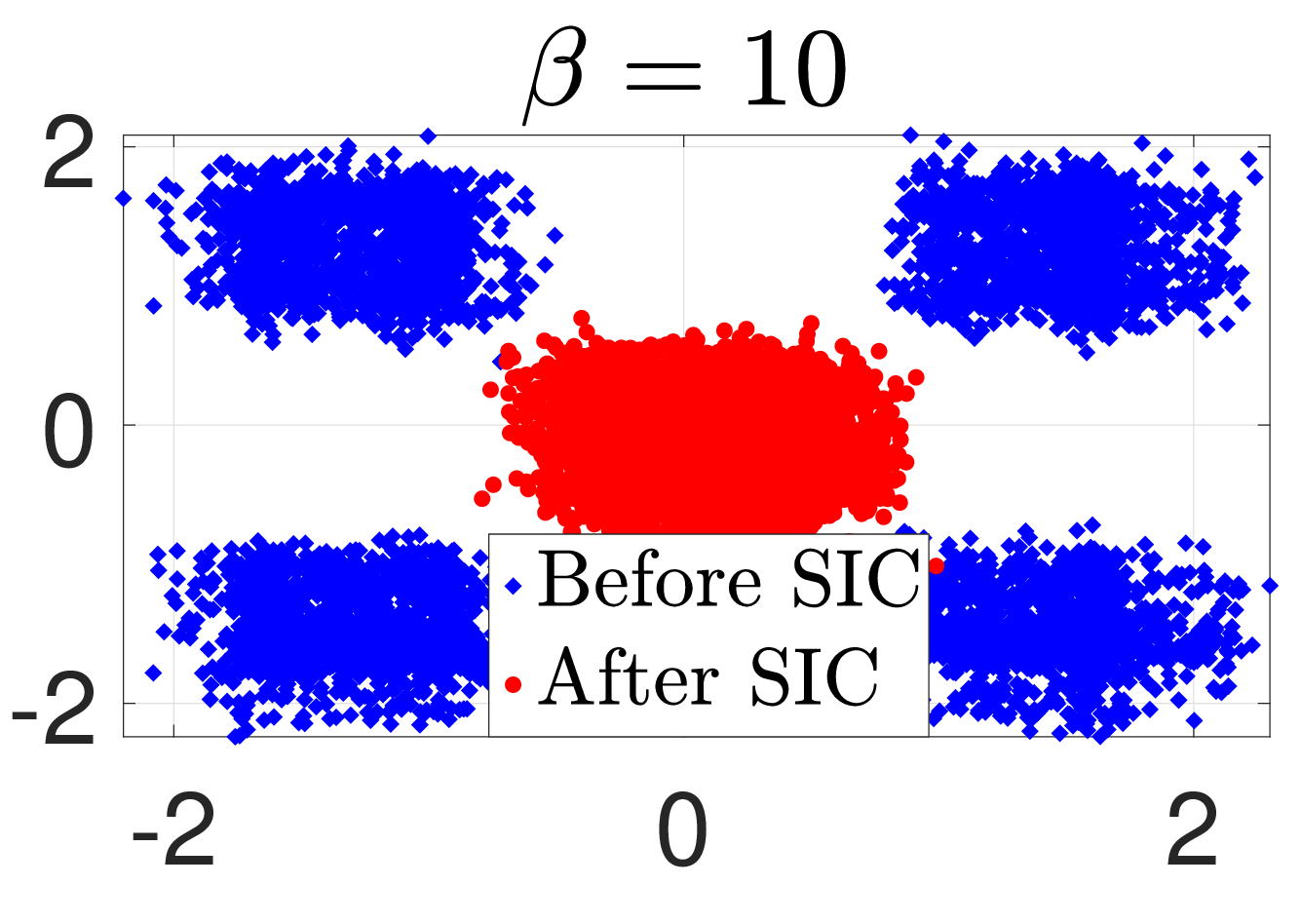}
    \label{b10}}       
   \end{center}
\vspace{-0.3cm}
\caption{\textcolor{black}{Constellation diagrams of the received IQ samples before and after removing SIC for different values of $\beta$.}}\label{CD_beta}
\vspace{-0.2cm}
\end{figure}

\subsection{\textcolor{black}{A Cost-effective Implementation of the DTRTF Scheme}}

\textcolor{black}{The hardware implementation of DTRTF is divided into three phases, wherein Phase-1 corresponds to transmission in time-slot 1, where Alice and Charlie cooperatively transmit their symbols.
In Phase-2, Charlie removes his own transmitted symbol from the received symbol before decoding Alice's bits transmitted in time-slot 1.
Phase-3 corresponds to the transmission in time-slot 2, wherein Charlie incorporates Alice's bits into his transmitted  symbols.
Given that the \textcolor{black}{signaling} schemes of DTRTF for Phase-1 and Phase-3 are similar to LLCRTF, we only implement Phase-2. Particularly, in Phase-2.A, we emulate the signal conditions experienced at Charlie's full-duplex receiver by constructing the received waveform using the desired signal and a controlled self-interference component. Subsequently, in Phase-2.B, Charlie performs SIC using knowledge of his transmitted signal. Using this setup, we validate the performance of DTRTF without requiring dedicated RF components for analog cancellation.}

\textcolor{black}{In Phase-2.A, we establish communication between two Adalm-Pluto devices, where one acts as Alice (transmitter) and the other acts as Charlie (receiver), with Alice transmitting her bits using OOK. 
Subsequently, instead of physically transmitting and receiving simultaneously at Charlie, we embed the self-interference component directly into the transmitted waveform.
This is done by generating Charlie’s QPSK signal using a pseudorandom generator, whose seed is later used at the receiver for SIC purposes.
Therefore, the transmitted waveform consists of a superposition of Alice’s signal and a scaled QPSK signal, where the scaling parameter, denoted by $\beta$, emulates the residual self-interference after analog cancellation.
}

\textcolor{black}{In Phase-2.B, Charlie employs the seed used to generate the QPSK signal to perform SIC. By varying $\beta$ at the transmitter, we emulate two practical regimes at the receiver, namely effective and poor analog-cancellation, corresponding to small and large $\beta$, respectively. In the first regime, AGC keeps the received signal within the linear operating region of the RF front-end, thereby enabling accurate digital cancellation. Conversely, in the second regime, strong self-interference causes AGC to drive the RF front-end towards nonlinear operation and saturation, thereby degrading digital cancellation performance.}

\textcolor{black}{We conduct experiments for $\beta\!=\! 0.4, 0.6, 0.7, 10$, with average interference powers of 0.16, 0.36, 0.49 and 100, respectively, and average signal power of 0.5.
Fig. \ref{CD_beta} shows the constellation diagrams of received IQ samples, before and after SIC, for both regimes.
For first regime, after SIC, Figs. \ref{b4}-\ref{b7} show two distinct clusters for the two bits (in red). In contrast, for second regime, Fig. \ref{b10} shows a single overlapping cluster, as  noise power exceeds  signal power. Thus, reliable decoding of  bits is only possible in first regime.
The decoded bits at Charlie are stored for incorporation in Phase 3.}
\textcolor{black}{We then feed the IQ samples of DTRTF into  ML-classifiers, and observe that similar to LLCRTF, for DTRTF, at low $P_{FA}$, high probabilities of detection cannot be achieved, thereby demonstrating that DTRTF cannot be detected with a high-probability  by a data-driven adversary equipped with ML-classifiers.}

\section{Future Research Directions}\label{SS}
The LLCRTF scheme performs poorer than DTRTF due to the absence of an additional reliability layer, which is impractical under its low-latency constraint. Under tighter latency requirements, outer codes or multiple-receive antennas at Bob may enhance performance.
Another direction is to extend the reliability and covertness analysis to $M$-QAM and to multiple-receive-antenna configurations.
From the channel standpoint, our strategies apply when the helper’s link is quasi-static with arbitrary coherence time, provided Bob can perform coherent detection. For fast-fading channels with one-slot coherence time, non-coherent modulation is required, motivating countermeasure design for that regime.
\textcolor{black}{We point out that when the jamming energy results in a probability of error of less than $0.5$ on the $f_{AB}$ band, there may exist values of $\alpha$ for which the proposed strategies become detrimental compared to the jamming attack, and therefore, a possible future direction is to characterize the corresponding break-even point of $\alpha$ in terms of the probability of error at Alice.}

\bibliographystyle{IEEEtran}

\bibliography{ref.bib}

\end{document}